\documentclass[useAMS,usenatbib]{mn2e}
\usepackage{graphicx}
\usepackage{graphics,float}                         
\usepackage{epsf}
\usepackage{amsmath}                                
\usepackage{amssymb}
\usepackage{picinpar}                               
\usepackage{relsize,setspace,subfigure}
\usepackage{tabularx}
\usepackage[innercaption]{sidecap}                  
\usepackage[figuresright]{rotating}
\usepackage{url}
\usepackage{aas_macros}
\usepackage{natbib}

\onecolumn

\title[Extended Void Merging Tree Algorithm]{Extended Void Merging Tree Algorithm for Self-Similar Models}
\author[Russell]{Esra Russell$^{1,2}$\thanks{E-mail: esrarussell@iyte.edu.tr}
\\
$^{1}$Kapteyn Astronomical Institute, University of Groningen, PO
Box 800, 9700 AM Groningen, The Netherlands.
\\
$^{2}$Department of Mathematics, Izmir Institute of Technology, 35430 Gulbahcekoyu/Urla, Izmir, Turkey.}

\begin{document}
\date{Accepted .... Received ...; in original form ...}

\pagerange{\pageref{firstpage}--\pageref{lastpage}} \pubyear{2013}

\maketitle

\label{firstpage}

\begin{abstract}
In hierarchical evolution, voids exhibit two different behaviors related with their surroundings and
environments, they can merge or collapse. These two different types of void processes can be described by the
two-barrier excursion set formalism based on Brownian random walks. In this study, the analytical approximate
description of the growing void merging algorithm is extended by taking into account the
contributions of voids that are embedded into overdense region(s) which are destined to vanish due to
gravitational collapse. Following this, to construct a realistic void merging model that consists of both
collapse and merging processes, the two-barrier excursion set formalism of the void population is used. Assuming
spherical voids in the Einstein de Sitter Universe, the void merging algorithm which allows us to consider the
two main processes of void hierarchy in one formalism is constructed. In addition to this, the merger rates,
void survival probabilities, void size distributions in terms of the collapse barrier and finally, the void
merging tree algorithm in the self-similar models are defined and derived.
\end{abstract}


\section{Introduction}
The present-day Universe shows a complex pattern of structures, called the Cosmic Web \citep{bond}. Galaxy
redshift surveys find that voids are dominant features of the large scale structure
\citep{kir,gehu,dac,shect,ei,plib,joveer,wk,vdWbond,PhDWeygaert,mauro,strauss,fisher,saunders,2004MNRAS.355..747J,rienerwin}.
Differing from overdense haloes, voids evolve out of the underdensities in the primordial gravitational
fluctuations. Due to their internal weak gravity driven by a negative density profile, they expand with respect
to the background Universe rather than collapse. According to the void based description of the cellular
morphology of the Cosmic Web, voids are important components of hierarchical structure formation since the
growth of the large scale structure may be driven by the expansion of voids
\citep{icke,PhDWeygaert,weygaert2003}. In addition to this, numerical calculations and
$N$-body simulations show that the voids tend to be spherical in time
\citep{centme,fuji,be85,regge,dubinski1993,wk,colb,Tavasoli2013} and this tendency is explained by the
Bubble Theorem \citep{icke}.

Although there are differences between haloes and voids due to their origins, voids evolve hierarchically
\citep{regge,wk,gott,sw,colb,sw,paranjape,sutter,2013MNRAS.428.3409A,sutter2013} in a similar fashion to dark
matter haloes. An important step towards understanding void hierarchy is provided by the void simulations of
\cite{dubinski1993} that are based on the analytical model of isolated
spherical voids \citep{blumenthal}. Another significant contribution is the work of \cite{sahni1994}. They
applied the Lagrangian adhesion model description to the evolution of voids, resulting in revealing the
unfolding void evolution. In their study, they found a strong correlation between the
size of voids and the primordial gravitational potential at their centre. Another important result is shown in
\cite{sahni1994,Ceccarelli,sutter2013}; that large voids have more
substructures than small size voids, and as voids grow older they become progressively emptier indicating less
substructures within them. This result that voids have substructures is in agreement with
$N$-body simulations, indicating that the interiors of voids are filled
with subvoids, galaxies and even filaments \citep{wk,gott,mathis,benson}. \cite{sutter2013} point out
that this tendency of a large void becoming emptier or erasing the substructures in itself is strongly related
with choosing the tracking density in the simulations. As a result, a lower tracking density indicates less
substructures in a void. The effect of these substructures on void evolution is important as they
interact with their surroundings and have an impact on void evolution. Based on these theoretical and numerical
results, \cite{sw} point out that contrary to overdense regions, the evolution of voids is dictated by two
processes depending on their environment: they can merge into larger voids and/or voids embedded in
overdense regions can collapse. \cite{Ceccarelli} find a similar
behavior as \cite{sw} in numerical simulations and classify voids into S (small)-Type and R (raising)-Type.
\cite{sw} suggest that the hierarchical evolution of these two different types of void processes can be
described by the two-barrier excursion set formalism. Representing merging and collapsing dynamical behaviors of
void populations, these two barriers encapsulate what it takes to let a void merge
and collapse. In linear theory, the merging of a spherical void is represented by a constant critical density,
$\delta_{v}=-2.81$ in the EdS Universe, while the collapse of a spherical void is given by a fixed critical
density value, $\delta_{c}=1.686$. Later on, \cite{piran2006} introduce an analytical model of void size in the
galaxy distribution by applying the EPS formalism. A numerical study of void hierarchy based on the Hierarchical
Cosmic Spine method is introduced by \cite{2013MNRAS.428.3409A}. \cite{2013MNRAS.428.3409A} apply this method
to cosmological $N$-body simulation in order to analyze voids in hierarchical space. In their study they stress
out the importance of small scale voids in order to obtain the dynamics around a halo. Based on the study of
\cite{sw}, \cite{eszrarussell} construct a void merging tree algorithm of the growing spherical void population,
based on the one-barrier excursion set theory as an adaption of the halo merging algorithm of \cite{lace}.
Following up on \cite{sw}, \cite{eszrarussell} obtain an
approximate analytical description of the merging tree based on excursion set theory, by assuming only growing
spherical voids in the EdS Universe. This formalism leads to a considerably modified view of the
evolution of voids. However, to define a more realistic void hierarchy model, it is essential
to consider the dominant environmental influence on the evolution of voids by following \cite{sw}. Environmental
influence strongly affects void sizes and their distribution. Following this, \cite{sutter2013} provide a
general framework in order to make connections between dark matter voids and galaxy voids from the same
cosmological simulation by using the hierarchical tree structure of voids.

The different aspects of void populations have been discussed by observational and numerical
studies
\citep{regge,dubinski1993,wk,El-Ad1997,plib,gott,Hoyle02,benson,hoylevogley2004,colb,tk2008,tully2008,2008MNRAS.387..933C,kraan,tinker2009,Lavaux2010,kreckel2011,Lavaux2012,Kreckel2012AJ,Panvogeley2012,sutter,Bos2012,2012ApJ...761..187S,Beygu2013,
2013MNRAS.428.3409A,Tavasoli2013}.
Today we know that the void size range is approximately $5-135\quad h^{-1}$Mpc \citep{sutter}. However, there
are some studies claiming that voids can have very small sizes, $0.7 - 3.5\quad h^{-1}$Mpc
\citep{Karachentsev2004,tk2008,TikhonovKlypin2009}. Based on these studies, \cite{tikhonovstefan2009} compare
the observed spectrum of minivoids in the local volume within an $8$ Mpc size sphere in the Local Group with
the spectrum of minivoids determined from the simulations in the cold dark matter (CDM)
and warm dark matter (WDM) models. They show the difference between the observed spectra of minivoids in the
$\Lambda$WDM and $\Lambda$CDM models. \cite{Viel2008} make a preliminary attempt to link the population of voids
in the transmitted Lyman-$\alpha$ flux to the underlying gas density,
temperature and dark matter density. The use of Lyman-$\alpha$ high resolution spectra is important in the sense
that it explores a new regime at scales, redshifts and densities which are currently not probed by other
observables: the scales are of the order of a few to tens of Mpc, the redshift range is between $z = 2$ and
$4$, while the densities are around the mean density.

There are more attempts to understand the dynamical, thermal and chemical evolution of the void population, and
the interplay between galaxies and the intergalactic medium \citep{Shang2007,D'Aloisio2007}. Motivated by the
empirical evidence for significant preheating of at least parts of the intergalactic medium at $z\sim 3$,
\cite{Shang2007} make a simple model for the spatial distribution of preheated regions. The model assumes
spherical ionized bubbles around collapsed dark matter haloes and allows these spheres to merge into larger
superbubbles. The number of voids that such ionized bubbles would produce in Lyman-$\alpha$ absorption spectra
of background quasars is predicted. \cite{D'Aloisio2007} present analytic estimates of galaxy
void sizes at redshifts $z\sim 5-10$ using the excursion set formalism. Another interesting study on void
excursions addresses the issue of void formation in modified gravity models. Some of these involve a scalar
field that couples to matter and introduces a fifth force. This force leads to a universal
enhancement of gravity \citep{FarrarP2004,GubserP2004,GubserP2004b},
\citep{Farrar2007,Brookfield2008,Hellwing2009,Clampitt2012}. \cite{Clampitt2012} investigate the fifth force in
voids in chameleon models by using void statistics based on the excursion set formalism. They point out that
driven by the outward pointing fifth force, individual voids in chameleon models expand faster and grow larger
than voids in the $\Lambda$CDM Universe.

In this study, a model is formulated to construct a merger tree formalism consisting of collapse and merging
processes as an extension of the growing void merging tree of \cite{eszrarussell} by treating the void in cloud
problem. In this model, the size of voids by using recent void catalogs is defined and a void merging tree
formalism is obtained taking into account the merging process of void populations as well as the collapse
process. To do this, the Lacey and Cole's merging tree algorithm \citep{lace} of dark matter haloes in the EdS
universe for spherical voids is adopted. Since the Lacey and Cole's merging tree algorithm \citep{lace} of dark
matter haloes provides an approximated dark matter solutions, in this study we only focus on the self-similar
spectra to construct a void merging algorithm. As a result, the
new void merging tree algorithm is extended to the two-barrier excursion set formalism in order to describe void
merging and collapse in one model in the self-similar models.

\section{Cosmology}
In this section, the normalization of the cosmological models is explained. These models are used in the
two-barrier excursion set to obtain the quantities used for the void merging tree. Here,
the power law power spectra are taken into account with different spectral index and these spectra are
approximated to a favoured $\Lambda$CDM by following \cite{sw}. The normalization of the spectra in the self-similar models
is relatively easier to calculate given the form of the power spectrum,

\begin{eqnarray}
P(k)\approx k^{n},
\end{eqnarray}
\noindent
where $n$ refers to the spectral index or the slope of the power spectrum. The variance of the self-similar
spectra as a function of mass, volume and size is given by,

\begin{eqnarray}
\sigma^2(M)=S(M)={{\delta^{2}}_{v}}{\left(\frac{M}{M_{*}}\right)^{-\alpha}}={\delta^{2}_{v}}{\left(\frac{V}{V_{*}}\right)^{-\alpha}}
={{\delta^2}_{v}}\left(\frac{R}{R_{*}}\right)^{-3\alpha},
\label{relationchd}
\end{eqnarray}
\noindent
in which $\alpha$ is defined as $(n+3)/3$. Here, $M_{*}$, $V_{*}$ and $R_{*}$ are the characteristic mass,
volume and radius. The characteristic mass of the self-similar spectra is given by,

\begin{eqnarray}
M_{*}(z)=\left(\frac{\sigma_{8}}{|\delta_{v}|}\right)^{6/n+3}.
\end{eqnarray}
\noindent
To normalize the self-similar power spectra, the mass function in mass variance in equation
(\ref{relationchd}) is chosen as $M=M_{*}$. Hence, the characteristic mass variance is equal to the linear
underdensity
$\sigma^{2}(M_{*})\sim |\delta_{v}|$. By following this definition of characteristic mass in equation
(\ref{relationchd}), the comoving radius $R$ of a void region for self-similar models is given by \cite{sw},

\begin{eqnarray}
\label{voidcharacteristicsizeCH4}
R =R_{*}\left(\frac{\sigma_{8}}{|\delta_{v}|}\right)^{2/n+3}.
\end{eqnarray}
\noindent
Note that in our calculations, the characteristic void size is chosen as $R_{*}=8$ $h^{-1}$Mpc in the
self-similar models. This allows us to fix the variance $\sigma$ at the scale $M=M_{*}$. Then, the void
excursion set of the toy models is obtained in terms of physical scales.

In this study, spherical contraction and expansion representing the collapse and merging void processes are
taken into account. These dynamical
processes are described with respect to linear theory. According to linear theory, an object of linear density
$\delta$ will collapse or shellcross
at redshift $z$ when its value dominates $\delta_{c}(z)$ or $\delta_{v}(z)$,

\begin{eqnarray}
\delta_{c}(z)&=& \frac{{\delta_{c}(0)}}{D(z)}=\delta_{c}(0)\left(1+z\right)\label{halodensitycontrast}\\
\delta_{v}(z)&=& \frac{|{\delta_{v}(0)}|}{D(z)}=|\delta_{v}(0)|\left(1+z\right),
\label{voiddensitycontrast}
\end{eqnarray}
\noindent
where $\delta_{c}(z=0)=1.686$ and $\delta_{v}(z=0)=-2.81$ are the linearly extrapolated densities in the linear
regime. Note that the growth factor $D(z)$ encapsulates the geometry of the collapsing/expanding objects to
denote the linear density perturbation. The growth
factor normalized to the present-day Universe $D(z=0)=1$. In linear theory, the growth factor $D(z)$ has the
same form for overdense and underdense regions, which is $D(z)= 1/\left(1+z\right)$ in the EdS Universe.
Here, by taking into account the fact that the shellcrossing and collapse barriers are only redshift
dependent in the EdS Universe as defined in equations (\ref{halodensitycontrast}) and
(\ref{voiddensitycontrast}), the linear densities are used as
time variables following the algorithm of \cite{lace}.

\section{Two Barrier EPS Formalism}
\begin{figure}
\centering
\includegraphics[scale=0.7]{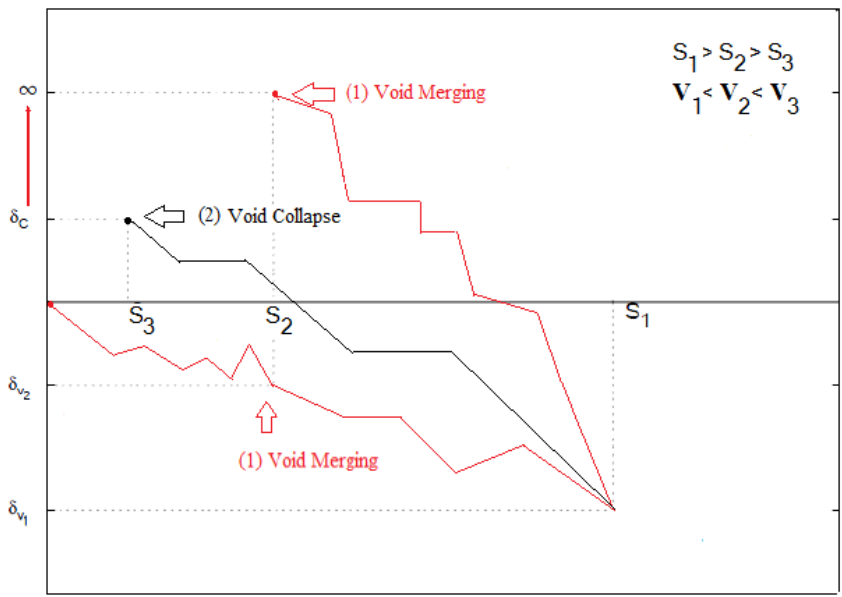}
\caption{Illustration of the excursion set theory formalism representing two important void processes in terms
of volume scales and two barriers; collapse $\delta_{c}$ and shellcrossing $\delta_{v}$. The void merging and
collapse processes are indicated by red and black lines.}
\label{fig:voidprocesses}
\end{figure}

The analytical evaluation of the two-barrier random walk problem takes into account a distribution function
$f_{v}(M)$ on a mass scale $M$ \citep{sw},

\begin{eqnarray}\label{mff}
f_{v}(M) dM \approx
\frac{1}{\sqrt{2\pi}}\frac{{\nu_{v}}}{\sigma^2}\exp\left[-\frac{\nu^{2}_{v}}{2}\right]
\exp\left[-\frac{|\delta_{v}|}{\delta_{c}}\frac{\Upsilon^2}{4{\nu^{2}_{v}}}-\frac{2\Upsilon^4}{\nu^{4}_{v}}\right]dM,
\label{massfraction2a}
\end{eqnarray}
\noindent
corresponding to a fractional underdensity function $\nu_{v}(M)$ which is defined as,

\begin{eqnarray}
\nu_{v}(M)\equiv\frac{\delta_{v}}{\sigma(M)}= \frac{\delta_{v}}{\sqrt{S}},
\label{voiddensityparametera}
\end{eqnarray}
\noindent
where $\delta_{v}$ is the void threshold density, the mass variance function is $\sigma(M)$ and the mass scale
function is $S$. In equation
(\ref{massfraction2a})
$\Upsilon$ is the void and cloud parameter. It is defined as,
\begin{eqnarray}
\Upsilon\equiv\frac{|\delta_{v}|}{\left(\delta_{c}+|\delta_{v}|\right)}.
\label{Dparametera}
\end{eqnarray}
\noindent
The void and cloud parameter $\Upsilon$ has a key importance, since it shows the effect of the overdense
regions/haloes on the void population. Therefore,
the mass fraction equation (\ref{massfraction2a}) represents the hierarchical evolution of voids that are
dominated by two different behaviors: merging or
collapsing. These two behaviors are represented by two barriers in the excursion set formalism; the merging
barrier is defined by the linear collapse
$\delta_{c}$ and shellcrossing $\delta_{v}$ densities. Since void and cloud parameter $\Upsilon$ is a
time-dependent function and consists of the ratio of
the barriers, it is rearranged as,

\begin{eqnarray}
\label{betaa}
\Upsilon=\frac{1}{\gamma(z_{c},z_{v})+1},\phantom{a}
\gamma(z_{c},z_{v})\equiv\frac{\delta_{c}(z_{c})}{|\delta_{v}(z_{v})|},\phantom{a}\gamma(z_{c},z_{v})=\frac{\delta_{c0}}{|\delta_{v0}|}\frac{1+z_{c}}{1+z_{v}},
\end{eqnarray}
\noindent
and a new parameter $\gamma$ is defined which is the barrier ratio $\delta_{c}/|\delta_{v}|$. For a given
collapse redshift $z_{c}$ when collapse happens
and a shell crossing redshift $z_{v}$ when merging occurs, the parameter $\gamma$ becomes constant. Note that
the mass fraction function can be
transformed into mass, volume or size scales by following \cite{sw},

\begin{eqnarray}f_{v}(S)d S\propto f_{v}(M)d M = f_{v}(V)d V\propto f_{v}(R)d R.
\nonumber
\end{eqnarray}
\noindent
This leads us to construct void evolution in terms of volume $S(V)$ and size scales $S(R)$ implicitly.
Considering this and substituting the barrier ratio by using equation (\ref{betaa}) in the two-barrier mass
distribution equation (\ref{massfraction2a}), the distribution function is obtained in terms of volume/size
scale as follows,

\small
\begin{eqnarray}
f_{v}(S) dS \approx \frac{1}{\sqrt{2\pi}}\frac{{\delta_{v}}}{S^{3/2}}\exp\left(-\frac{\delta^2_{v}}{2S}\right)
\exp\left(-\frac{1}{4}\frac{1}{\gamma}\frac{1}{\left(1+\gamma\right)^2}\frac{S}{\delta^2_{v}}-2\frac{1}{\left(1+\gamma\right)^4}
\frac{S^{2}}{\delta^4_{v}}\right)dS.
\label{massfractionSchap4}
\end{eqnarray}
\normalsize
\noindent
The volume/size distribution equation (\ref{massfractionSchap4}) shows the hierarchical evolution of voids that
are dominated by
two processes; merging and collapsing in terms of the EPS formalism. Fig. \ref{fig:voidprocesses} illustrates
the two main void processes in the context
of the EPS formalism. As is seen from the figure, a spherical void with volume scale $S_{1}=S(V_{1})$ starts its
evolution at barrier $\delta_{v_{1}}$ and
its volume grows due to a merging event when the random walk of the density function crosses a new barrier
$\delta_{v_{2}}$ with volume scale $S_{2}=S(V_{2})$
($V_{2} > V_{1}$). At this point, the void processes are classified into two main groups. The first one is the
void merging/growing gradually.
According to this, there are two possible scenarios to explain a gradual merging event: i) if a void is not
embedded in an overdense region, it will merge
gradually (lower red line in Fig. \ref{fig:voidprocesses}), or an embedded void is large enough that it is not
affected by collapse regions. This indicates
that in the context of the two-barrier formalism, the collapse barrier is larger than the shell crossing barrier
($\delta_{c}\gg \delta_{v}$, which leads
to the collapse barrier moving to $\delta_{c}\rightarrow\infty$) (upper red line in Fig.
\ref{fig:voidprocesses}). In Fig. \ref{fig:voidprocesses}, the
second void process is void collapse. The random walk of voids that are of a relatively small size compared to
their large volume counterparts, merge
until reaching the collapse barrier $\delta_{c}$ and when they reach this barrier they are squeezed at overdense
boundaries and collapse under
gravitational collapse (black lines).

\begin{figure}
\centering
\includegraphics[width=0.65\textwidth]{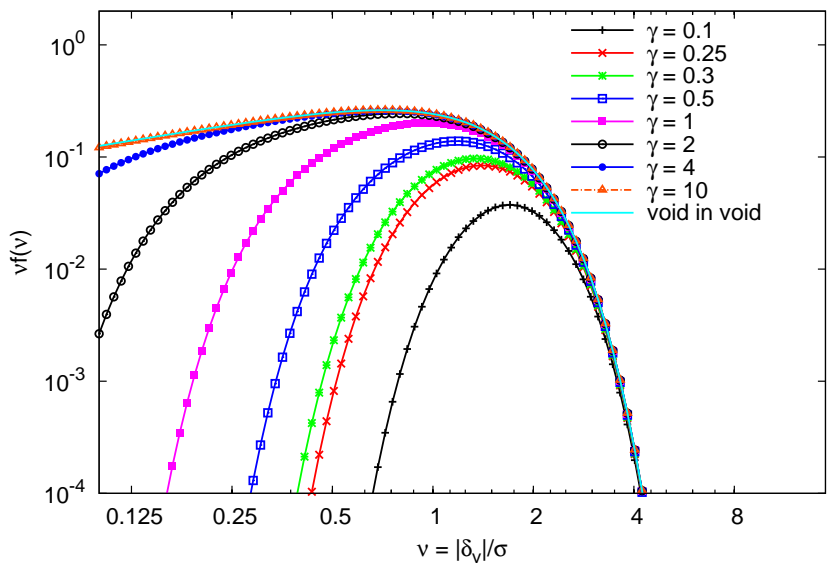}
\caption{The scaled fraction of the void population in terms of different barrier ratios $\gamma$.
When the ratio increases $\delta_{c}\gg\delta_{v}$, the distribution turns into the void in void distribution
(most upper line) in the limit of $\gamma=10$. When the ratio value reaches $\gamma=0.5$ or higher values than
this,
the distribution is well peaked around the characteristic mass $\nu=1$ ($\sigma=|\delta_{v}|$).}
\label{fig:voidcriteriagamma}
\end{figure}

These two evolutionary paths can be described in terms of the barrier ratio or void in cloud parameter since
these parameters have the key importance of indicating which process is dominant over the other one. When the
value of the overdense barrier $\delta_{c}$
becomes higher than the underdense barrier $\delta_{c}\gg \delta_{v}$ or $\delta_{c}\rightarrow\infty$, the
barrier ratio becomes infinite, which
makes the void and cloud parameter zero, $\Upsilon\rightarrow 0$. Due to this fact,
the second exponential term in equation (\ref{massfractionSchap4}) disappears. This means that the contribution
of embedded/minor voids
in the distribution equation (\ref{massfractionSchap4}) is unimportant. At this limit, the two-barrier
distribution reduces to a single barrier at
$\delta_{v}$ and then the abundance of voids is given by the void in void process which is analogous to the
cloud in cloud process
\citep{sw,Damico2011,eszrarussell}. Fig. \ref{fig:voidcriteriagamma}
shows the distribution function of the scaled density with respect to different barrier ratio parameters
$\gamma$.
As is seen for larger values $\gamma\gg 0.5$, the mass fraction function approach to the void in void problem of
the one-barrier excursion set, indicates a gradually merging void process. However the void distribution
function with the barrier ratio $\gamma\leq 0.5$
has more contributions from the embedded voids, which leads to the void in cloud problem. In this case, the
distribution function shows two cut-offs
at small and large fractional underdensities $\nu$, in other words at small and large void volumes, indicating
embedded voids are unlikely to have volumes larger or smaller than these cut-off values. \cite{sw} point out
that the distribution function is well peaked around $v\sim 1$
($\sigma\sim|\delta_{v}|$)
for the barrier ratio $\gamma\geq 0.25$. They also mention that the distribution function is not correct for the
barrier ratio $\gamma\leq 0.25$.

These basic descriptions and formulations frame the centre of this paper. Here, the main goal is to construct a
void merging algorithm that takes into account the two void processes into one hierarchical model by using the
two-barrier volume/size distribution function of \cite{sw} based on the excursion set formalism.

Before giving the details of the void merging algorithm based on the two-barrier excursion set, it is crucial to
treat the void in cloud problem in the
context of excursion set theory. The problem of the void in void process is based on the derivation of the EPS
formalism. The EPS formalism is a random
walk (Brownian motion) excursion process that is conditioned to be positive $\delta >0$ in terms of increasing
scales $S$. Here, a problem arises in the void in cloud process, in that random trajectories of this process
change from positive $\delta > 0$ to negative $\delta < 0$ values with increasing scales
$S$. In other words, random trajectories of a void in cloud process make their first crossing at a negative
barrier, then by crossing $\delta=0$, reach the positive collapse barrier (see void collapse random walk in
Fig.\ref{fig:voidprocesses}).

In the following subsection, a simple method is introduced based on the idea of constructing a sequence of
random walks converging almost entirely to a
Brownian motion \citep{philippe}. This simple method allows us to treat the complex evolution of the void
processes in the EPS theory based on the scaled
void distribution function.

\subsection{Method to Treat Void in Cloud Problem}
Here, the definition of Brownian motion and its classification are briefly introduced. Then, the interpretation
of this classification is shown  in terms of the EPS formalism. After defining the random walk property of the
void in cloud, the details of the method and then application of this simple
method to the two-barrier void distribution are given in order to obtain a merging algorithm of void populations
based on the halo merging algorithm of \cite[][hereafter LC93]{lace}.

\cite{presc} derive an analytical formalism to infer the number density of collapsed objects at a given redshift
and mass interval based on combining
Gaussian statistics of the linearly extrapolated density field with the non-linear evolution described by the
spherical model. However, their formalism
indicates a problem that is named as the cloud in cloud problem since they did not take into account low density
areas representing small embedded
structures. \cite{bo} propose a solution to this problem by taking into account the probability that a
subsequent filtering of larger scales results
in having linear density contrast larger than the collapse barrier $\delta > \delta_{c}$ at some point.
\cite{bo} identify the mass fraction of matter
in virilized objects with mass greater than $M$ in which the initial density contrast lies above a critical
overdensity when smoothed on some filter of
radius greater than or equal to $R_{f}(M)$. The mass density function is then given by the rate of first
upcrossing of the critical overdensity level as
one decreases $R_{f}$ at a constant position $R$ \citep{bo}. The shape of the mass function depends on the
choice of filter function. The simplest case
is sharp-$k$ space filtering, in which the field performs a \emph{Brownian random walk} as each increment to
$\delta(S)$ when $S$ is increased, which comes
from a new set of Fourier modes in a thin spherical shell in $k$-space. Thus for a Gaussian random field it is
not correlated with any of these previous
steps. As a consequence of this, these trajectories $\delta(S)$ are governed by a simple diffusion equation in
which $\delta(S)$ increases with $S$.

\subsection{Brownian Random Walk Characteristics}
\begin{figure}
\centering
\includegraphics[width=0.65\textwidth]{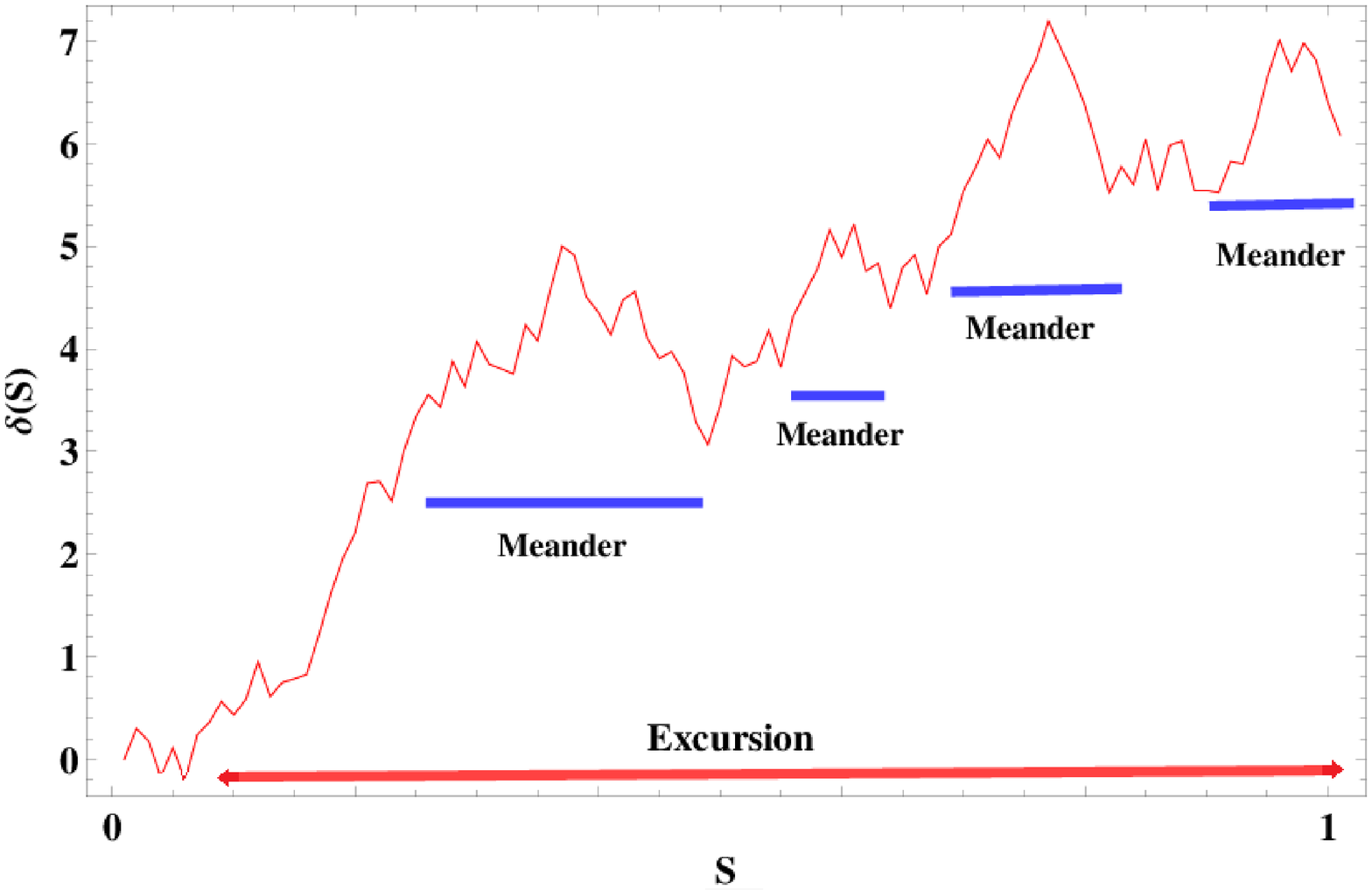}\\
\includegraphics[width=0.65\textwidth]{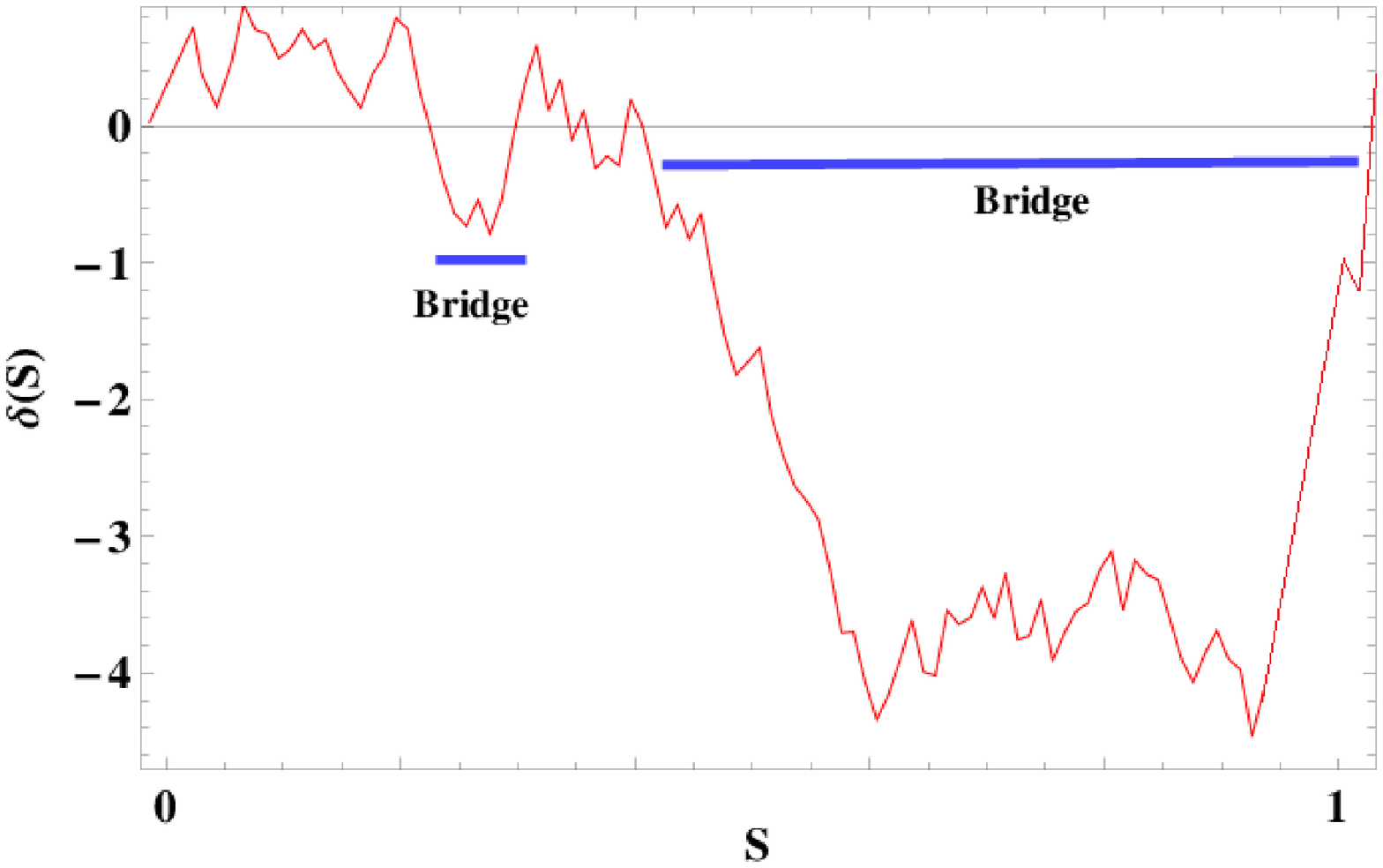}
\caption{Three characteristics of the Brownian random walk: excursion, meander (upper) and bridge (lower).
(Upper) The excursion characteristic of
Brownian motion is a positive definite random walk. It is shown in red and it increases in each increment
$\delta(S_{1})< \delta(S_{1}-S_{2})$. Meander
is always a positive definite local maximum in the positive definite excursion. However a Brownian bridge has
negative values ($\delta (S) < 0$) for
positive definite steps ($S > 0$). (Lower) Brownian bridges are defined as minima in a Brownian random walk.
Note that the excursion characteristic of
random walks can be described as a positive definite bridge.}
\label{fig:walks}
\end{figure}

Here, the definition and properties of a Brownian random walk are given in terms of the EPS formalism:

\emph{\textbf{Definition:}}
A standard ($1$-dimensional) Brownian motion with respect to a filtration $R_{f}$ is a collection of random
variables $\delta(S)$, $S\geq 0$ satisfying the following \citep{pitman99}:

\begin{itemize}
  \item $\delta(S=0) = 0$;
  \item if $S^{\prime} < S$, then $\delta(S)-\delta(S^{\prime})$ is a measurable random variable,
  independent of the previous increment,
  with a Gaussian distribution;
  \item with probability one, $S \mapsto\delta(S)$ is a continuous function.
\end{itemize}
There are three characteristics of the Brownian random walk; bridge, excursion and meander:

\emph{\textbf{Theorem:}}
There exists a family of random walks in the interval [$0$, $S$] where all numbers in the interval are integers
for every $S$ \citep{philippe}:
\begin{enumerate}
\item a Brownian bridge is defined as a random walk of cumulative density contrast $\delta(S)$ which has
    length $S$ and is conditioned to return to
     $0$ at scale $S$ in the interval (Fig. \ref{fig:walks}). In the EPS formalism, a Brownian bridge can be
     described as minima of the excursion characteristic with negative barriers which form underdense
     regions.
\item a Brownian excursion is defined as a cumulative random walk of density contrast, $\delta(S)$ which has
    length $S$ and is conditioned to
    stay positive in the interval (Fig. \ref{fig:walks}).
\item a Brownian meander is defined as a cumulative random walk of density contrast, $\delta(S)$ which has
    length $<S$ and $\delta(S)$
    is conditioned to stay positive in the interval (Fig. \ref{fig:walks}). In the EPS formalism, meander
    represents local maxima with positive
    barriers in a Brownian excursion, indicating collapse regions.
\end{enumerate}
\noindent
In the context of the EPS/excursion set theory, the cloud in cloud problem can be related to the excursion
characteristic of the Brownian random walk,
while the void in cloud problem admits the combination of two characteristics of the Brownian random walk, that
of a meander with a positive barrier,
indicating a collapse region, and a bridge with a negative barrier, indicating a merging region.

\begin{figure}
\centering
\includegraphics[width=0.65\textwidth]{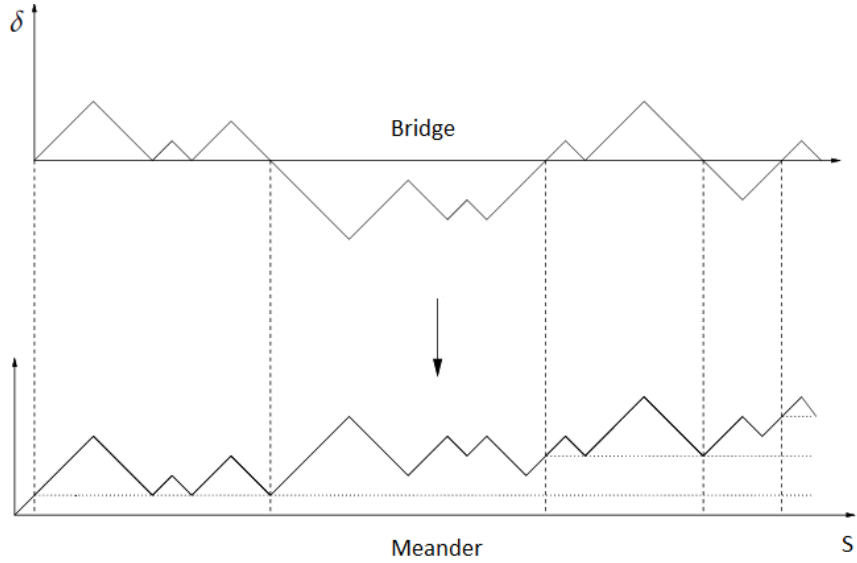}
\caption{Method of transforming a bridge to a meander.
There is bijection (one to one correspondence) between the negative valued walk (bridge) and the positive valued
local maxima (meander) characteristics of Brownian random walks. To obtain a positive valued meander/excursion
of certain length from a bridge, one needs to replace each negative excursion (barrier) of the bridge by its
symmetric positive excursion. Then, the last negative step of this symmetric positive excursion should be
replaced by a positive step. The final step is to add the first positive step to the whole path
\citep{philippe}.}
\label{fig:f1distributiona}
\end{figure}

These definitions of the conditioned Brownian motions have been rigorous in many ways
\citep{wim,Fpitman93,pitman99,philippe}. There is a simple and an effective method developed by \cite{philippe}
that allows one to transform a negative definite bridge into a positive
definite meander/excursion. Recall that there is a bijection between the bridge and the meander. To construct a
meander/excursion of length $2S+1$
from a bridge of length $2S$, one must replace each negative excursion of the bridge by its symmetric positive
excursion and replace the last negative
step of this symmetric positive excursion by a positive step. Finally a first positive step must be added to the
whole path (barrier uplifting)
(Fig. \ref{fig:f1distributiona}).

\subsection{Void in Cloud Problem: From Bridge to Excursion}
Following the definitions of the Brownian characteristics of the random walk and the method to obtain a positive
definite excursion, here a simple
treatment to the void in cloud random walk is defined and discussed to describe it in terms of the EPS
formalism.

\begin{figure}
\centering
\begin{tabular}{cc}
\includegraphics[width=0.45\textwidth]{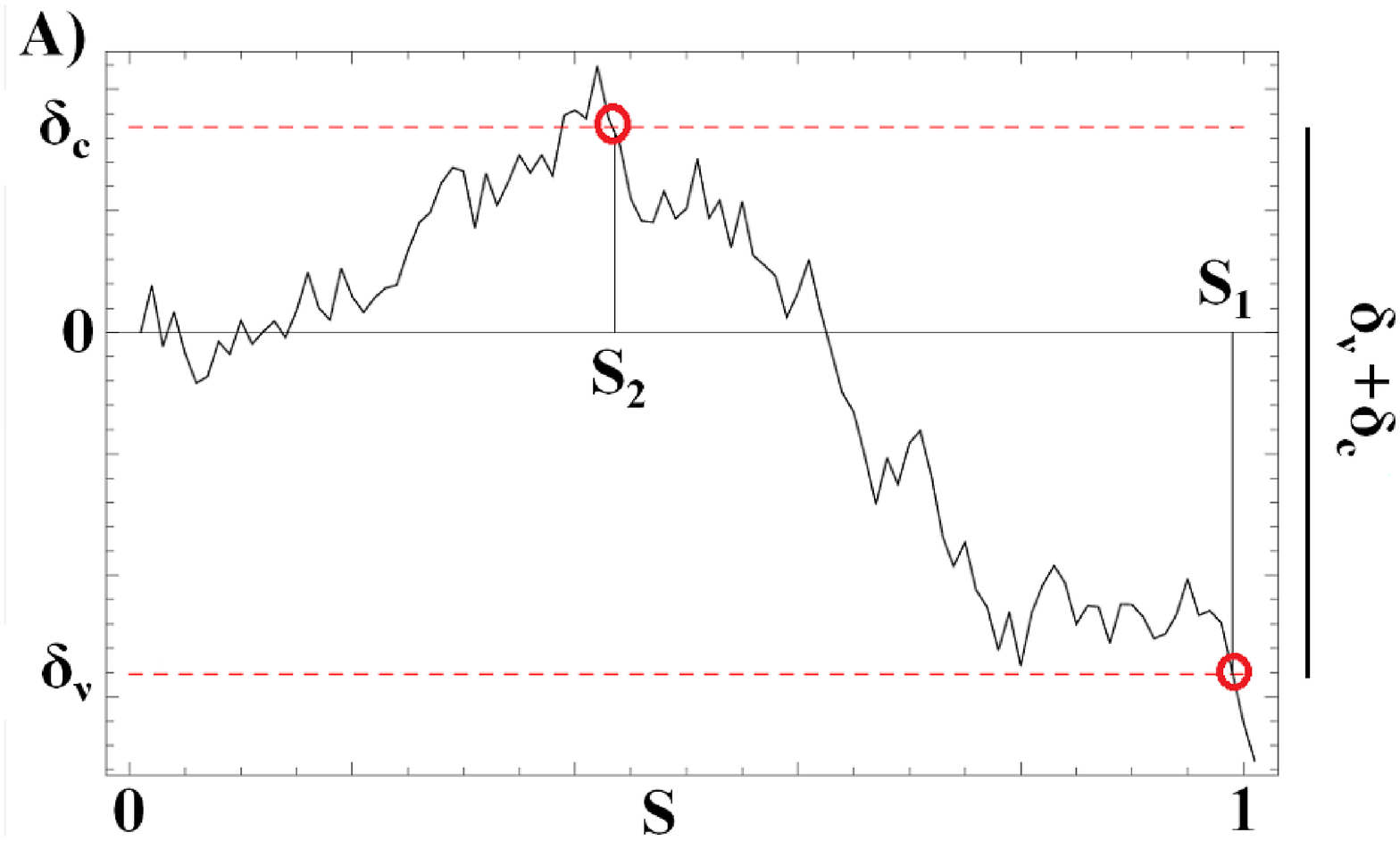}
\includegraphics[width=0.45\textwidth]{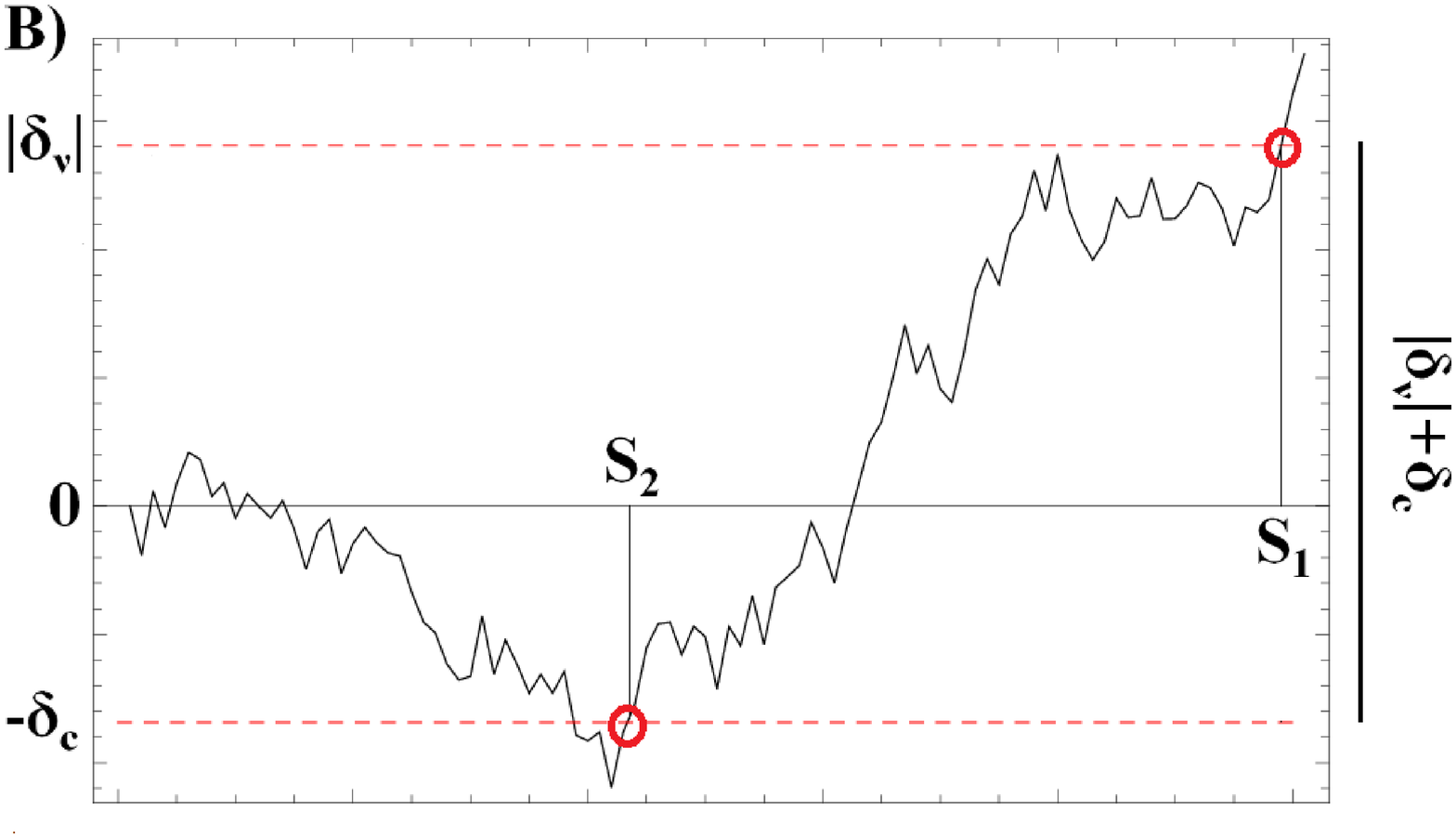}\\
\includegraphics[width=0.45\textwidth]{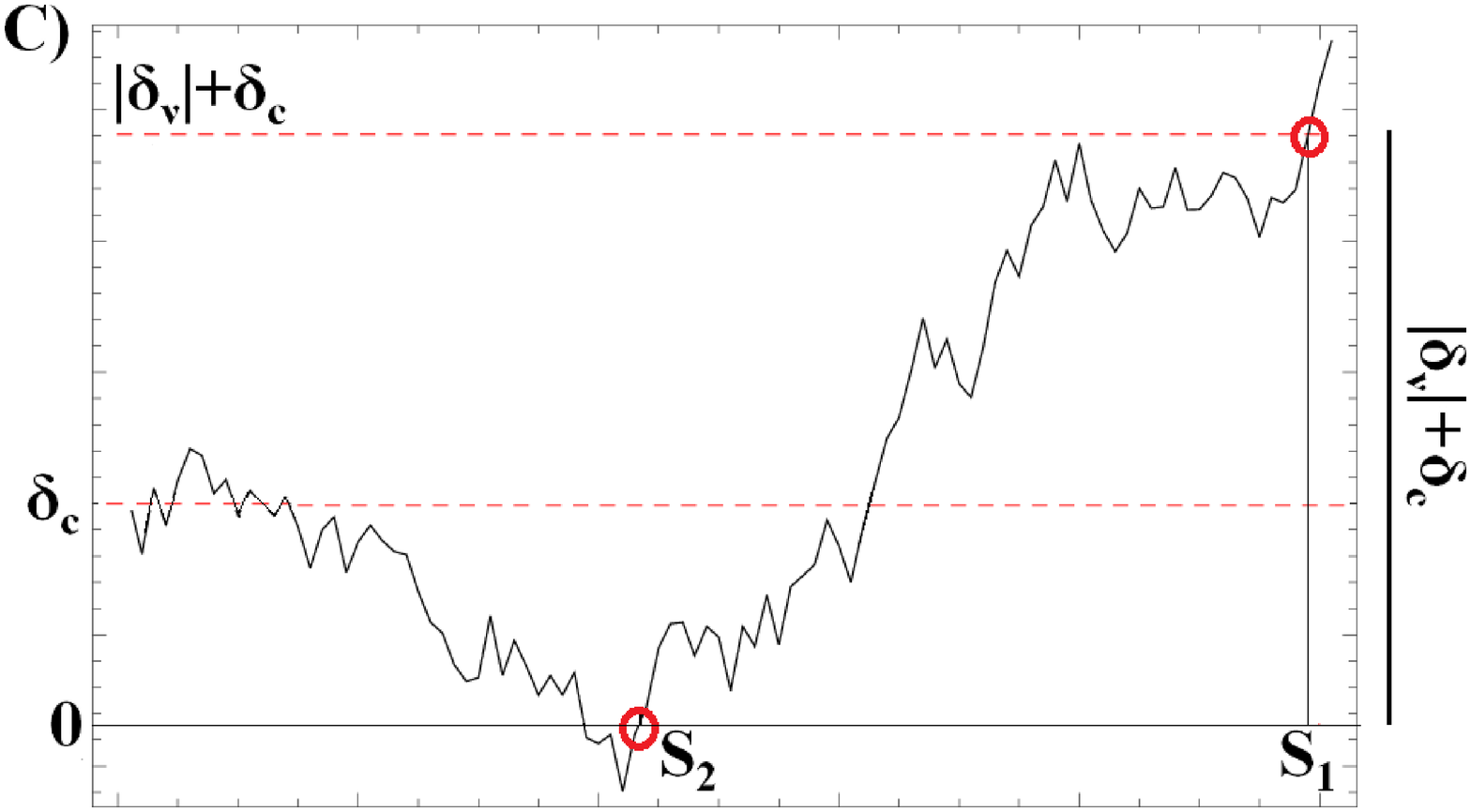}
\includegraphics[width=0.45\textwidth]{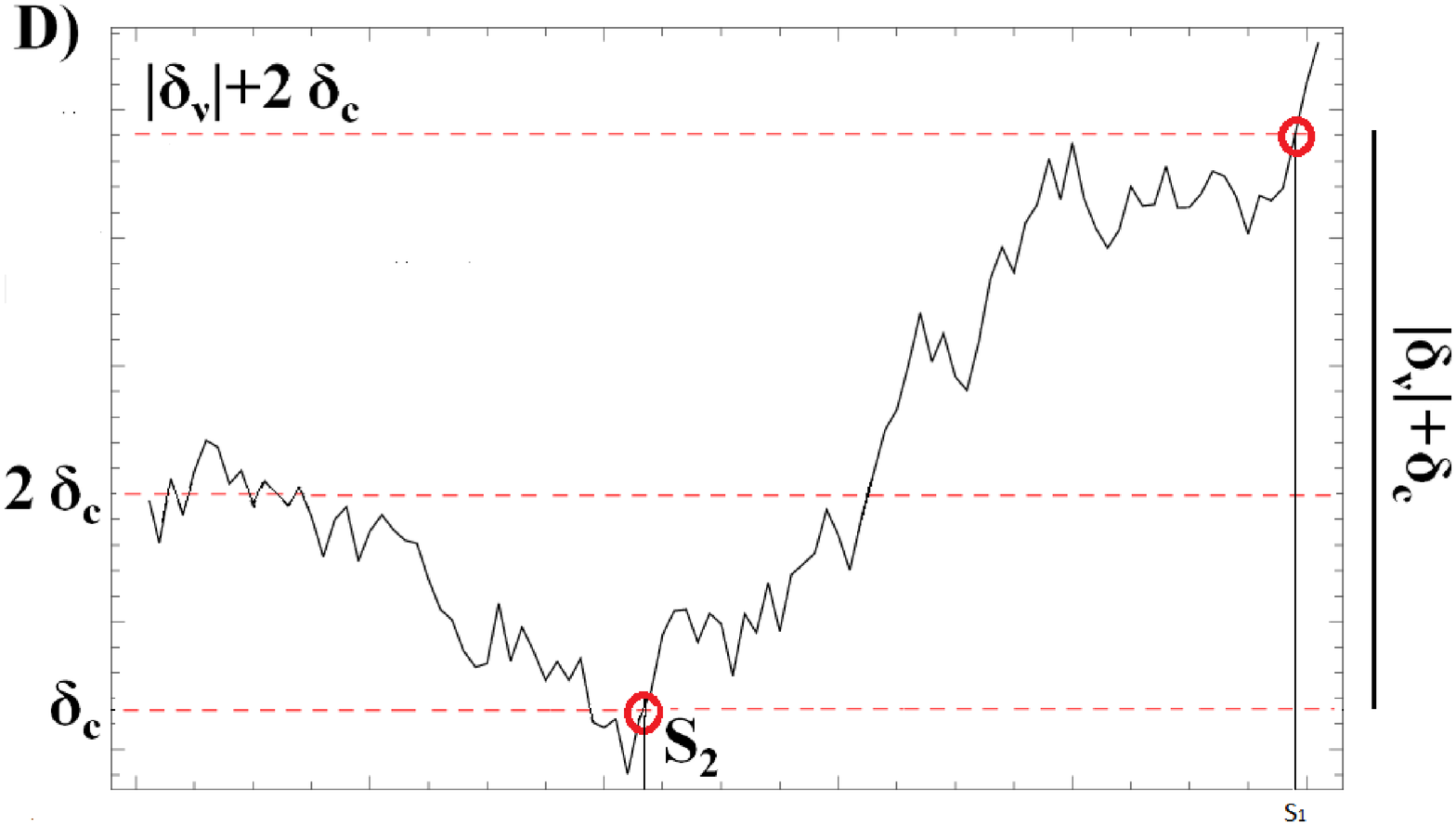}
\end{tabular}
\caption{Steps of constructing excursion with positive barriers from the void in cloud random walk with a
negative definite bridge characteristic.}
\label{fig:f1distributionaa}
\end{figure}

As mentioned before, the void in cloud random walk has one negative signed deep bridge on large volume scales
(see panel $A$ in Fig. \ref{fig:f1distributionaa}) with a negative barrier ($\delta_{v}$) and one positive
signed meander (or excursion) on the small mass scales with a positive
barrier $\delta_{c}$. Note that as of now, the volume distribution function is adopted in terms of volume
scale $S(V)$ to describe volume evolution in the context of void hierarchy. The distribution of the void
population is given by \cite{sw} as follows,

\begin{eqnarray}
f_{v}dS\approx
\frac{1}{\sqrt{2\pi}}\frac{{\delta_{v}}}{S^{2/3}}\exp\left[-\frac{\delta^{2}_{v}}{2S}\right]
\exp\left[-\frac{\delta_{v}}{\delta_{c}}\frac{S}{4{\delta^{2}_{v}}\left(\frac{\delta_{c}}{\delta_{v}}+1\right)^2}-2
\frac{S^2}{{\delta^{4}_{v}}\left(\frac{\delta_{c}}{\delta_{v}}+1\right)^4}\right]dS.
\label{A}
\end{eqnarray}
\noindent
Here, this distribution function is taken with a partially negative random walk and makes this a complete
positive excursion without changing its probability distribution. To do this by following the remark of Theorem,
each excursion of the whole walk is replaced by their symmetric excursion. This happens by taking the mirror of
the distribution around the $x$-axis ($S$) (see $B$ in Fig.~\ref{fig:f1distributionaa}).
However here there is still a negative bridge crossing the negative collapse barrier which is not allowed in the
EPS formalism. To solve this problem, the
barriers are shifted up by $2\delta_{c}$ with an accompanying shift in the probability distribution (see $C$-$D$
in Fig.~\ref{fig:f1distributionaa}).
Then, none of the barriers have a negative value and the resulting distribution function is given by,

\begin{eqnarray}
f_{v}dS\approx
\frac{1}{\sqrt{2\pi}}\frac{|\tilde{\delta}_{v}|}{S^{2/3}}\exp\left[-\frac{\tilde{\delta}_{v}^{2}}{2S}\right]
\exp\left[-\frac{1}{\tilde{\gamma}}\frac{1}{\left(\tilde{\gamma}+1\right)^2}\frac{S}{4{\tilde{\delta}^{2}_{v}}}-2
\frac{S^2}{\tilde{\delta}^{4}_{v}\left(\tilde{\gamma}+1\right)^4}\right]dS,
\label{CD}
\end{eqnarray}
\noindent
where a new barrier ${\tilde{\delta}_{v}}$ is defined. This new barrier is given by,

\begin{eqnarray}
{\tilde{\delta}_{v}}\equiv|{\delta_{v}}|+2\delta_{c},
\end{eqnarray}
\noindent
while the new barrier height ratio $\tilde{\gamma}$ is defined as,

\begin{eqnarray}
\tilde{\gamma}\equiv\frac{\delta_{c}}{|\tilde{\delta}_{v}|}.
\end{eqnarray}
\noindent
Hereafter the new barrier ratio is called the parametrized barrier ratio.
Therefore, a positive excursion random walk (Brownian random walk) of the void evolution is obtained. This
allows us to form the void mass/volume scale function that mimics the excursion set formalism of haloes. By
applying this method which constructs an excursion set from a negative defined bridge:

\begin{enumerate}
\item The distance between the barriers is not changed (see Fig. \ref{fig:f1distributiona}). In addition to
    this,
the scales where the random walk meets the barriers is not changed either. The only change is that the
distribution with negative barriers becomes
a completely positive Brownian random walk, like the excursion set formalism \citep[see][]{bo}.

\item Since the distance of the barriers and the volume scales where the trajectories make their first
    upcrossing are not changed by the barrier
shifting up process, the probability distribution does not change. This allows us to use the two-barrier
fraction function given by \cite{sw}.
\end{enumerate}
\noindent
Now the questions are: how can we obtain a merger tree algorithm from this distribution, and what is the
physical interpretation of this distribution from the void perspective?

\begin{figure}
\centering
\includegraphics[width=0.65\textwidth]{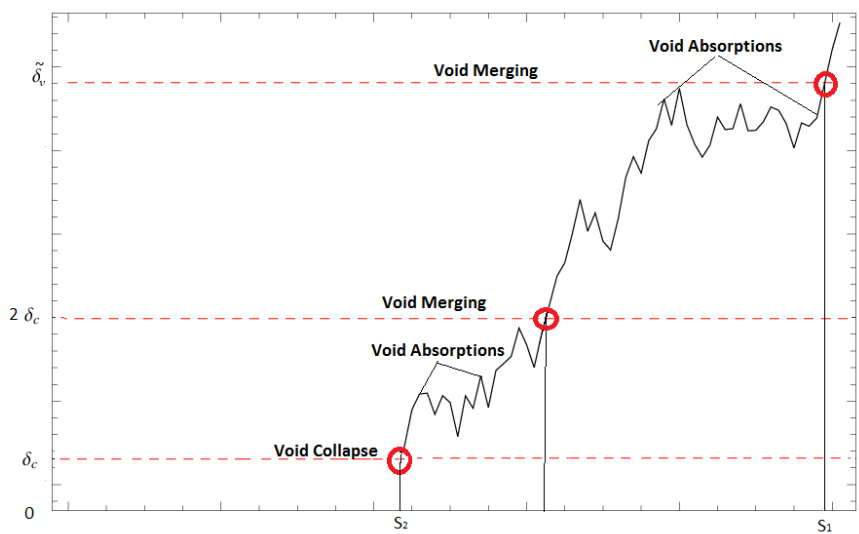}
\caption{Excursion set interpretation of the void in cloud process. A given trajectory $\delta(S)$ describes the
history of an embedded/minor void that
starts merging
with volume scale $S_{1}=S(V_{1})$ at shell crossing/merging barrier $\tilde{\delta}_{v}$ and later on collapses
at barrier $\delta_{c}$ corresponding
volume scale $S_{2}$.}
\label{fig:f1distributionF}
\end{figure}

Fig. \ref{fig:f1distributionF} shows the random walk of the void in cloud which is converted into a positive
excursion set by following the method, that is introduced above. This method has a key role since it allows one
to treat the random walk of the void in cloud as a cloud in cloud random walk.
As a result, in Fig. \ref{fig:f1distributionF}, a given trajectory
$\delta(S)$ describes the merging history for a given void which starts its evolution with volume scale $S_{1}$
at barrier $\tilde{\delta}_{v}$, and later it will merge into other voids with larger volume at smaller barrier
values $< \tilde{\delta}_{v} $ and eventually will collapse
at barrier $\delta_{c}$ corresponding to volume scale $S_{2}$. Note that in Fig. \ref{fig:f1distributionF}, the
random walk makes horizontal jumps when
$S$ decreases. These jumps correspond to sudden jumps in the volume of the void. The small steps in between two
barriers correspond to adding only a
small amount of volume to the void. These events are called void absorption events, since bigger voids absorb
small ones.
Note that after several merges with other voids, the void will reach the end of its lifetime when its random
walk crosses the collapse barrier\
$\delta_{c}$. This means that the collapse barrier represents the end of void merging events.

\subsection{Redshift and Size Constraints on Voids}
The parameters $\gamma$ and $\tilde{\gamma}$ are defined as the ratio of the collapse and underdense barriers of
the random walks,

\begin{eqnarray}\label{beta1}
\gamma=\frac{\delta_{c}}{|\delta_{v}|} \phantom{a}\text{and}\phantom{a}
\tilde{\gamma}\equiv\frac{\delta_{c}}{\tilde{\delta}_{v}}=\frac{\delta_{c}}{|{\delta_{v}}|+2\delta_{c}}.
\end{eqnarray}
\noindent
The relation between the barrier ratios is given by,

\begin{eqnarray}\label{ratiorelation}
\tilde{\gamma}=\frac{\gamma}{2\gamma+1}.
\end{eqnarray}
\noindent
The backbone of the constraints of the extended void merging tree is formed by the fact that the earliest
merging barrier $\delta_{v}$ must be bigger than the $2\delta_{c}$ in order to obtain a merging event and at the
end of the excursion process the walk will cross the collapse barrier,

\begin{eqnarray}\label{conditiondeltavc}
\tilde{\delta}_{v}=|\delta_{v}|+2\delta_{c}\phantom{a} \text{and}\phantom{a} \tilde{\delta}_{v}
> |\delta_{v}| > 2\delta_{c}.
\end{eqnarray}
\noindent
Note that $\tilde{\delta}_{v}$ stands for the first
merging event with the biggest volume scale $S(V_{1})$ indicating a void with smallest volume in the EPS
formalism. Taking into account $|\delta_{v}| > 2\delta_{c}$, it is possible to obtain relations between merging
$z_{v}$ and collapse $z_{c}$ redshifts,

\begin{eqnarray}\label{redshiftrelation}
z_{v} > 2 \frac{\delta_{c0}}{\delta_{v0}}\left(1+z_{c}\right)-1.
\end{eqnarray}
\noindent
In this study, in all calculations for plots, the collapse redshift is chosen as present-day redshift $z_{c}=0$.
As a result, for this collapse redshift $z_{c}=0$, a lower limit of merging redshift is obtained as $z_{v}>0.2$.
The condition (\ref{conditiondeltavc}) puts a constraint on the barrier ratio $\gamma$ as $\gamma < 0.5$.

Another constraint on the barrier ratio $\gamma$ resulting from the merging redshift comes from the theoretical
study of \cite{sw}. In their study, \cite{sw} point out that two barrier mass fraction function (\ref{mff}) is
accurate for values of $\gamma\geq 1/4$. This allows us to obtain an upper merging redshift value as $z_{v}\leq
1.4$. Hence, the constraints on the barrier ratio and the merging redshift are given as,

\begin{eqnarray}
0.25\leq \gamma < 0.5\phantom{a} \text{and} \phantom{a}0.2 < z_{v} \leq 1.4\phantom{a} \text{at}\phantom{a} z_{c}=0.
\end{eqnarray}
\noindent
One may show that,

\begin{eqnarray}
\tilde{\gamma}=\frac{\gamma}{2},
\end{eqnarray}
\noindent
by using the relation between $\gamma$ and $\tilde{\gamma}$ (\ref{ratiorelation}) in the limit of $|\delta_{v}|$
tends to  $2\delta_{c}$  ($\gamma\rightarrow 0.5$). As a consequence of this relation, the interval of the
parametrized barrier height ratio $\tilde{\gamma}$ can be obtained as,

\begin{eqnarray}
0.125\leq \tilde{\gamma} < 0.25.
\end{eqnarray}
\noindent
In addition to this, from the relations (\ref{beta1}) and (\ref{ratiorelation}), it is obtained that when the
collapse barrier is moved towards infinity $\delta_{c}\rightarrow \infty$, $\gamma$ tends to infinity. As a
result, from the relation (\ref{massfractionSchap4}), the new barrier ratio becomes $\tilde{\gamma}\geq 0.5$.
Therefore, the distribution turns into the one-barrier distribution of the void in void problem in Fig.
\ref{fig:voidcriteriagamma}. Based on this, two different void behaviors are modelled by the following
constraints:

\begin{enumerate}
\item Void collapse: Contribution of the small voids into the distribution is dominant:
"Voids are squeezed",

\begin{eqnarray}\label{constraintgammazv}
0.125\leq \tilde{\gamma} < 0.25,\phantom{a} \text{and}\phantom{a} 0.2 < z_{v} \leq 1.4\phantom{a} \text{for}
z_{c}=0.
\end{eqnarray}

\item Void merging: When the ratio of collapse and merging barriers is high, $\delta_{c}\gg |\delta_{v}|$,
the new barrier height
parameter becomes $\tilde{\gamma}\gg 0.5$. The two-barrier excursion set theory is reduced to the
one-barrier void in void problem. Therefore, voids do not
vanish due to collapsed regions where they are embedded and their evolution is dominated only by merging
events \citep{eszrarussell}.
\end{enumerate}
\noindent
An important result on the void volume scale is given by \cite{sw}. This criterion is based on the statement
that there are no large scale voids embedded in large scale haloes, on the scales where,

\begin{eqnarray}
\sigma \ll \left(\delta_{c}+|\delta_{v}|\right),
\end{eqnarray}
\noindent
and the collapse barrier $\delta_{c}$ does not have any effect on the void population. If this statement is
rearranged in terms of the void size, a size criterion on void sizes that do not feel the environmental effects
can be obtained as,

\begin{eqnarray}
R \gg \ R_{*}\left({\gamma+1}\right)^{-\frac{2}{n+3}},
\phantom{a} \gamma > 0.5,
\end{eqnarray}
\noindent
where the characteristic radius $R_{*}$ is chosen as $R_{*}=8\quad h^{-1}$Mpc. This size criterion leads to a
rough classification between void sizes. Some voids are embedded in the overdense regions and eventually
collapse (with $0.25\leq \gamma < 0.5$ or $0.125\leq \tilde{\gamma} < 0.25$) while other voids that satisfy this
size criterion and expand, merge continuously ($\gamma > 0.5$ or $\tilde{\gamma} > 0.25$). In the limit of the
above redshift and the size criterion their EPS formalism should satisfy the one-barrier EPS formalism of the
growing voids \citep{eszrarussell}. In a way, the last case is just a theoretical statement since it is not
possible for some voids to merge and expand forever.

In the following section, the evolutionary paths of these two void groups are investigated in terms of the EPS
formalism by adapting the LC93 merging tree algorithm.

\section{Merging Tree Algorithm of Growing and Embedded/Minor Voids}
\begin{figure}
\centering
\includegraphics[width=0.65\textwidth]{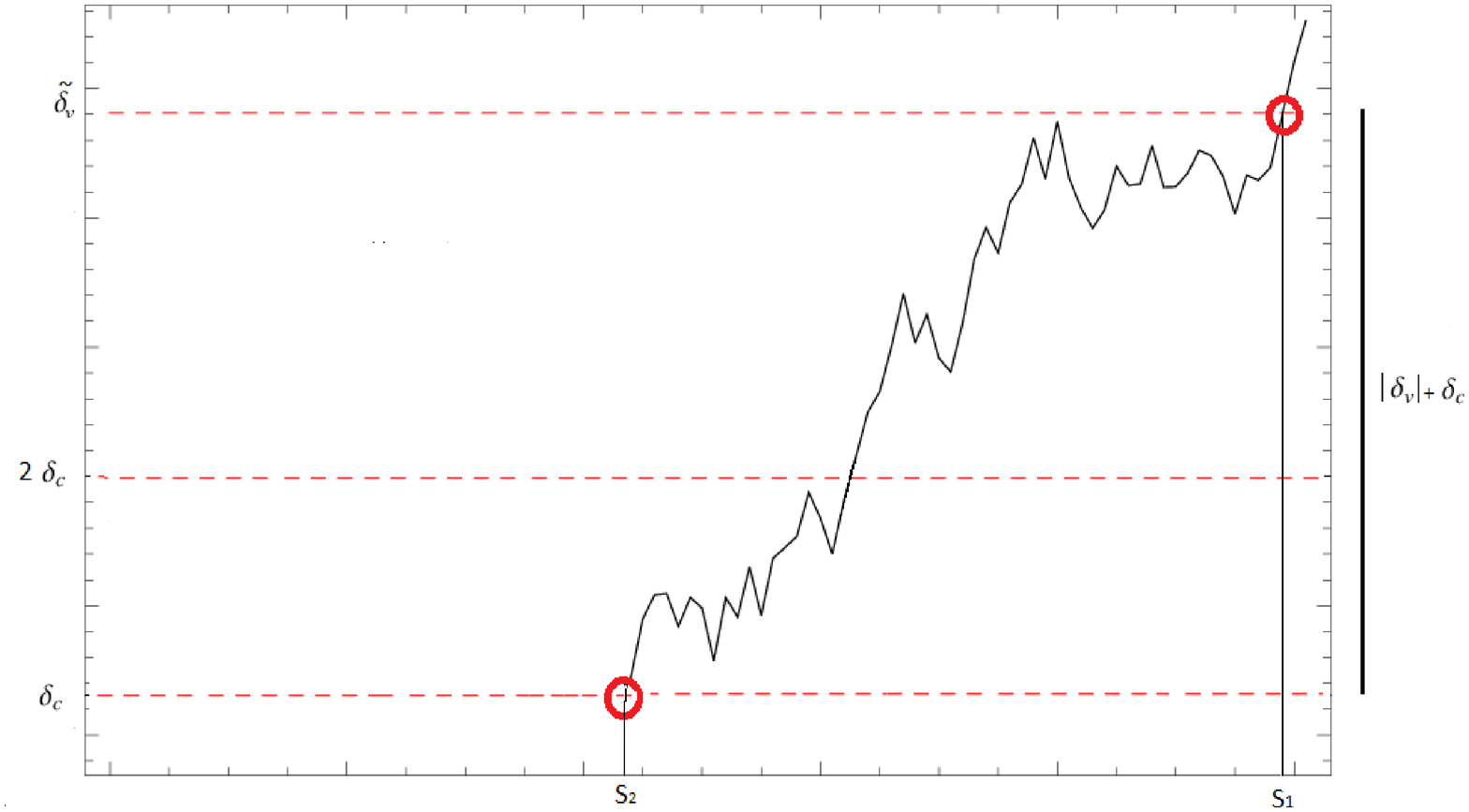}
\caption{Subset of trajectories making their first upcrossing of a barrier $\delta_{c}$ at $S_{2}$ and then
continue until crossing a second barrier
$\tilde{\delta}_{v}$ at various scales $S_{1} > S_{2}$. These trajectories represent voids which at the time
corresponding to $\tilde{\delta}_{v}$ have
volumes corresponding to $S_{1}(V)$, which by the later time correspond to the scale $\delta_{c}$ at volume
scale $S_{2}(V)$ where they will vanish due to
collapse.}
\label{fig:f1distributionE}
\end{figure}

Here, the aim is to determine the merger probability per unit time for a void of a given volume and time. Therefore
the subset of trajectories is considered,
depicted in Fig. \ref{fig:f1distributionE}. These trajectories make their first upcrossing of a barrier
$\delta_{c}$ at $S_{2}$ and then continue until they eventually
cross a second barrier of height $\tilde{\delta}_{v}$ at various scales $S_{1} > S_{2}$.

From the point of merging history of voids, these trajectories
represent voids which at the time
corresponding to $\tilde{\delta}_{v}$ have relatively large volume scales $S_{1}$s. Later on, these trajectories
cross the collapse barrier  $\delta_{c}$ where voids die after several times merging. Note that in between the first merging barrier and the last collapse barrier, many merging barriers can
be replaced as long as the last barrier is the collapse one, ${\delta}_{c}$.

\begin{figure}
\centering
\begin{tabular}{cc}
\includegraphics[width=0.45\textwidth]{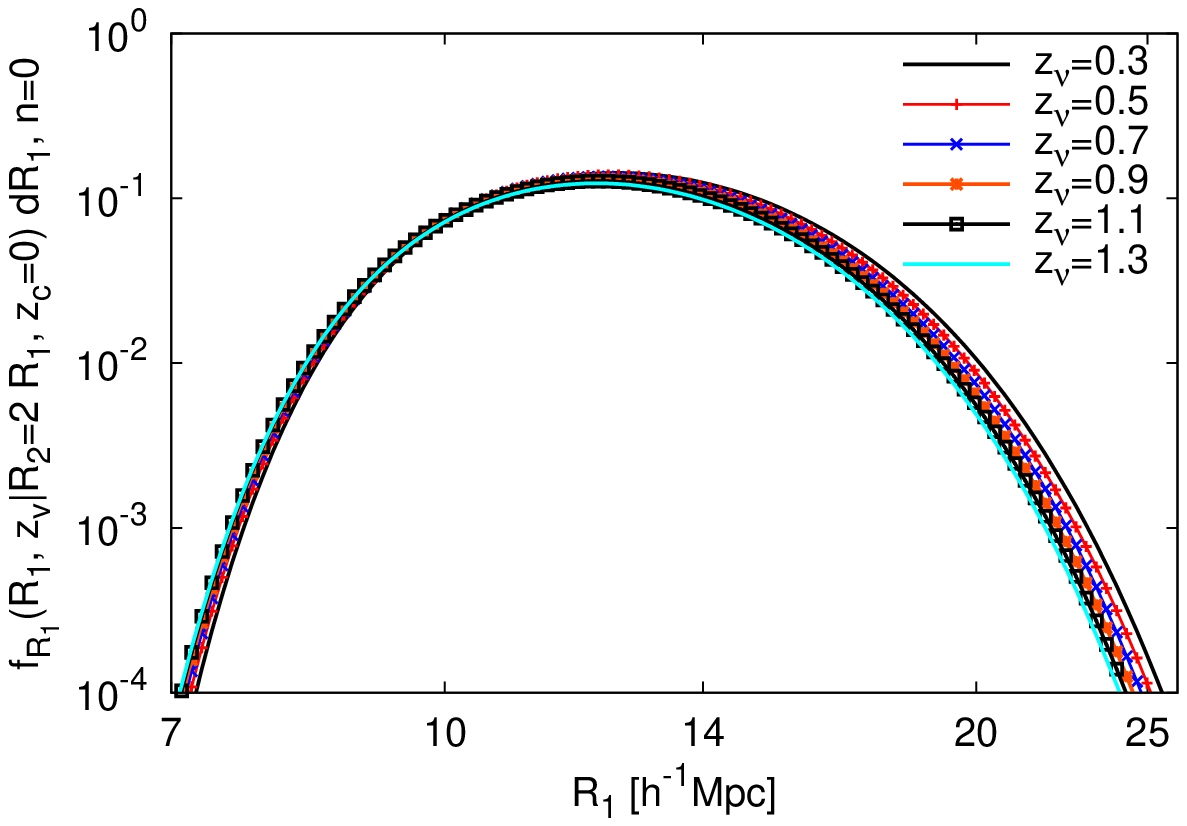}
\includegraphics[width=0.45\textwidth]{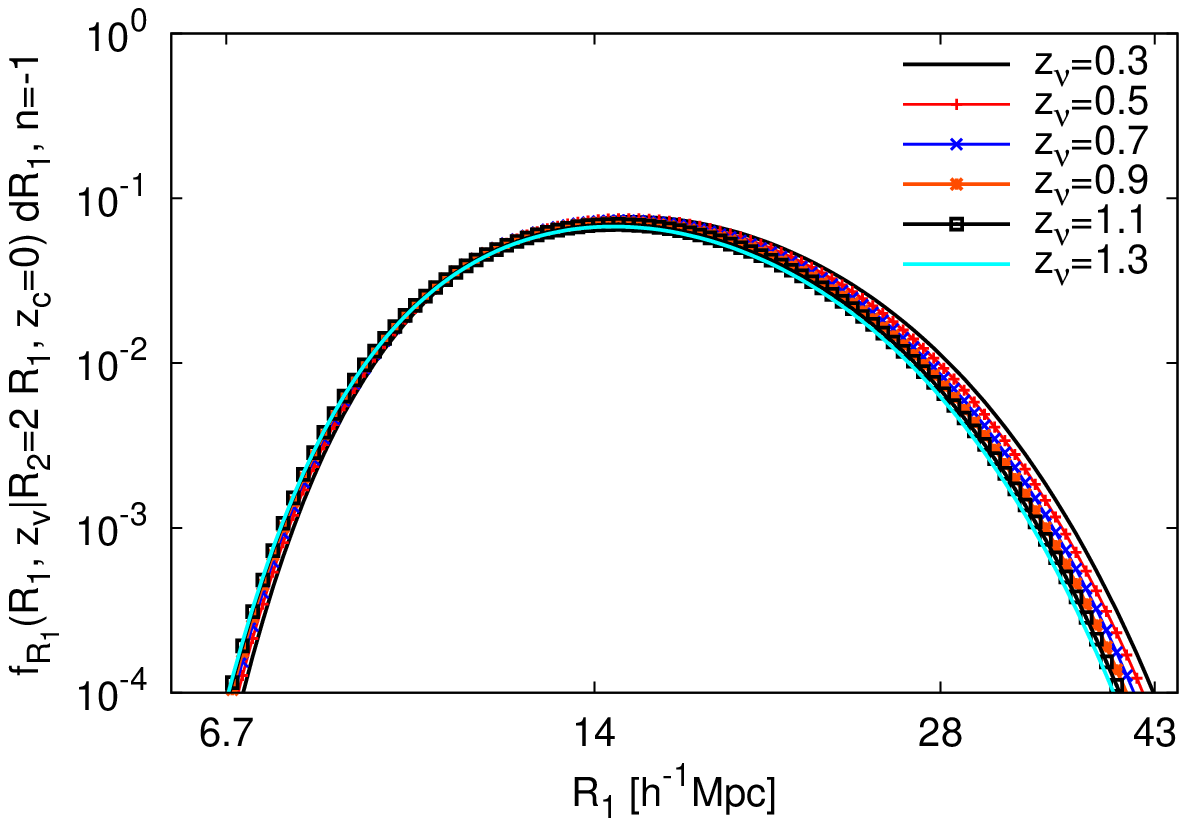}\\
\includegraphics[width=0.45\textwidth]{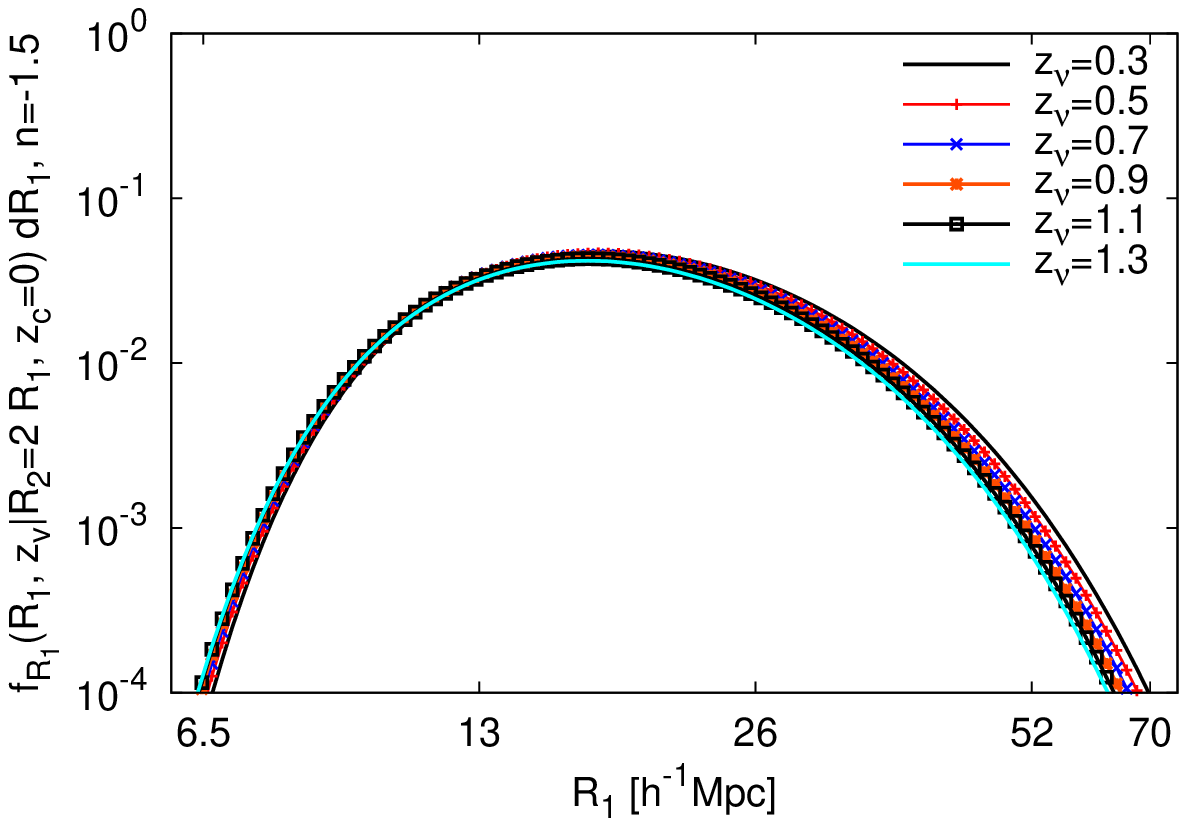}
\includegraphics[width=0.45\textwidth]{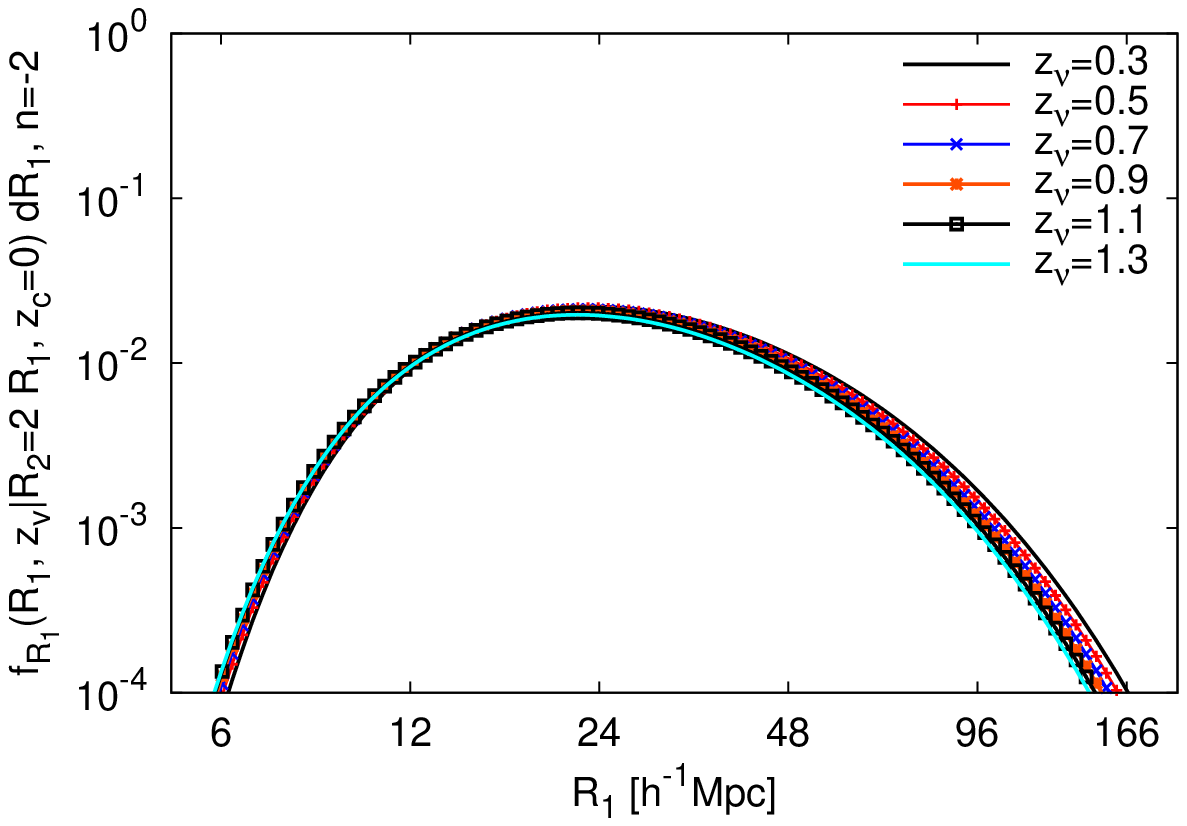}
\end{tabular}
\caption{Size distribution of voids embedded in overdense regions at given redshift $z_{v}$ in terms of
$R_{1}$ [$h^{-1}$Mpc] for the self-similar $n=0$, $-1$, $-1.5$, $-2$ at redshifts $z_{v}=0.3, 0.5, 0.7, 0.9,
1.1, 1.3$. Later on voids with size $R_{1}$ at redshift $z_{v}$ will double their size, $R_{2}=2R_{1}$ at
collapse redshift $z_{c}=0$ when they will vanish due to the presence of a collapse region.}
\label{fig:smallradiusdistributionselfsim}
\end{figure}

The scale fraction function derived in equation (\ref{CD}) is,

\begin{eqnarray}
f_{v}dS\approx
\frac{1}{\sqrt{2\pi}}\frac{\tilde{\delta}_{v}}{S^{2/3}}\exp\left[-\frac{\tilde{\delta}_{v}^{2}}{2S}\right]
\exp\left[-\frac{1}{\tilde{\gamma}}\frac{1}{\left(\tilde{\gamma}+1\right)^2}\frac{S}{4{\tilde{\delta}^{2}_{v}}}-2
\frac{S^2}{\tilde{\delta}^{4}_{v}\left(\tilde{\gamma}+1\right)^4}\right]dS.\nonumber
\end{eqnarray}
\noindent
The conditional probability $f_{S_{1}}(S_{1},\tilde{\delta}_{v}\big|S_{2},\delta_{c}) d{S_{1}}$ in which one of these
trajectories makes its first upcrossing at
$\tilde{\delta}_{v}$ in the interval $S_{1}+dS_{1}$ can be obtained directly from equation (\ref{CD}) but with a difference that the source of the trajectories moved from the origin to the
point $(S_{2},\delta_{c})$ (by following the algorithm derived by LC93). Then, the
conditional probability density of a void whose trajectory is in the interval $S_{1}+ dS_{1}$ making its first
up crossing at $\tilde{\delta}_{v}$ which collapses at the point ($S_{2},\delta_{{c}}$) is,

\begin{eqnarray}
f_{S_{1}}(S_{1},\tilde{\delta}_{v}\big|S_{2},\delta_{c})
d{S_{1}}&=&
\frac{1}{\sqrt{2\pi}}\frac{\tilde{\delta}_{v}-\delta_{c}}{\left({S_{1}}-{S_{2}}\right)^{3/2}}\exp\left(\frac{-\left(\tilde{\delta}_{v}-\delta_{c}\right)^{2}}
{2\left(S_{1}-{S_{2}}\right)}\right)\exp\left(-\frac{1}{4}\frac{1}{\tilde{\gamma}}\frac{1}{\left(1+\tilde{\gamma}\right)^2}\frac{S_{1}-S_{2}}
{\left(\tilde{\delta}_{v}-\delta_{c}\right)^2}\right)\nonumber\\
&&\exp\left(-2\frac{1}{\left(1+\tilde{\gamma}\right)^4}
\frac{\left(S_{1}-S_{2}\right)^2}{\left(\tilde{\delta}_{v}-\delta_{c}\right)^4}
\right)dS_{1}.
\label{probabilitys1a}
\end{eqnarray}
\noindent
Here, $\tilde{\delta}_{v} > \delta_{c}$ ($\approx z_{v} > z_{c}$) and the volume scale is $S_{1} > S_{2}$. Note
that high volume scale indicates small
size radius. Hence, the void size fraction derived from equation (\ref{probabilitys1a}) for the self-similar
models becomes,

\small
\begin{eqnarray}
&&f_{R_{1}}(R_{1},\tilde{\delta}_{v}\big|R_{2},\delta_{c})
d{R_{1}}=
\frac{3\alpha}{\sqrt{{2\pi}}}\frac{\tilde{\delta}^{2}_{v}}{{R_{*}}}
\left(\frac{R_{1}}{R_{*}}\right)^{-3\alpha-1}
\frac{\tilde{\delta}_{v}-\delta_{c}}{\left({\tilde{\delta}^{2}_{v}\left(\frac{R_{1}}{R_{*}}\right)^{-3\alpha}}-
{{\delta}^{2}_{c}\left(\frac{R_{2}}{R_{*}}\right)^{-3\alpha}}\right)^{3/2}}
\exp\left(\frac{-\left(\tilde{\delta}_{v}-\delta_{c}\right)^{2}}{2\left(\tilde{\delta}^{2}_{v}{\left(\frac{R_{1}}{R_{*}}\right)^{-3\alpha}}-
{{\delta}^{2}_{c}\left(\frac{R_{2}}{R_{*}}\right)^{-3\alpha}}\right)}\right)\nonumber\\
&&\times\exp\left(-\frac{1}{4}\frac{1}{\tilde{\gamma}}\frac{1}{\left(1+\tilde{\gamma}\right)^2}
\frac{\left(\tilde{\delta}^{2}_{v}{\left(\frac{R_{1}}{R_{*}}\right)^{-3\alpha}}-
{{\delta}^{2}_{c}\left(\frac{R_{2}}{R_{*}}\right)^{-3\alpha}}\right)}
{\left(\tilde{\delta}_{v}-\delta_{c}\right)^2}\right)\exp\left(-2\frac{1}{\left(1+\tilde{\gamma}\right)^4}
\frac{\left(\left({\tilde{\delta}^{2}_{v}\left(\frac{R_{1}}{R_{*}}\right)^{-3\alpha}}-
{{\delta}^{2}_{c}\left(\frac{R_{2}}{R_{*}}\right)^{-3\alpha}}\right)\right)^2}
{\left(\tilde{\delta}_{v}-\delta_{c}\right)^4}
\right)dR_{1}.
\label{probabilitys1size}
\end{eqnarray}
\normalsize
\noindent
Depending on the value of the barrier ratio, equation (\ref{probabilitys1size}) displays the size distribution
of the embedded voids in one formula for the self-similar models. Fig.~\ref{fig:smallradiusdistributionselfsim}
shows embedded void size distributions for self-similar ($n=0$, $-1$, $-1.5$, $-2$) respectively at
given redshift $z_{v}$ values satisfying the constraint equalities (\ref{constraintgammazv}) at collapse
redshift $z_{c}=0$. Note that, in all models, the collapse redshift is chosen as $z_{c}=0$ when minor voids
collapse and vanish after merging at a given redshift $z_{v}$. In Fig.~\ref{fig:smallradiusdistributionselfsim},
the peak of a size distribution at a given redshift indicates a specific size. Embedded/minor voids with this
size value dominate the distribution at a given redshift. In all self-similar models peaks of the size
distributions slightly move towards smaller sizes with increasing redshift $z_{v}$. This result may indicate
that there are more smaller minor voids and they fill the Universe at high redshifts. This is what it is
expected from the bottom-up hierarchical scenarios that small size voids merge together and construct larger
size voids. In addition, in the self-similar models, the peaks of the size distribution decrease with decreasing
spectral index while the size range of minor voids increases drastically when the spectral index decreases. For
example, the size distribution $f_{R1}$ in the model with spectral index $n=-2$ at redshift $z=0.3$ shows
$6-166$ $h^{-1}$Mpc void size range while one with spectral index $n=0$ at the same redshift has $7-25$
$h^{-1}$Mpc size range. Note that decreasing spectral index corresponds to an increase in hierarchical
clustering. Therefore, one may say that the probability of detecting an embedded void with size $50$ $h^{-1}$Mpc
is most likely in the model with spectral index $n=-2$ than in the self-similar model with index $n=0$. On the
other hand, in all models, size distribution shows two cutoffs corresponding to the smallest and the largest
size values at a given redshift $z_{v}$. As is seen in Fig.~\ref{fig:smallradiusdistributionselfsim}, these
cutoff values have the smallest probability values. While the small cutoff becomes smaller, the big cutoff moves
towards higher values by decreasing spectral index for a given redshift $z_{v}$. This behavior indicates that
the probability of seeing the largest and smallest size embedded voids at a given redshift and a given model is
unlikely.

\begin{figure}
\centering
\begin{tabular}{cc}
\includegraphics[width=0.45\textwidth]{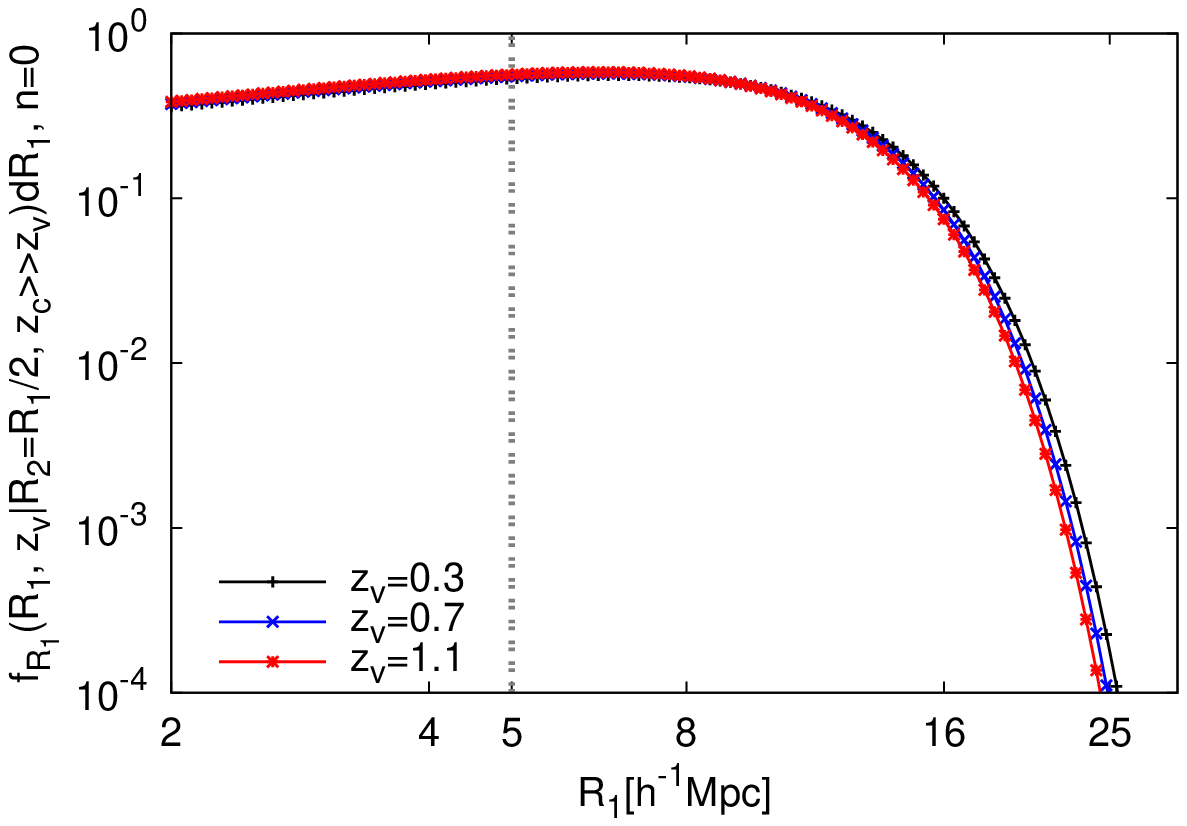}
\includegraphics[width=0.45\textwidth]{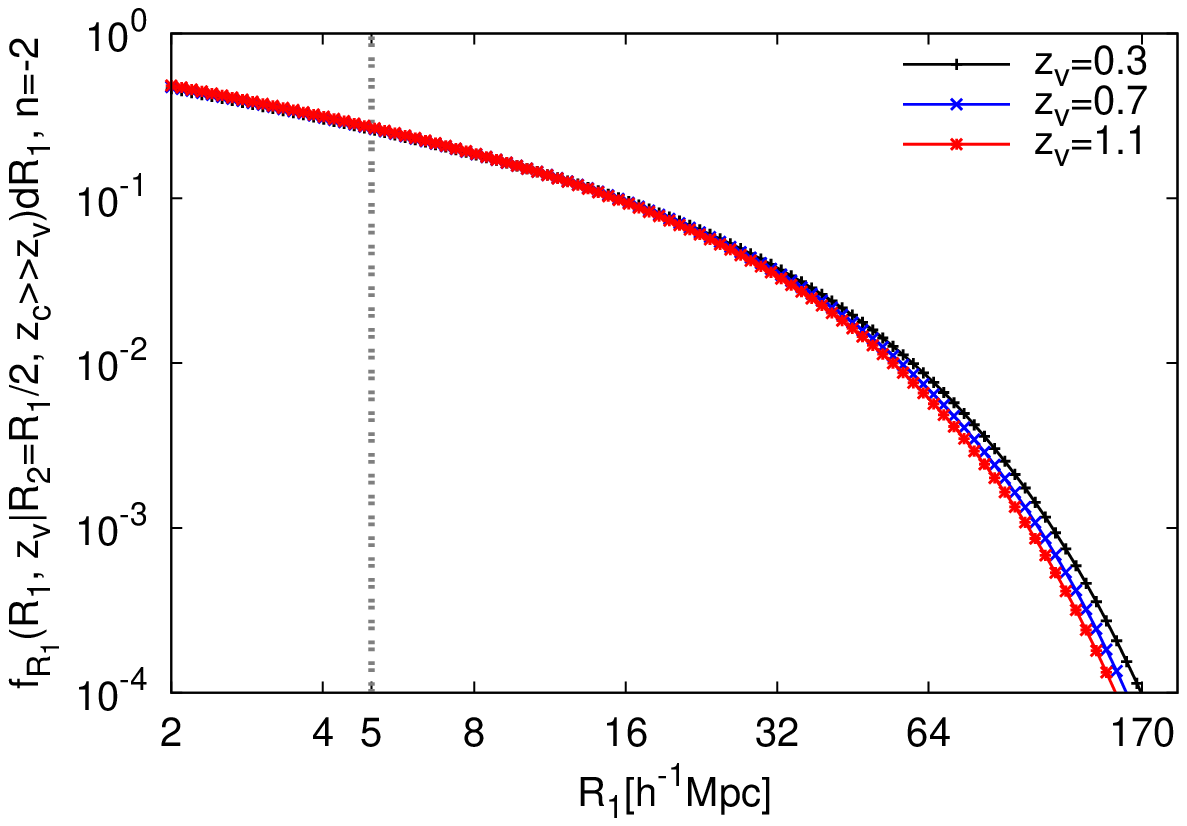}
\end{tabular}
\caption{Size distributions of embedded growing voids. These voids are not affected by collapse regions at given
redshifts $z_{v}=0.3, 0.7, 1.1$ for self-similar models $n=0, -2$. In the models, the collapse barrier is
inserted at very high redshift satisfying $z_{c}\gg z_{v}$, corresponding to the relation $\delta_{c}\gg
\delta_{v}$. The vertical line at size $5$ $h^{-1}$Mpc represents the limit of non-linearity corresponding to voids smaller than this size indicating non-linearity effects.}
\label{fig:largeradiusdistributionselfsim}
\end{figure}

In Fig. \ref{fig:largeradiusdistributionselfsim}, the growing void distributions are plotted from minor void
size distribution equation (\ref{probabilitys1size}) by choosing $\delta_{c}\gg \delta_{v}$ (void in void
limit). In this limit, embedded/minor voids are not affected by their surroundings. Therefore they merge continuously. As is seen in Fig. \ref{fig:largeradiusdistributionselfsim}, differing from the
embedded/minor voids, the growing embedded void probability does not show the same feature as the minor void
distribution. Instead, maxima in the growing embedded void distribution increase with decreasing spectral index
(from $n=0$ to $n=-2$) towards small size voids. This behavior is consistent with the one-barrier EPS formalism
of growing voids constructed by \cite{eszrarussell}.
However, like the minor void case, the maxima in the size distribution increase with increasing redshift value.
In Fig. \ref{fig:largeradiusdistributionselfsim}, the vertical line at $5$ $h^{-1}$Mpc stands for the limit of
non-linear effects. Both void populations with sizes smaller than $5$ $h^{-1}$Mpc are highly effected by
non-linear effects. Therefore they tend to have deformed or elliptical shapes. Note that in this study, the
spherical void merging tree model is investigated. As a result, non-linear effects are neglected in the extended
void merging algorithm when they are smaller than the $5$ $h^{-1}$Mpc limit. On the other hand, in Fig.
\ref{fig:smallradiusdistributionselfsim}, the size distribution of embedded voids in all self-similar models
does not present void sizes smaller than this limit value. Therefore, the extended void size distribution based
on the two-barrier EPS formalism is free of non-linear effects.

\begin{figure}
\centering
\begin{tabular}{cc}
\includegraphics[width=0.45\textwidth]{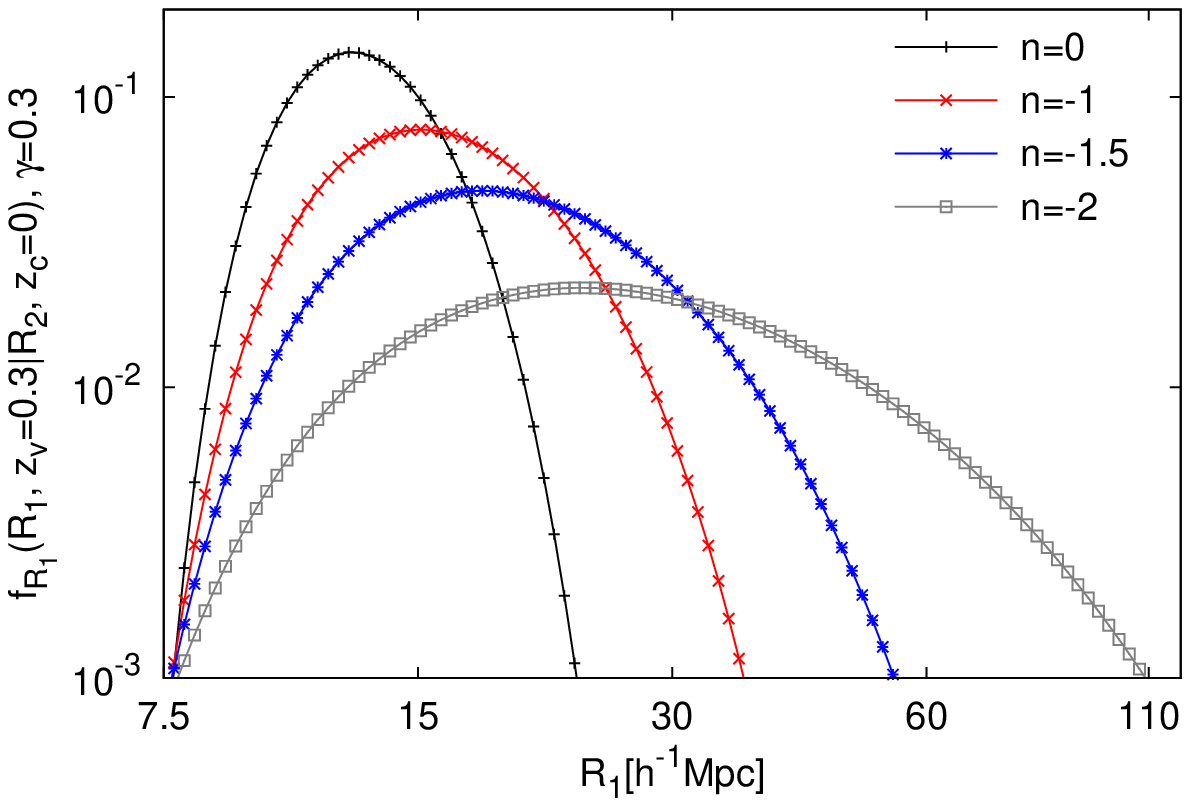}
\includegraphics[width=0.45\textwidth]{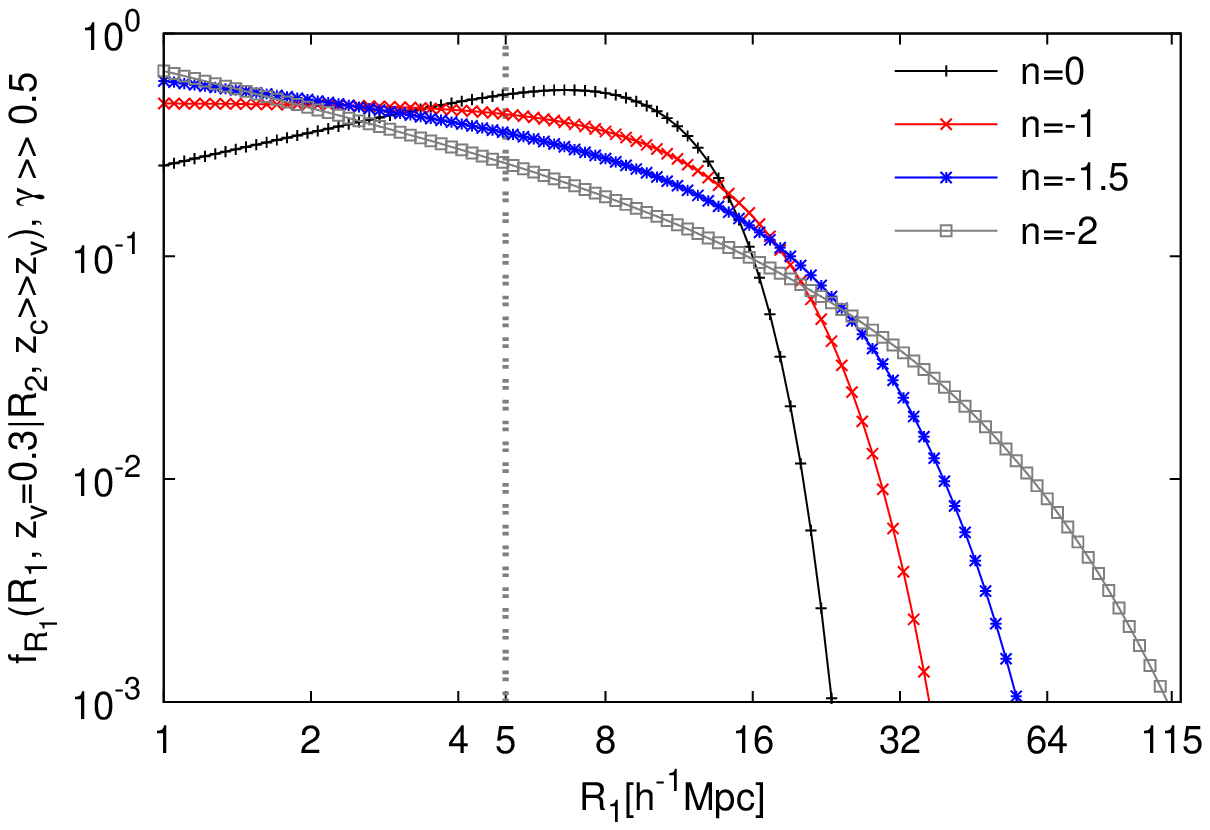}
\end{tabular}
\caption{Size distribution of embedded/minor voids $\tilde{\gamma}=0.15$ (or ${\gamma}=0.3$) (left) and growing
voids with the barrier ratio $\tilde{\gamma}\gg 0.25$ (right) from equation (\ref{probabilitys1size}) in terms
of self-similar with spectral index $n=0$, $-1$, $-1.5$, $-2$ models. In all models merging redshift is chosen
as $z_{v}=0.3$ for minor voids while the collapse redshift is moved towards late redshift $z_{c} >> z_{v}=0.3$. The vertical line corresponds to $5$ $h^{-1}$Mpc size limit of non-linearity. Evolution of voids that are smaller than this size is formed by non-linearity effects.}
\label{fig:comparionFSL}
\end{figure}

To clarify the similarities and differences between minor and growing void size distributions that are derived
from equation (\ref{probabilitys1size}), in Fig. \ref{fig:comparionFSL}, the size distribution equation
(\ref{probabilitys1size}) is displayed for embedded/minor voids with the barrier ratio $\gamma=0.3$
$(\tilde{\gamma}=0.15)$ (left) and growing voids with $\gamma\gg 0.5$ ($\tilde{\gamma}\gg 0.25$) (right) for
self-similar with index $n=0$, $-1$, $-1.5$, $-2$ models are presented. According to this, the size distribution
equation (\ref{probabilitys1size}) of embedded/minor and embedded/growing voids indicate a broad range in void
sizes with decreasing spectral index. Generally speaking, the distribution of minor and growing voids decreases
with decreasing spectral index. Embedded/minor void distributions show two cut-offs at very small and large
sizes, while growing void population distributions indicate one cut-off at the largest sizes. In addition, in
the growing void population, the smallest size voids show the highest distributions. Taking into account that
decreasing spectral index indicates hierarchical clustering, obtaining larger size embedded voids in the
self-similar models with smaller spectral index is expected. Therefore, even though very large and small
embedded voids are possible to be seen in the embedded void population, they are rare compared to the embedded
voids that form the peaks of the distributions. On the other hand, the growing void population retains numerous
small voids while large voids are less numerous.

\begin{figure}
\centering
\begin{tabular}{cc}
\includegraphics[width=0.45\textwidth]{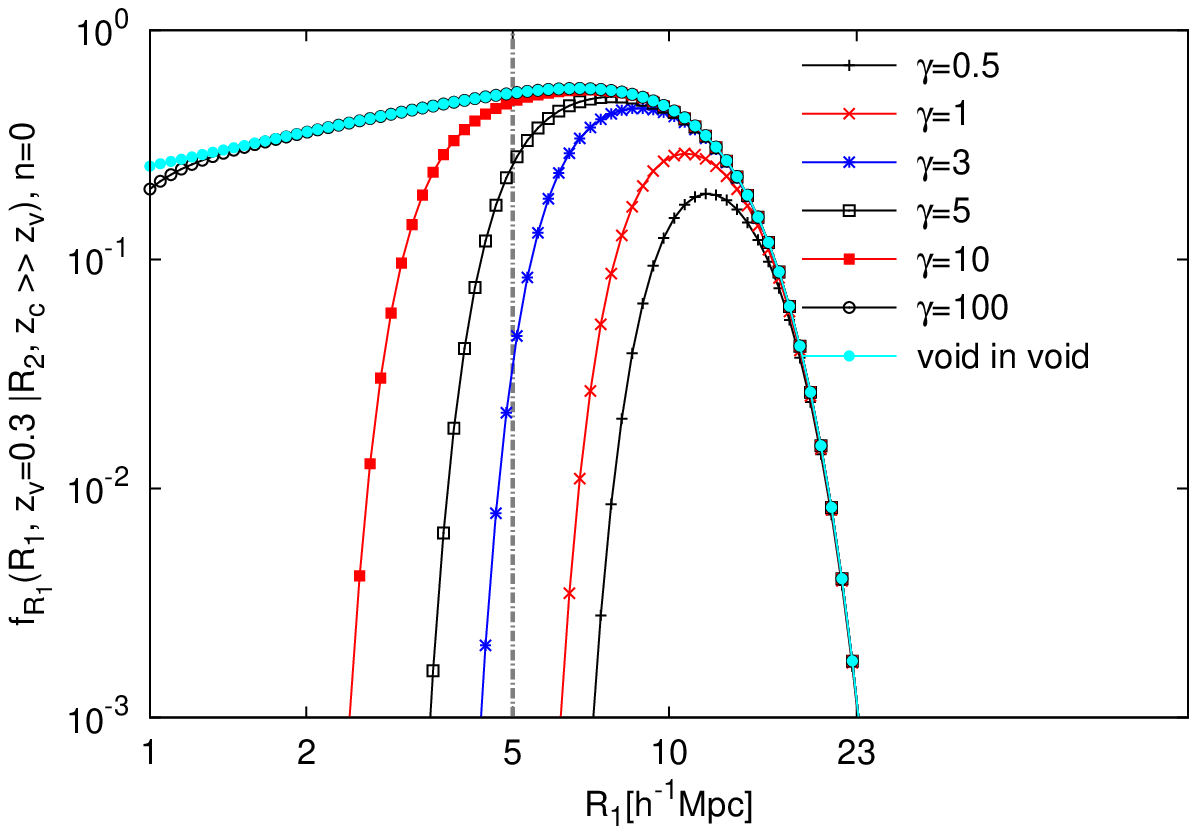}
\includegraphics[width=0.45\textwidth]{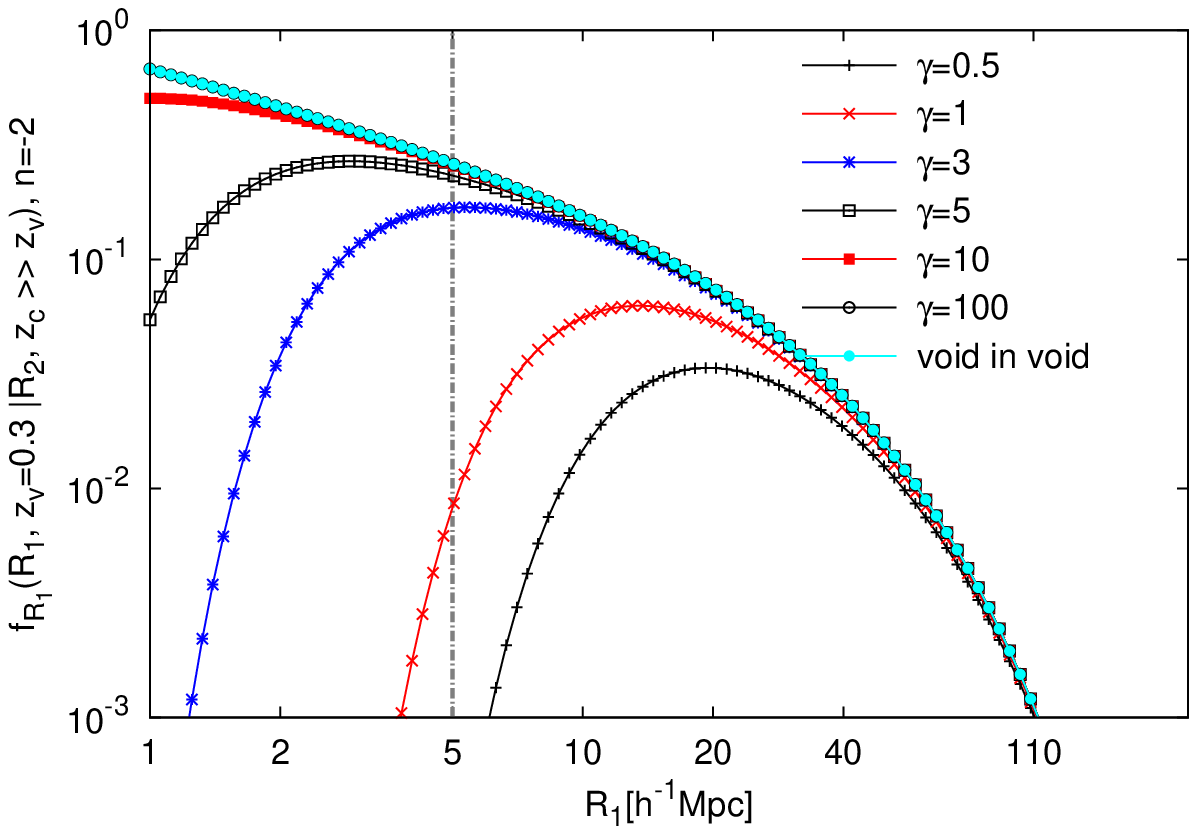}\\
\includegraphics[width=0.45\textwidth]{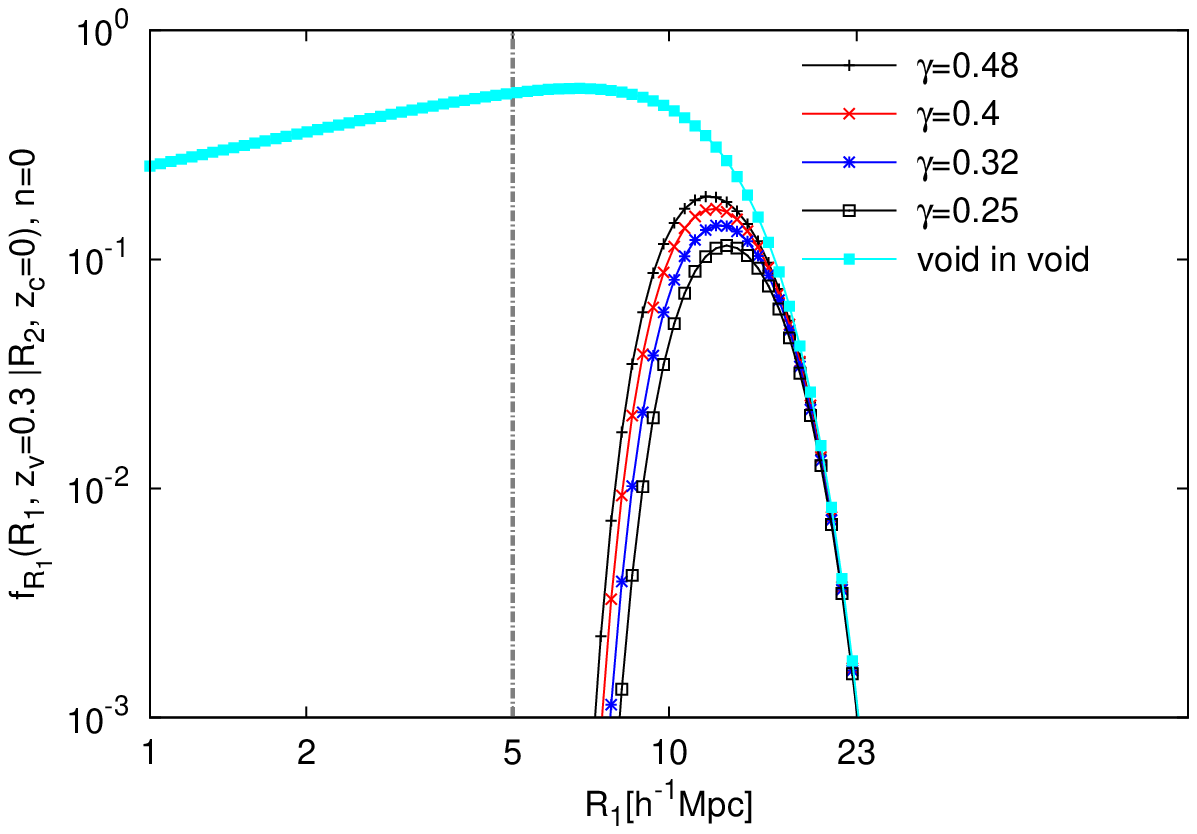}
\includegraphics[width=0.45\textwidth]{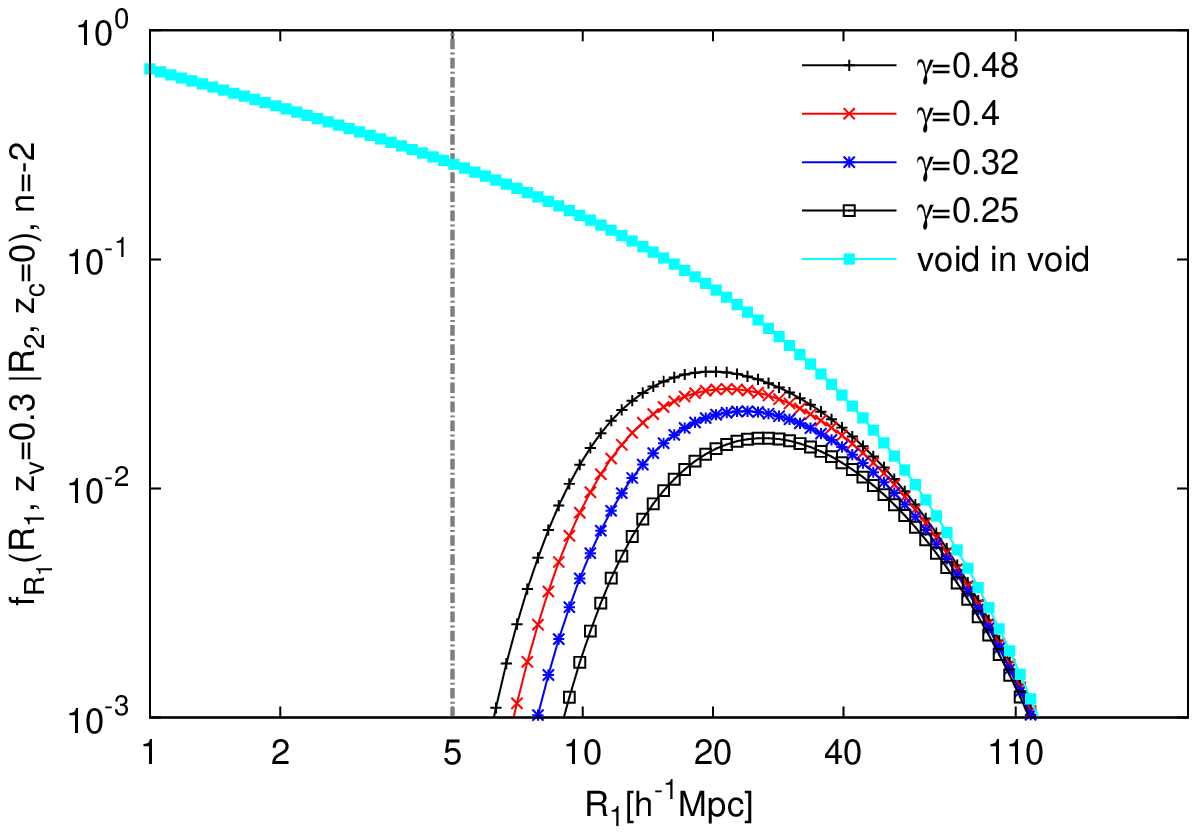}
\end{tabular}
\caption{The barrier ratio dependence of the size distribution of growing (upper) and minor (lower) voids in the
two-barrier EPS formalism for self-similar models with index $n=0$ and $-2$. The two-barrier growing void size
distribution is presented for the barrier ratio $\gamma\geq 0.5$ and then $z_{c}\gg z_{v}=0.3$ while the barrier
ratio dependence of the size distribution of minor voids is displayed by choosing $\gamma < 0.5$, $z_{v}=0.3$ and
$z_{c}=0$.}
\label{fig:gammadependence}
\end{figure}

As was pointed out before, the barrier height ratio has the key importance of defining hierarchical evolution of
the void population by making a significant difference between minor and growing voids in the two-barrier
excursion set. The barrier ratio dependence of the size probability distribution function (\ref{probabilitys1a})
is shown in Fig. \ref{fig:gammadependence}. The vertical line displayed in all panels is presented, with size
$R_{1}=5$ $h^{-1}$Mpc. Voids with sizes smaller than this value may show non-linear effects. In this study the
non-linear effects are not considered. That is why voids with size $R_{1} \geq 5$ $h^{-1}$Mpc are considered. In
addition, in all panels, the void in void size distribution is inserted to show how the value of the barrier
ratio affects the size distribution.

In Fig. \ref{fig:gammadependence}, in the upper panels, growing void size distribution is presented for
self-similar models with index $n=0$ and $n=-2$. In both upper panels, the barrier ratio is chosen as
${\gamma}\geq 0.5$ and $z_{c}\gg z_{v}$. When the barrier height ratio increases, the conditional size
distribution starts behaving like the size distribution of growing voids that are not affected by their
environments as much as minor voids are. Particularly, when barrier ratio reaches large values ${\gamma}=100$
indicating that collapse barrier is very large compared to merging barrier $\delta_{c}\gg \delta_{v}$ ($z_{c}\gg
z_{v}$). As a result, placing merging barrier $\delta_{c}$ at very large redshift values, it is ensured that the
distribution function becomes dominated by only merging events. Therefore, contribution of minor voids can be
negligible in the size distribution. In the lower panels of Fig. \ref{fig:gammadependence}, the minor void size
distribution is displayed for the same models as in the upper panels. In the case of the minor void population
size distribution, the barrier ratio is chosen as ${\gamma}< 0.5$ and $z_{c}=0$ and $z_{v}=0.3$. In minor and
growing void size distributions, the peaks shift to small sizes when the barrier height ratio increases. While
this shifting peak behavior is not striking for minor voids with ${\gamma}< 0.5$, for growing voids with
increasing barrier ratio $\gamma \geq 0.5$ the shifting behavior becomes prominent. Also, peaks become flattened
with increasing barrier height ratios. Therefore, there is a large diversity of sizes for growing voids. The
peaks become flatter when the barrier ratio approaches the one-barrier $\tilde{\delta}_{v}$
($\delta_{c}\gg\tilde{\delta}_{v}$). Therefore, the distribution turns into the "void in void" one.

\begin{figure}
\vspace{70mm}
\hspace{5mm}
\includegraphics[scale=0.86]{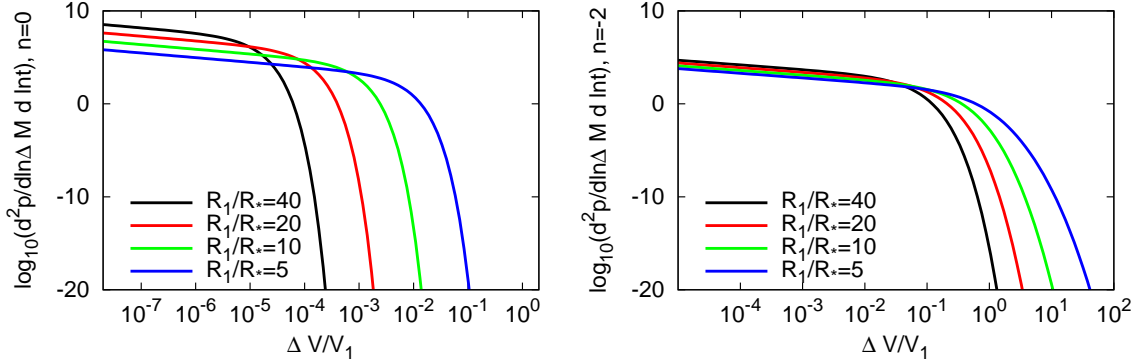}
\vspace{-70mm}\caption{The merger rates of voids given by equation (\ref{mergingratea}) for the self-similar
models with the index $n=0$ and $-2$ in terms of different radius ratios. They have the curve which is the
highest on the left of each plot $R_{1}/R_{*}= 40$ and successive curves are $R_{1}/R_{*}= 20, 10, 5$.}
\label{fig:selfmergerrates}
\end{figure}

Another probability density function that can be derived from the random walks is the probability of a
trajectory first up crossing $\delta_{c}$ then $\tilde{\delta}_{v}$ at $S_{1}$,
\small
\begin{eqnarray}
f_{S_{2}}(S_{2},\delta_{{c}}\big|S_{1},\tilde{\delta}_{v})
d{S_{2}}&&=\frac{1}{\sqrt{{2\pi}}}\frac{\delta_{c}\left(\tilde{\delta}_{v}-\delta_{{c}}\right)}{\tilde{\delta}_{v}}
{\left[\frac{S_{1}}{S_{2}\left({S_{1}}-{S_{2}}\right)}\right]^{3/2}}
\exp\left(-\frac{\left(\delta_{{c}}S_{1}-{\tilde{\delta}_{v}}{S_{2}}\right)^2}{{2S_{1}}S_{2}\left(S_{1}-{S_{2}}\right)}\right)\nonumber\\
&&\times
\exp\left(-\frac{1}{4}\frac{1}{\tilde{\gamma}}\frac{1}{\left(1+\tilde{\gamma}\right)^2}\left(\frac{S_{1}-S_{2}}
{\left(\tilde{\delta}_{v}-\delta_{{c}}\right)^2}+\frac{S_{2}}{\delta^{2}_{c}}-\frac{S_{1}}{\tilde{\delta}^{2}_{v}}\right)\right)\exp\left(-2\frac{1}{\left(1+\tilde{\gamma}\right)^4}\left(\frac{\left(S_{1}-S_{2}\right)^2}
{\left(\tilde{\delta}_{v}-\delta_{{c}}\right)^4}+\frac{S^2_{2}}{\delta^{4}_{c}}-\frac{S^2_{1}}{\tilde{\delta}^{4}_{v}}\right)\right)dS_{2}.
\label{probabilitys2a}
\end{eqnarray}
\normalsize
\noindent
This helps one to obtain merging rates of the two types of void processes. To do this, first the mean transition
rate is derived by taking the limit $\delta_{c}\rightarrow \tilde{\delta}_{v}$,

\begin{eqnarray}
\lim_{\delta_{c}\rightarrow\tilde{\delta}_{v}}\frac{{d^2} f_{S_{2}}(S_{2},\delta_{{c}}\big|S_{1},\tilde{\delta}_{v})dS_{2}}
{dS_{2}d\delta}\equiv \lim_{\delta_{c}\rightarrow\tilde{\delta}_{v}}\frac{{d^2} p}
{dS_{2}d\delta},
\end{eqnarray}
in which the functional $p$ is $p\equiv f_{S_{2}}(S_{2},\delta_{{c}}\big|S_{1},\tilde{\delta}_{v})dS_{2}$.
This limit leads to a distribution function being obtained which changes in terms of barrier steps
$d{\delta}=\tilde{\delta}_{v}-\delta_{{c}}$ and as a result of the limit $\delta_{{c}}\rightarrow
\tilde{\delta}_{v}$ corresponding to going back in time to obtain merging events, in equation
(\ref{probabilitys2a}) the last two exponentials indicating the tail of the faction disappear. Therefore,
the distribution function is reduced to the following form,

\begin{eqnarray}
\frac{{d^2} p}{d{S_{2}}d\tilde{\delta}_{v}}\left({S_{1}}\rightarrow
{S_{2}},\tilde{\delta}_{v}\right) &=& \frac{1}{\sqrt{{2\pi}}}
{\left[\frac{S_{1}}{S_{2}\left({S_{1}}-{S_{2}}\right)}\right]^{3/2}}\exp\left[-\frac{\tilde{\delta}^2_{v}}{2}
\frac{\left(S_{1}-{S_{2}}\right)}{S_{1}S_{2}}\right]d{S_{2}}d{\delta}.
\label{meanrateba}
\end{eqnarray}
\noindent
Equation (\ref{meanrateba}) is used by LC93 for overdense haloes in order to derive the merging rates of haloes and by \cite{eszrarussell} for
gradually merging voids. Note that equation (\ref{meanrateba}) is the same as the form of the one-barrier void merging distribution of \cite{eszrarussell}.
The only difference comes from the behaviors of the barriers; while one of the barriers is the collapse barrier
$\delta_{c}$ instead of a second void merging barrier $\tilde{\delta}_{v_{2}}$. Equation (\ref{meanrateba})
also can be interpreted as one or more merging void event(s) depending on expansion/underdense $\tilde{\delta}_{v}$ and collapse/over\-dense $\delta_{c}$ barriers.
While any finite interval of $\Delta \delta =\tilde{\delta}_{v}-\delta_{c}$ at $\Delta{S}$ shows the cumulative
effect of more than one merger, an infinitesimal interval $d\delta$ at $d{S}$ indicates a single void merger
event. Hence, equation (\ref{meanrateba}) represents the probability of
a void with volume scale $S_{1}$ merging at later times with another void of volume $\Delta V =V_{2}-V_{1}$.
Thus the rate of merging, divided by the volume $\Delta V$ of the void that it is being merged with, is,

\begin{eqnarray}
\frac{d^2 p}{d\ln \Delta V d\ln t}({V_{1}}\rightarrow V_{2},|t)
&=&\sqrt{\frac{2}{\pi}}\frac{\Delta
V}{V_{2}}\frac{\tilde{\delta}_{v}(t)}{\sqrt{S_{2}}}\left|\frac{d\ln\sqrt{S_{2}}}{d\ln
V_{2}}\right|\left|\frac{d\ln{\tilde\delta_{v}}}{d\ln
t}\right|\frac{1}{\left(1-S_{2}/S_{1}\right)^{3/2}}\exp\left[-\frac{\tilde{\delta}^2_{v}}{2}\left(\frac{1}{S_{2}}
-\frac{1}{S_{1}}\right)\right].
\label{mergingratea}
\end{eqnarray}
\noindent
The merger rates of embedded/minor and growing voids are formulated by the same equation. Consequently, their
merging rates show the same behavior. Fig. \ref{fig:selfmergerrates} presents merger rates of embedded/minor
voids of a given size in terms of self-similar models. According to this, in Fig. \ref{fig:selfmergerrates}
the merger rates of the self-similar models decrease with increasing spectral index.
This behavior is less prominent in voids with smaller size ratios ${R/R_{*}}\leq\quad 10 h^{-1}$Mpc than their
large size counterparts ${R/R_{*}}> 10\quad h^{-1}$Mpc. Due to this result, large size voids contribute to the
merging events more than smaller size voids.

\section{Timescales in The Merging Tree Algorithm}
In the following sections, survival and formation times of the two-barrier void hierarchy are defined and
formulated. Recall that here the linear over- and underdensities are used as time parameters following LC93.
This is only allowed in the EdS Universe since linearly extrapolated densities are constant and only time
dependency comes from the linear growth factor $\Upsilon$.

\subsection{Void Collapse and Failure Times}
As is shown in the previous section, the merging rates of embedded voids have the same merging features as
growing voids. The question is: What is the main distinction of these populations if it is not their merging
character? The answer is their survival probabilities which form a separation
between the two groups. Due to their environmental characteristics, it is expected that their survival times
should be different since one of the groups vanishes under its gravitational collapse and the other one merges
gradually. However in reality, before they collapse, we cannot really distinguish between them.

By following the definition of the survival time of growing voids by \cite{eszrarussell}, the void survival
time is defined as the time $\delta_{surv}$ when the volume gets doubled $2V$ due to merging. The survival
probability function of a void succeeding to merge and double in size, is given by,

\begin{eqnarray}
\underbrace{P(S>S_{2},{{\delta}_{v_{2}}}|S_{1},{\delta}_{v_{1}})}_{\text{probability of survival
beyond}\phantom{a}S_{2}} = 1-\underbrace{P(S<S_{2},{\delta}_{v_{2}}| S_{1},{\delta}_{v_{1}})}_{\text{probability
of a void failing before reaching}\phantom{a} S_{2}},
\end{eqnarray}
\noindent
where the survival probability distribution $P(S>S_{2},{\delta}_{v_{2}}|S_{1},{\delta}_{v_{1}})$ varies between
one and zero indicating survival and death processes. Although the same survival profile from both void
populations is expected, the survival probability of the void population shows two different
behaviors based on their survival size distribution, which is strongly related with their environmental
properties. For example for embedded voids, it is expected that the probability of a void failing before
reaching its double size should be related with the collapse event due to the presence of
overdense regions in their environment. The effect of gravitational collapse of overdense regions slows down the
void merging process and forces the void evolution to stop, or may even make them vanish. Here, these
two different behaviors are obtained depending on the barrier ratios. As is
mentioned before, the value of the barrier ratio is the key element of the two-barrier excursion set
since it makes a distinction between merging and collapsing behaviors of the void population. In this sense, the
barrier ratio $\tilde{\gamma}$ or ${\gamma}$ can be considered as an environmental indicator. When the ratio of
the two barriers is higher than the value $\delta_{c}/\tilde{\delta}_{v}\geq 0.5$, the collapse barrier
becomes larger $\delta_{c}\gg\tilde{\delta}_{v}$. As is known, this is the indication of gradual merging events
\citep{sw,eszrarussell}. In the limit of $\tilde{\gamma}< 0.25$ (or ${\gamma} < 0.5$), the two barriers get
closer. Due to the collapse barrier, the merging event cannot be a continuous event. By taking
into account these two possibilities, here two survival probability functions of voids are obtained, whereby
they merge gradually and first they merge and later on collapse.

\emph{\textbf{Survival probability of growing voids:}}
The survival probability of voids that merge at the volume scale
$S_{2}$ and survive and continue merging after this certain scale, is defined as the probability of a void
making its transition from
$S < S_{2}$ to $S > S_{2}$ at barrier $\delta > {\delta_{v_{2}}}$ and its explicit form given by
\cite{eszrarussell},

\small
\begin{eqnarray}
P(S> S_{2},{{\delta}_{v_{2}}}|S_{1},{\delta}_{v_{1}})&=&
1-\int^{S_{2}}_{0}f_{S_{2}}\left(S^{\prime}_{2},{\delta}_{v_{2}}|
S_{1},{\delta}_{v_{1}}\right)dS^{\prime}_{2}\nonumber\\
&=&\sqrt{\frac{2}{\pi}}{\delta}_{v_{1}}-{\delta}_{v_{2}}\left|\frac{{\delta}_{v_{2}}}{{\delta}_{v_{1}}}\right|
\left[\sqrt{\frac{S_{2}}{S_{1}\left(S_{1}-S_{2}\right)}}e^{-\frac{{\delta}^{2}_{v_{1}}}{2}\left(\frac{1}{S_{1}}
-\frac{1}{S_{2}}\right)}+\sqrt{\frac{\pi}{2}}\frac{1}{{\delta}_{v_{1}}}\left(\frac{{\delta}^2_{v_{1}}}{S_{1}}-1\right)
\times\text{erf}\sqrt{\frac{{\delta}^2_{v_{1}}}{2}\left(\frac{1}{S_{1}}-\frac{1}{S_{2}}\right)}\right].
\end{eqnarray}
\normalsize
\noindent
Since the collapse barrier tends to be infinite, this implies the collapse event occurred in the past. Therefore,
minor void contribution to the survival probability becomes negligible. As a
result, voids continue their merging behavior gradually. Therefore, the two-barrier problem is reduced to a
one-barrier problem and void evolution can be modelled with only one type of barrier $\delta_{v}$. It is also possible to derive the distribution of times \citep{eszrarussell},

\begin{eqnarray}
{F}_{{\delta}_{v_{2}}}&=& -d{\delta}_{v_{2}} \left(\frac{\partial P(S>
S_{2},{\delta}_{v_{2}}|S_{1},{\delta}_{v_{1}})}{\partial {\delta}_{v_{2}}}\right) =
d{\delta}_{v_{2}}
\left(\frac{\partial P(S< S_{2},{\delta}_{v_{2}}|S_{1},{\delta}_{v_{1}})}{\partial
{\delta}_{v_{2}}}\right)\nonumber\\
&=&\sqrt{\frac{2}{\pi}}\left|1-2\frac{\delta_{v_{2}}}{\delta_{v_{1}}}\right|
\left[\sqrt{\frac{S_{2}}{S_{1}\left(S_{1}-S_{2}\right)}}e^{-\frac{\delta^{2}_{v_{1}}}{2}\left(\frac{1}{S_{1}}-\frac{1}{S_{2}}\right)}
+ \sqrt{\frac{\pi}{2}}\frac{1}{|\delta_{v_{1}}|}\left(\frac{\delta^2_{v_{1}}}{S_{1}}-1\right)
\text{erf}\sqrt{\frac{|\delta^2_{v_{1}}|}{2}\left(\frac{1}{S_{1}}-\frac{1}{S_{2}}\right)}\right].
\label{failurerateGV}
\end{eqnarray}
\noindent
\cite{eszrarussell} defines the survival time distribution of a growing void population as the probability of a
void with volume
scale $S_{1}=S(V_{1})$ being incorporated into a system of volume larger than the corresponding scale
$S_{2}=S(V_{2})$ during the time interval
$d{\delta}_{v_{2}}$ by adapting the survival times of dark matter haloes of LC93. As is pointed out in
\cite{eszrarussell}, the negative sign of the survival time distribution, equation (\ref{failurerateGV}), is
known as the conditional failure rate or hazard function
in statistical mathematics. Then equation (\ref{failurerateGV}) measures the failure rate of void radii that
could not merge at a given redshift or measures
the risk of voids not merging/growing for a given size with respect to a time interval.

\begin{figure}
\centering
\begin{tabular}{cc}
\includegraphics[width=0.45\textwidth]{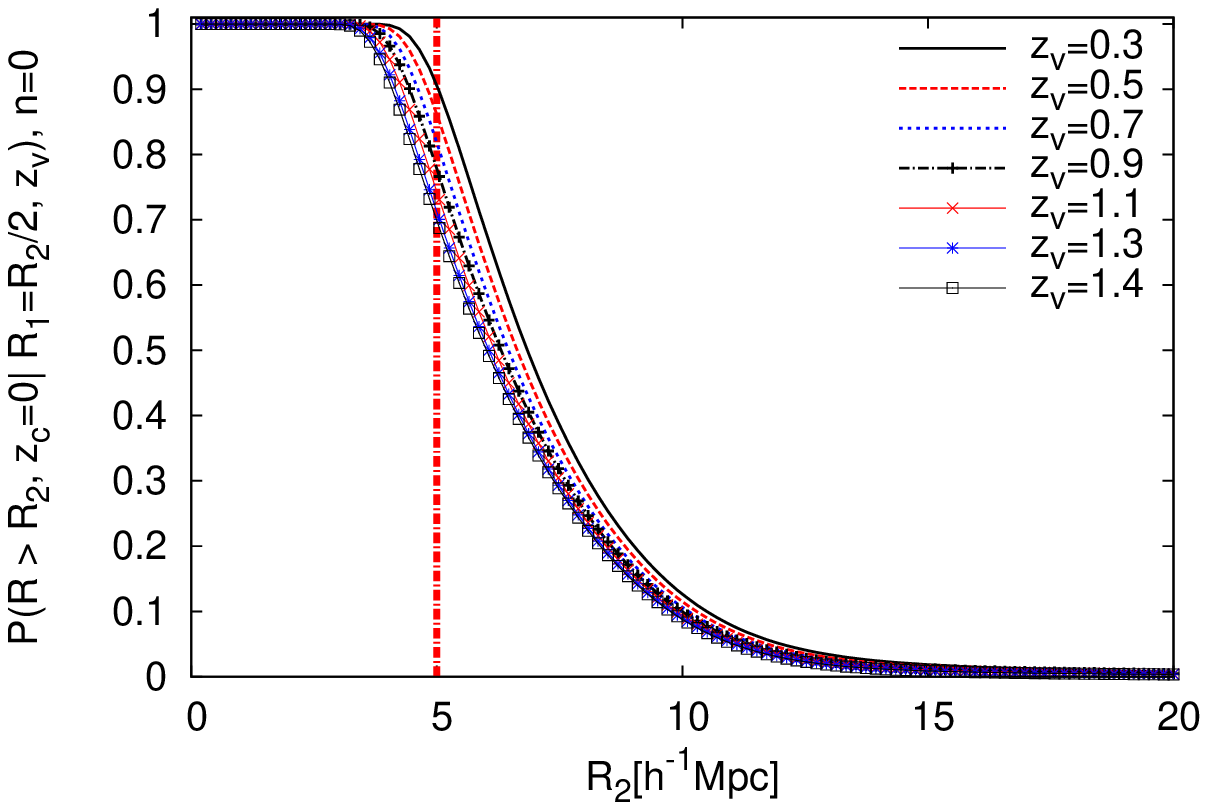}
\includegraphics[width=0.45\textwidth]{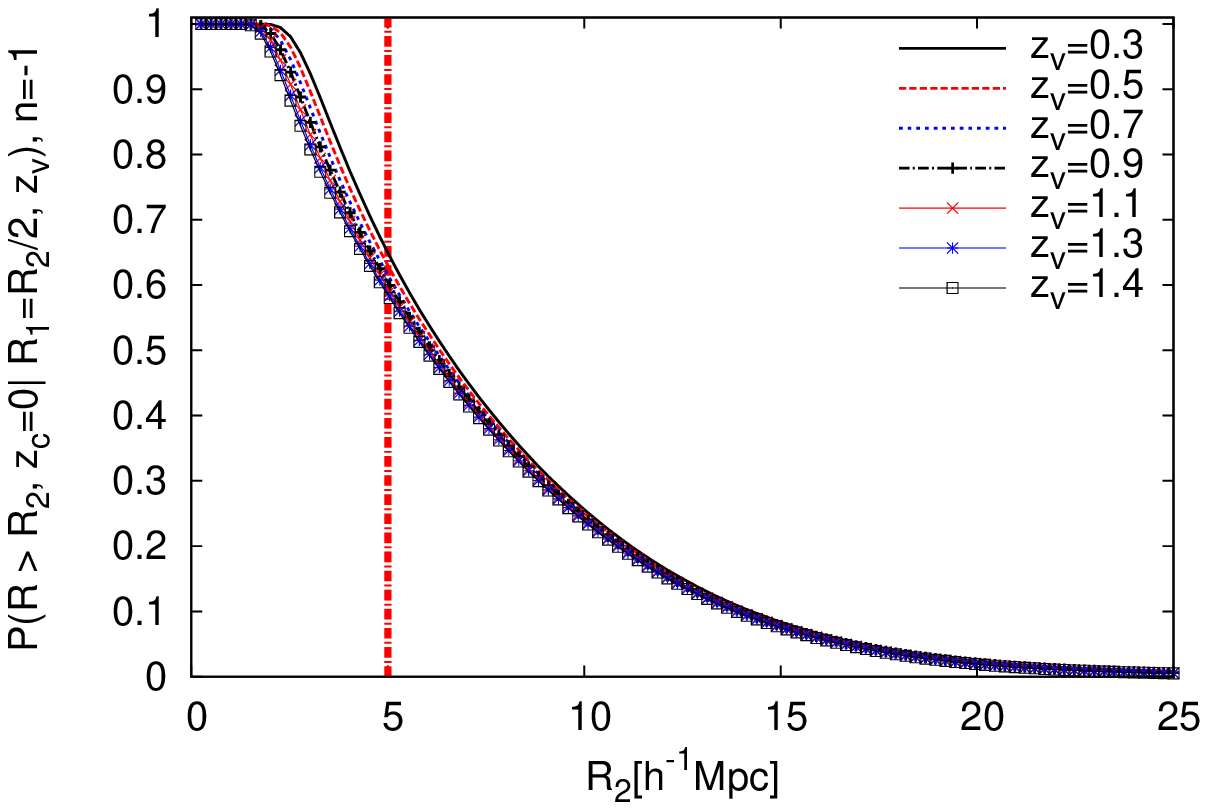}\\
\\
\\
\\
\\
\\
\\
\\
\\
\\
\\
\\
\\
\\
\\
\\
\includegraphics[width=0.45\textwidth]{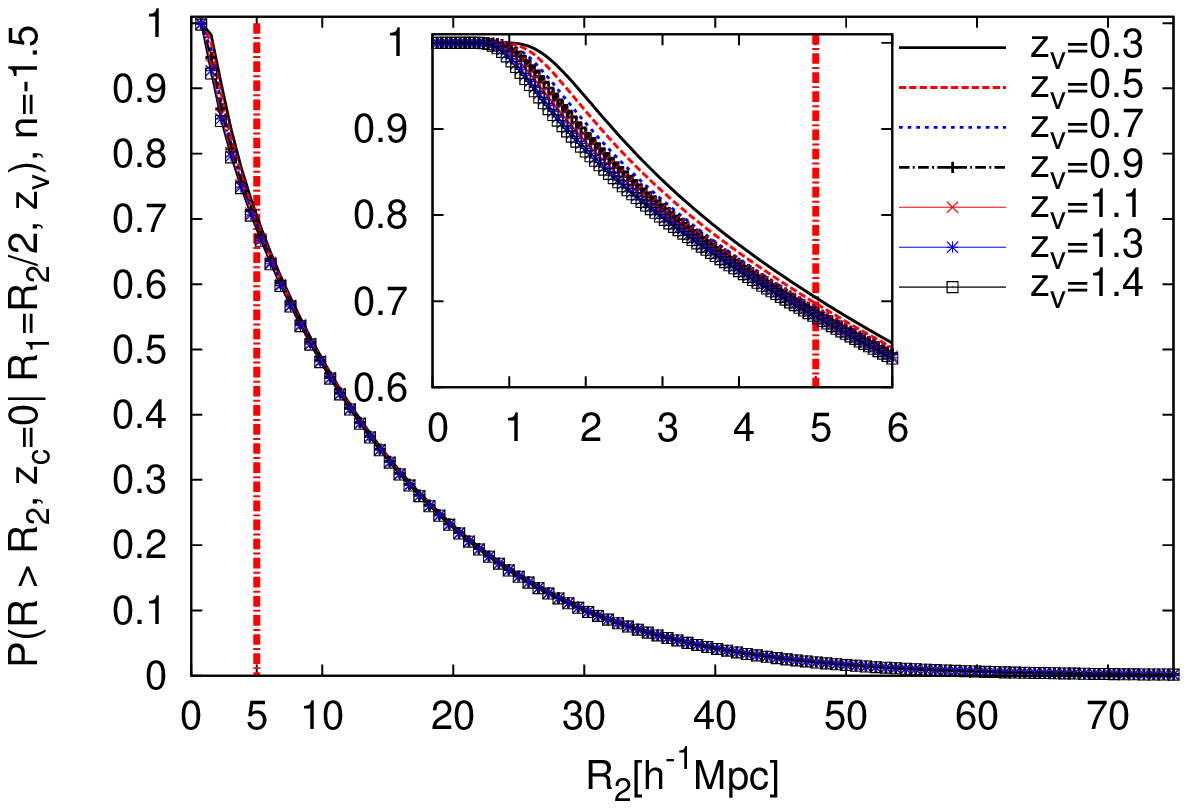}
\end{tabular}
\vspace{-55mm}\caption{The collapse size distribution of embedded voids for self-similar models with index
$n=0$, $-1$, $-1.5$. In all models, the collapse probability of an embedded void is displayed in terms of radius
$R_{1}$ at a given redshift $z_{v}$ which later on collapses at $z_{c}=0$ when it doubles in size $R_{2}=2$
$R_{1}$.}
\label{fig:analyticsurvivalvolumessmall}
\end{figure}

\emph{\textbf{Collapse probability of embedded/minor voids:}}
The concept of survival probability of embedded/minor voids is slightly different than their growing
counterparts. Unlike growing voids, embedded voids collapse at the volume scale $S_{2}$ instead of surviving.
Therefore, the survival embedded void probability is named as the collapse probability, since it is the chance of an embedded void to collapsing. Then the collapse probability is defined as the probability of a void
making its transition from $S < S_{2}$ to $S > S_{2}$ at barrier ${\delta_{c}}$ and vanishing at this collapse
barrier $\delta_{c}$. Similarly to the survival probability of a growing void, the mathematical description of
the collapse probability of an embedded/minor void is given by,

\begin{eqnarray}
\underbrace{P(S\geq S_{2},{{\delta}_{c}}|S_{1},\tilde{\delta}_{v})}_{\text{probability of collapsing
beyond}\phantom{a}S_{2}} = 1-\underbrace{P(S\leq S_{2},{\delta}_{c}|
S_{1},\tilde{\delta}_{v})}_{\text{probability of a void failing before reaching}\phantom{a} S_{2}}.
\label{Pcollmathdef}
\end{eqnarray}
Similar to the survival probability of growing void population, the collapse probability of the embedded voids
$P(S\geq S_{2},\delta_{c}|S_{1},\tilde{\delta}_{v})$ varies between one and zero, corresponding to success in
collapse and failing. Hence, the collapse probability by using equation (\ref{Pcollmathdef}) and equation (\ref{CD}) is derived analytically as,

\begin{eqnarray}
P_{coll}\left(S\geq\phantom{a}S_{2},{\delta_{c}}|S_{1},\tilde{\delta}_{v}\right)&=&
1-\int^{S_{2}}_{0}f_{S_{2}}\left(S^{\prime}_{2},{\delta}_{c}| S_{1},{\tilde{\delta}_{v}}\right)dS^{\prime}_{2}
\nonumber\\
&=&1-
e^{-\frac{2}{3}\frac{S_{1}}{\tilde{\delta}^{2}_{v}}-{y}_{n}\frac{S^2_{1}}{\tilde{\delta}^{4}_{v}}}
\text{erf}\left(\frac{\delta_{c}}{\sqrt{2S_{1}}}\left(\frac{{S_{1}}-2{S_{2}}}{
{\sqrt{S_{2}\left({S_{1}}-{S_{2}}\right)}}}
\right)\right),
\label{solprobssmall}
\end{eqnarray}
\noindent
where ${y}_{n}$ (in which $n=0$, $-1$, $-1.5$, $-2$) is a model-dependent numerical value that admits the
following values for the self-similar models,

\begin{eqnarray}
{y}_{0}= 2.77,\phantom{a} {y}_{-1}= 2.98,\phantom{a}{y}_{-1.5}=3.307,\phantom{a}{y}_{-2}=3.86.
\label{y}
\end{eqnarray}
\noindent
Here, the spherical model is adopted which leads to the barrier ratio ${\gamma}= 0.5$
\citep{sw} and binary merging which gives the relation of scales depending on the self-similar model
$S_{2}\approx 2^{-\alpha}S_{1}$. The scales should be set as $S_{1}= S(V)$,
$S_{2}= S(2V)$ due to the binary merging method. The dynamical variables of the hierarchical
evolution are defined by the linear over- and underdensities
$\tilde{\delta}_{v}=\tilde{\delta}_{v}(t)$ and ${\delta_{c}}= \delta(t_{surv})$. Equation
(\ref{solprobssmall}) also defines the probability that a void
merges into a system of volume larger than the corresponding volume scale $S_{2}$ at collapse barrier
$\delta_{c}$. Then at collapse barrier $\delta_{c}$, this void collapses. Fig.
\ref{fig:analyticsurvivalvolumessmall} shows the collapse probabilities of embedded voids
for given redshifts, when they incorporate into radius $R_{2}$ for the self-similar models. According to this,
in all models, the collapse probability shows similar behavior in terms of redshift. In other words, in a given
model the probability distributions in terms of different redshifts slightly deviate from each other. In
addition, in all models the collapse probabilities decrease
with increasing size. This is an expected result since embedded/minor voids are to be affected by
their environments. Embedded voids most likely collapse.  Note that the radius range of the collapse probability
distribution with value $P(R_{2})=1$ indicates $100\%$ success in collapsing. After the maximum collapse
radius at a given redshift, the probability distribution starts decreasing towards large radii. Then it reaches
the value zero indicating no collapse for voids in this radius at this given redshift. As a general feature of
the models, the smallest size voids have the highest collapse probability. In addition, The size range of voids
with highest collapse probability decreases with decreasing spectral index and also decreases towards higher
redshift values in a given model. For example, in the self-similar models with index $n=0$, $-1$, $-1.5$, voids
with radius up to
$\leq 4.5\quad h^{-1}$Mpc, $\leq 2.5\quad h^{-1}$Mpc and $\leq 1.5\quad h^{-1}$Mpc
are destined to collapse at present-day $z_{c}=0$. All voids with radii above these
limits in the related model are less likely to collapse. The collapse radii with zero probability
$P(R_{2})=0$ define the lowest radius limit of immortal voids that
continuously grow/merge without collapsing. The radii of embedded voids that do not
collapse ever, depending on the models are $\geq 65\quad h^{-1}$Mpc for $n=-1.5$ ,
$\geq 35\quad h^{-1}$Mpc for $n=-1$ and $\geq 20\quad h^{-1}$Mpc for $n=0$.

\begin{figure}
\vspace{60mm}
\centering
\includegraphics[scale=0.8]{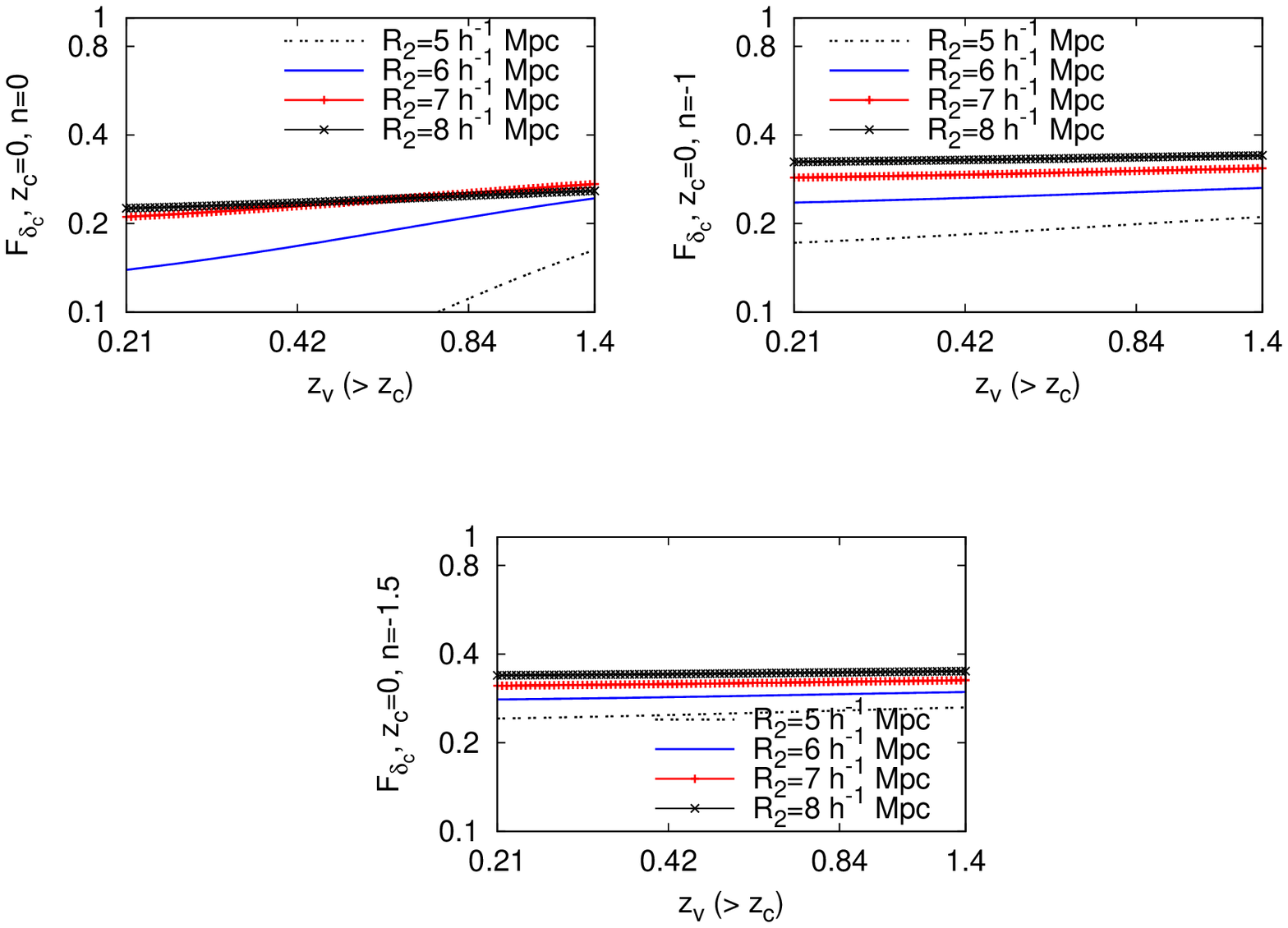}
\caption{The failure rate of embedded voids in terms of merging redshift $z_{v}$ for given radii $5$, $6$, $7$,
$8$ $h^{-1}$Mpc in the self-similar models with index $n=0$, $-1$, $-1.5$.}
\label{fig:survtimesmall}
\end{figure}

Apart from the collapse probability, the collapse failure rate of the embedded/minor void population is also
defined. The collapse failure rate can be defined as the change of probability of a void that will not collapse
at a given redshift interval. In other words, some embedded voids fail to collapse at a given redshift or a
given size. The collapse failure rate is given by the negative signed derivative of the collapse probability in
terms of $\delta_{c}$. Hence, the collapse failure rate of embedded/minor voids is given by,

\begin{eqnarray}
{F}_{\delta_{c},{coll}}= d\delta_{c}
\left(\frac{\partial P(S \leq S_{2},\delta_{c}|S_{1},\tilde{\delta}_{v})}{\partial \delta_{c}}\right)
=\sqrt{\frac{2}{\pi
}}e^{-\frac{2}{3}\frac{S_{1}}{\tilde{\delta}^{2}_{v}}-y_{n}\frac{S^2_{1}}{\tilde{\delta}^{4}_{v}}}e^{-
\frac{\left(S_{1}-2 S_{2}\right)^2}{2 S_{1}S_{2}\left(S_{1}-S_{2}\right)}\delta^2_{c}}
\frac{\left(S_{1}-2 S_{2}\right)}{\sqrt{S_{1}S_{2}\left(S_{1}-2 S_{2}\right)}},
\label{solprobsmall}
\end{eqnarray}
\noindent
in which $y_{n}$ are the numerical values and given by values (\ref{y}). Equation (\ref{solprobsmall}) then
provides important information by showing at what redshift interval and what size of voids have chance to
collapse or survive. Fig. \ref{fig:survtimesmall} presents the collapse failure rates of embedded/minor voids in
terms of redshift $z_{v} (> z_{c}=0)$ for $5$, $6$, $7$ and $8$ $h^{-1}$Mpc size voids. These embedded voids
start merging at $z_{v}$ before reaching the collapse at redshift $z_{c}=0$ in self-similar models based on
equation (\ref{solprobsmall}). According to this, Fig. \ref{fig:survtimesmall} demonstrates that the collapse
failure rate slightly increases from $n=0$ to $n=-1.5$ for a given size void. In addition to this, the failure
rate becomes constant towards $n=-1.5$. According to this, relatively smaller size voids at high redshifts show
lower collapse failure rate than their larger radius counterparts. While this behavior is dominant in the self-similar-model $n=0$, it becomes constant in redshift. However, smaller size embedded voids present smaller
collapse failure rate. This shows that high redshifts keep slightly more small size voids rather than large
ones.

\begin{figure}
\centering
\begin{tabular}{cc}
\includegraphics[width=0.45\textwidth]{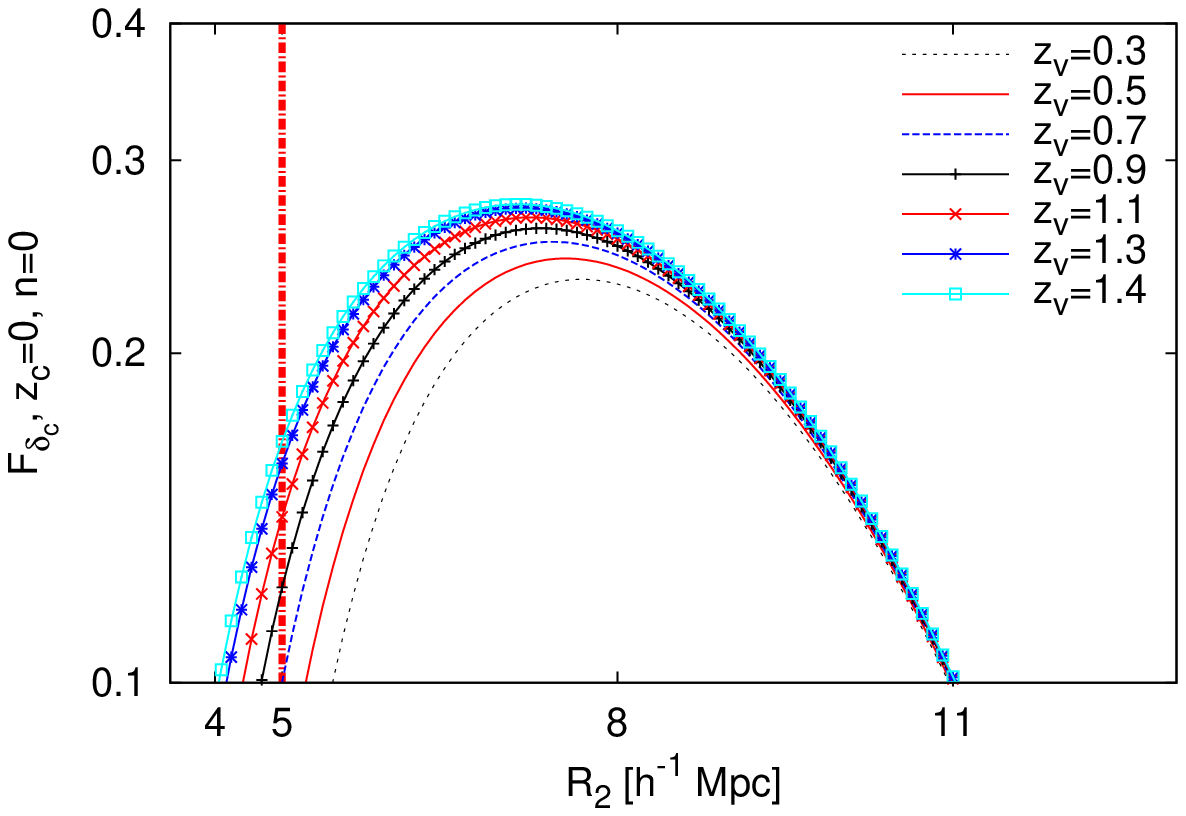}
\includegraphics[width=0.45\textwidth]{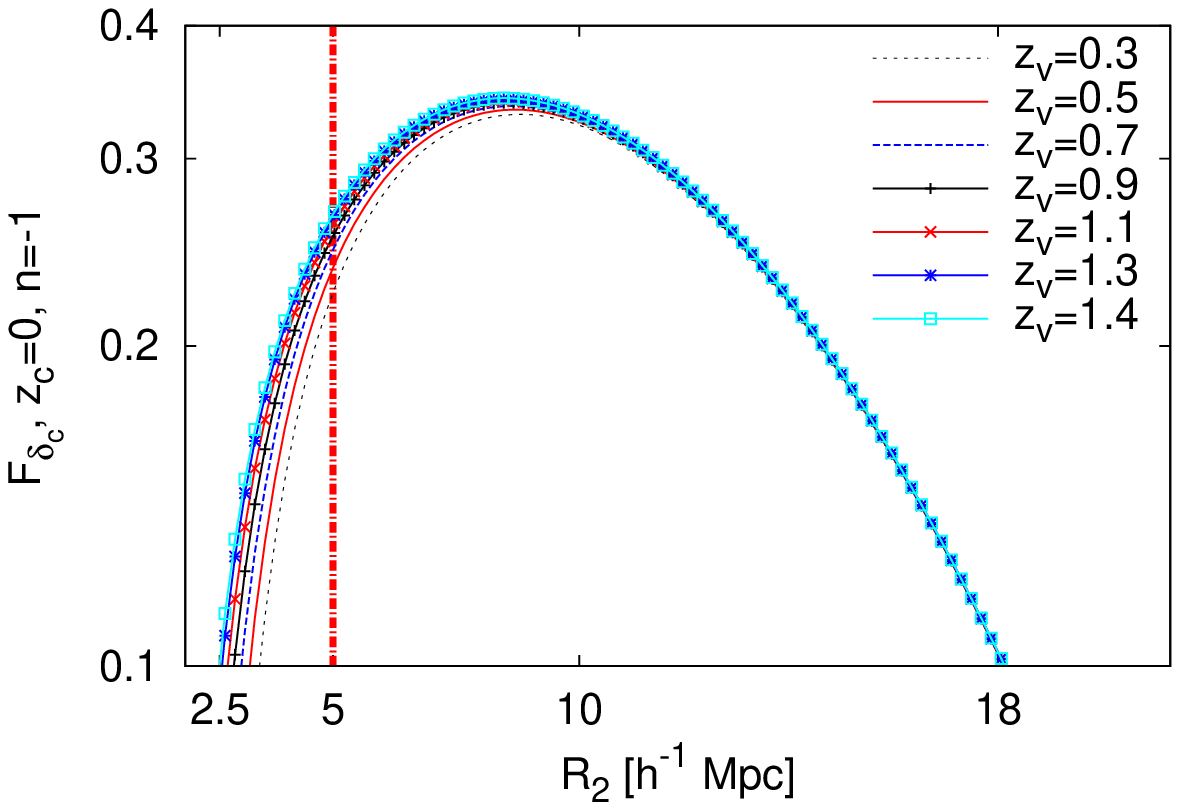}\\
\includegraphics[width=0.45\textwidth]{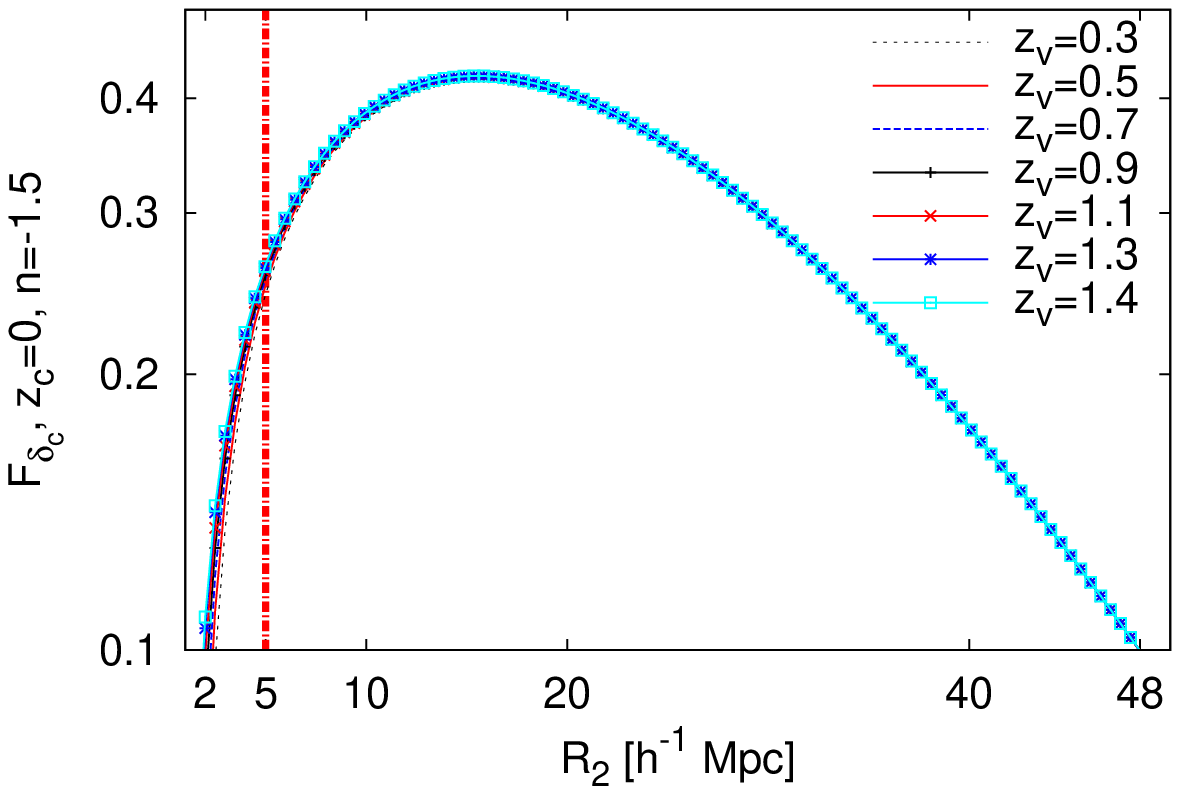}
\end{tabular}
\caption{The failure rate of embedded/minor void regions in terms of the incorporated radius
$R_{2}$ at given redshifts $z_{v}$ when voids start merging and they collapse at $z_{c}=0$ in the self-similar
models (with index $n=0$, $-1$, $-1.5$ from upper to lower panels respectively).}
\label{fig:survivaltimeR2}
\end{figure}

Apart from the collapse failure rate of the void population in terms of redshift, it would be interesting to see
the failure rate in terms of the size distribution for a given redshift range. Therefore,
Fig.~\ref{fig:survivaltimeR2} shows the instantaneous probability of failing rate in terms of void radius
$R_{2}$, in which a void doubles its size and collapses, for self-similar models with index $n=0$, $-1$, $-1.5$.
As is seen in the figure, the collapse failure rates are first increasing and then decreasing with increasing
void radius at a given redshift in each model. This special shape of the collapse failure rate model is called
the lognormal survival model in the context of survival analysis \citep{Kleinbaum}. In Fig.
~\ref{fig:survivaltimeR2}, the collapse failure rate of voids increases towards high redshifts. In the
self-similar models, the collapse failure rates at given redshifts approach each other with decreasing spectral
index. In addition, the failure rate increases with decreasing spectral index.
However, in all models, the collapse failure rate increases until reaching a certain radius at a given redshift,
then it starts decreasing. This means
that in each redshift there is a characteristic (turn around) void radius that behaves as a transition criterion
between collapsing and growing void radius. As a consequence of this, voids at their characteristic failure
radius at a given redshift most
likely dominate the void size distribution and they will dominate the given redshift. As is seen in Fig.
~\ref{fig:survivaltimeR2}, in the models, the peak of the failure rate increases with decreasing spectral index
and decreasing redshift.

\subsection{Volume Formation Time and Extended Void Merging Tree}
Formation time for voids in the two-barrier excursion set is the same as growing voids in the one-barrier
excursion set. Then, the formation time of a void
is defined as the redshift at which a progenitor void of the main void forms with half of the volume of the main
void. Following LC93, it is mentioned that after the formation time (or $S < S_{1}$), the choice of the largest
volume progenitor as the main progenitor defines a continuous track through the merging tree. It is obvious that
formation times have key importance in constructing a merging tree of
voids. However it is pointed out that obtaining formation times from random walks is problematic compared to
obtaining the survival times since the volume assigned for a particle by tracking its density $\delta$ is not
its actual volume but its approximate value (based on LC93). However this fact does not lead to any self
inconsistency in merger rates and survival times. In addition to this, it has been shown that the analytical
counting argument of generating merging histories provides similar results to LC93. The void counting method
from the void perspective is discussed in the following.

\subsubsection{Void Counting, Analytical Method to Extended Void Merging Tree}
The void counting is based on defining the number density of voids in a given volume range. This number density
evolves into a larger range at later times, which
then allows us to obtain the probability distribution of voids with volume ${V_{2}}$ that had a parent in the
volume range ${V_{2}}/2 < {V_{1}} < {V_{2}}$ at $\tilde{\delta}_{v}$ and this probability equals the probability
that its formation time is earlier than $\tilde{\delta}_{v}> {{\delta_{f}}}$. The counting method provides
analytical solutions in terms of self-similar models which can be extended into the CDM model.

\begin{figure}
\centering
\includegraphics[width=0.6\textwidth]{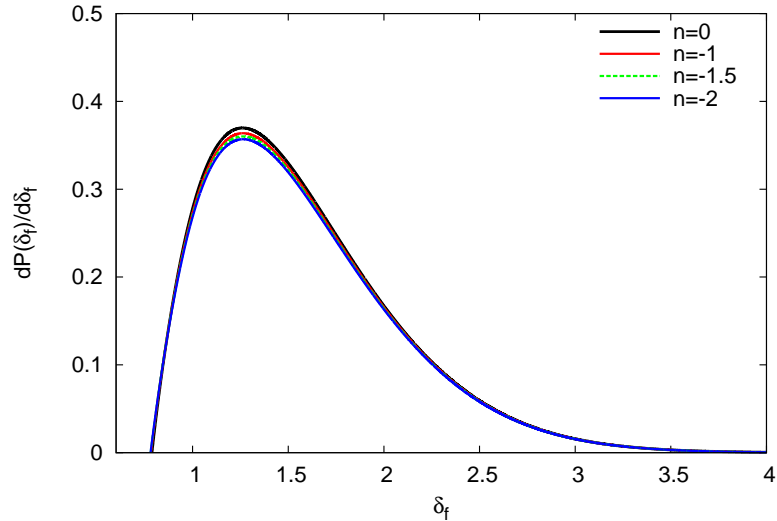}
\caption{Numerical solution of the probability distribution of merging voids. This shows the distribution as a
function of threshold height $\delta_{f}\sim z_{f}$ in terms of self-similar models. As is seen, the difference
between models is small and negligible
and the probability distributions of formation times form a peak around $\delta_{f}=1.25$ for all self-similar
models.}
\label{fig:smallProbmerger}
\end{figure}

The number density of voids $\left(V_{1}, V_{1} + dV_{1}\right)$ at time ${{\tilde{\delta}_{v}}}$, which are
incorporated into voids of volume $\left(V_{2}, V_{2}+dV_{2}\right)$ at time $\tilde{\delta}_{v}>{{\delta_{c}}}$
is,

\begin{eqnarray}
{d^2}n=\frac{d n}{d V_{1}}(V_{1},\tilde{\delta}_{v})d V_{1}
f_{S_{2}}\left(S_{2},{\delta_{c}}|S_{1},\tilde{\delta}_{v}\right){d S}_{2}.
\label{probhalofa}
\end{eqnarray}
\noindent
As long as $V_{2}\leq V_{1} < V_{2}/2$ each trajectory must connect unique voids because there cannot be two
paths each of which contain more than half of the final volume. However, it is possible that the volume of void
$V_{2}$ at ${{\delta_{c}}}$ has no progenitor of volume $< V_{2}/2$ at time ${\tilde{\delta}_{v}}$. The
probability that a void with volume $V_{2}$ at ${{\delta_{c}}}$ has a progenitor in the volume
range $V_{2} > V_{1} > V_{2}/2$ at time ${\tilde{\delta}_{v}}$ is then given by the ratio of voids that evolve
to another volume $V_{2}$ relative to the total void volume $V_{1}$,

\begin{eqnarray}
\frac{d P \left({V_{1}},{\tilde{\delta}_{v}}|{V_{2}},{\delta_{c}}\right)}{d{V_{1}}}=\left(\frac{d n({V_{1}})/ d
{V_{1}}}{d n({V_{2}})/ d {V_{2}}}\right)
{f_{S_{1}}}\left({S_{1}},\tilde{\delta}_{v}|{S_{2}},{{\delta_{c}}}\right)\left|\frac{d{S_{1}}}{d V_{2}}\right|,
\label{probhalovoidaa}
\end{eqnarray}
\noindent
which leads to,

\begin{eqnarray}
\frac{d P\left({V_{1}},{\tilde{\delta}_{v}}|{V_{2}},{\delta_{c}}\right)}{d{V_{1}}}d{V_{1}}=\left(\frac{{V_{2}}}
{{V_{1}}}\right) {f_{S_{1}}}\left({S_{1}},\tilde{\delta}_{v}|{S_{2}},{{\delta_{c}}}\right)d{S_{1}}.
\label{probhalovoidba}
\end{eqnarray}
\noindent
Integration of equation (\ref{probhalovoidba}) over the volume range ${V_{2}}/2 < {V_{1}} < {V_{2}}$ gives the
probability distribution of void ${V_{2}}$ that had a parent in this volume range at ${{\tilde{\delta}_{v}}}$,
which equals the probability that its formation time is earlier than this,

\begin{eqnarray}
P\left(\delta_{f} < {{\tilde{\delta}_{v}}}| V_{2},{{\delta_{c}}}\right)&=&P\left(V_{1}< V_{2}/2
{{\tilde{\delta}_{v}}}|V_{2},{{\delta_{c}}}\right)\nonumber\\
&=&\int^{S_{h}=S_{2}({V_{2}}/2)}_{S_{2}}\left(\frac{V_{2}}
{V_{1}}\right)f_{S_{1}}\left(S_{1},\tilde{\delta}_{v}|S_{2},\delta_{c}\right)d S_{1},
\label{probhalovoidca}
\end{eqnarray}
\noindent
where ${V_{2}}/{V_{1}}$ is the weighting factor and $S_{h}=S({V_{2}}/2)$. Volume ratios ${V_{2}}/{V_{1}}$ in
the probability density function $f_{S_{1}}\left(S_{1},\tilde{\delta}_{v}|S_{2}, \delta_{c}\right)$
define interval $\left[S_{h}, {S_{2}}\right]$. Therefore, in terms of self-similar models, the exact solutions
of the probability function for $n=0, -1.5, -2$ are derived for the spherical model (see
\ref{appendix:analyticsolutionsprob}). In addition, probability (\ref{probhalovoidca}) defines the distribution
of formation times as well,

\small
\begin{eqnarray}
P\left( >
\delta_{f}\right)=\int^{1}_{0}\frac{1}{\sqrt{2\pi}}\left[\tilde{S}\left(2^{\alpha}-1\right)+1\right]^{1/\alpha}\frac{\delta_{f}}{\tilde{S}^{3/2}}
\exp\left(-\frac{1}{2}\frac{\delta^2_{f}}{\tilde{S}}\right)\exp\left(-\frac{2}{9}\frac{{\tilde{S}}}{\delta^2_{f}}-\frac{32}{81}\frac{{\tilde{S}^2}}
{\delta^4_{f}}\right) d {\tilde{S}},
\label{reducedprobabilitya}
\end{eqnarray}
\normalsize
\noindent
by substituting the following transformations of $\tilde{S}$ and $\delta_{f}$, into the probability
distribution (\ref{probhalovoidca}),

\begin{eqnarray}
{\tilde{S}}\equiv\frac{S-S_{2}}{S_{h}-S},\phantom{a}
{\delta_{f}}\equiv\frac{\delta-{\delta_{c}}}{\sqrt{S_{h}-S_{2}}}.
\label{newparametersa}
\end{eqnarray}
\noindent
For embedded voids, obtaining model-dependent analytical solutions is not possible. Therefore, the numerical
solution of the merging probability (\ref{reducedprobabilitya}) is investigated. Fig.~\ref{fig:smallProbmerger}
indicates the probability densities in terms of barrier height $\delta_{f}$ which is approximately the formation
redshift $z_{f}$. This depicts the fact that the models have similar distributions and their difference is so
small that it can be negligible.

\section{Summary and Discussion}
In this study, the analytical growing void merging tree model of \cite{eszrarussell} based on the one-barrier
excursion set/EPS formalism is extended to the two-barrier
excursion set by taking into account collapsing subvoids as well as merging ones. To do this, excursion
Brownian walk is constructed from the bridge-meander random walk of the void in cloud process by using a method
given by \cite{philippe}. This method allows one to use the two-barrier fraction function of void populations
given by \cite{sw}. Then the merging tree algorithm of overdense regions derived by LC93 is modified into a void
merging one in terms of the spherical model. A new parameter called the barrier height ratio $\gamma$ is defined, and this parameter is transformed into a new barrier height ratio $\tilde{\gamma}$ after the void and cloud process is treated to the cloud in cloud process. These barrier height ratios have key importance since they define the collapse and merging behaviors of the distribution function of the void population. Assuming spherical voids and binary merging which gives the relation of scales depending on the self-similar model $S_{2}\approx 2^{\alpha}S_{1}$ and the barrier ratio $\tilde{\gamma}=0.25$, the void merging algorithm for void populations is obtained in terms of the two-barrier EPS formalism. According to this, the following results are found,

\begin{enumerate}
\item The size distribution of the embedded void populations are obtained by using the two-barrier EPS
formalism. Making the distinction between growing and collapsing voids based on a choice of the barrier
height ratio $\tilde{\gamma}$ (or ${\gamma}$) and the redshift constraints, it is shown that the size
distribution of self-similar models present two cutoff values indicating the largest and smallest embedded
voids at a given redshift. The cutoff sizes have the same lowest probabilities. In addition,
the peak of a size distribution at a given redshift indicates that this size of embedded voids dominate the
given epoch. The embedded voids with the highest distribution rate become smaller towards high redshifts.
This behavior corresponds to the fact that relatively small size embedded voids dominate the void
distribution of the Universe at high redshifts compared to low redshift values. This result agrees with
hierarchical scenarios \citep{sw,2013MNRAS.434.1192R,eszrarussell}. On the other hand, observing of the
largest spherical embedded voids is unlikely due to their low probability values in the self-similar models.
This theoretical result is in agreement with the observational and numerical study of
\cite{Tavasoli2013}. In Millennium I simulation mock data, \cite{Tavasoli2013} find that in
the simulation, large voids are less abundant.

However, \cite{sutter2013} point out the importance of the density of voids. \cite{sutter2013} show that
voids become larger, spherical and also small voids disappear when the tracer density is reduced. This
is particulary interesting since void density is chosen as in the linear theory $\delta_{v}=-2.81$ in this
study, based on \cite{sw}. Following up on the study of \cite{sutter2013}, the extended void merging
algorithm should be tested for different underdensities based on the observational and numerical studies, in
order to see how the embedded void algorithm changes.

In the extended void merging algorithm it also is shown that the size of dominant embedded voids increases
towards low spectral index while the size range of minor voids increases. This indicates that small scales
are most likely dominated by larger size embedded voids than larger scales in the self-similar models.

The one-barrier excursion set formalism of growing voids of \cite{eszrarussell} shows slightly lower
probability of sizes for a given model compared to the size distribution of growing voids that is obtained
from the two-barrier EPS formalism by choosing $\delta_{c}\gg \left|\delta_{v}\right|$. However, the size
distribution of growing voids of the two-barrier excursion set obtained from equation
(\ref{probabilitys1size}) for the barrier height ratio $\gamma \gg 0.5$ indicates smaller size voids than
the size distribution function that is derived from the one-barrier excursion set formalism
\citep{eszrarussell}. This is obviously because of the second barrier value in the conditional fractional
function. In the one-barrier excursion set formalism, the second barrier in the conditional distribution
function is a merging barrier $\delta_{v}$ while in the two-barrier approach the second barrier in the
conditional distribution function is the collapse barrier.

Fig. \ref{fig:CH4CH3} explicitly represents this distinction between growing and embedded void populations
with different spectral index. As was aforementioned, the void size distribution range increases with
decreasing spectral index. As is seen in Fig. \ref{fig:CH4CH3}, this behavior is more striking in a growing
void size distribution than in embedded ones.

\begin{figure}
\centering
\includegraphics[width=0.6\textwidth]{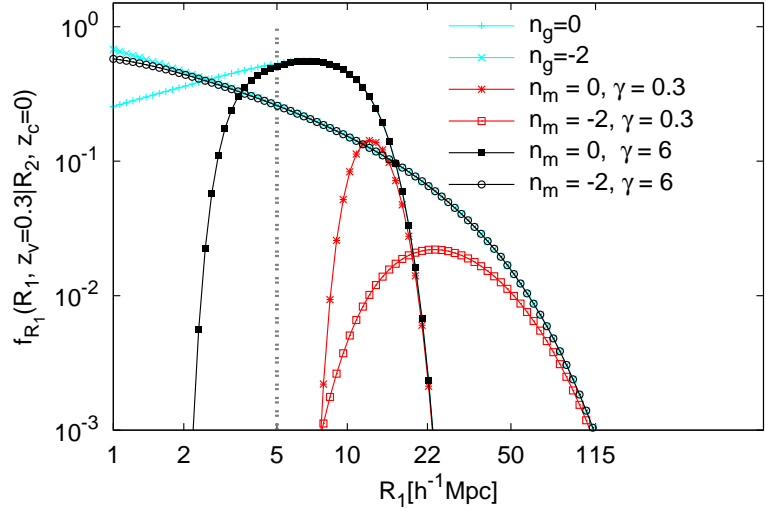}
\caption{Comparison of the size distribution of void populations in terms of self-similar models with
spectral indices $n_{g}=0$ and $-2$ for growing and $n_{m}=0$, and $-2$ for embedded/minor voids. In both
models of size distribution of embedded/minor voids start merging events at $z_{v}=0.3$ and collapse at
present-day redshift $z_{c}=0$. Growing voids only show merging behavior since the collapse barrier is moved
towards very high redshifts, in which the barrier ratio is chosen as $\gamma=0.6$ and the merging redshift
$z_{v}=0$.}
\label{fig:CH4CH3}
\end{figure}

\item It is found that the merger rates of embedded voids that are derived in
this paper and growing void populations of \cite{eszrarussell} are the same. Therefore, it can be concluded
that embedded and growing void populations have the same merger characteristics. However, at some point
embedded voids are destined to vanish or will be squeezed by the overdense regions while growing voids
continuously merge. As an extension of this result and \cite{eszrarussell}, it may be possible to generalize
that embedded/minor and growing voids have the same merger rate characteristic as dark matter
haloes.

\item The survival probability of the void population is defined and
formulated analytically in terms of the two-barrier excursion set formalism. The two-barrier survival
probability of the embedded/minor voids is named as the collapse probability. Then, a distinction between
the
survival probability of growing voids is made based on the one-barrier formalism and the collapse
probability of the two-barrier formalism which has contributions from both the growing and the embedded
void populations. According to this, while the survival probability of the growing void population is
defined by the distribution of voids that will continue merging after reaching double their
size, the collapse probability of embedded minor voids is defined as the probability of voids that merge
until the collapse barrier/redshift.

In addition, it is shown that the collapse probability of embedded voids, that will vanish under collapse
regions, indicates small void size ranges. This is due to the fact that large size/growing voids are not
strongly affected by collapse regions \citep{sw,D'Aloisio2007,eszrarussell,Ceccarelli}. Here,
model-dependent upper limits on the radius of the embedded voids that will collapse with the highest
probability $P_{coll}=1$ are obtained; these are $\leq 4.5$ $h^{-1}$Mpc, $\leq 2.5$ $h^{-1}$Mpc and $\leq
1.5$ $h^{-1}$Mpc for self-similar models with the index $n=0$, $-1$, $-1.5$ respectively. Besides this,
lower limits on the radius of embedded voids are calculated. According to this, all voids with radii above
these limits in the related model continuously grow/merge without collapsing. These lower limits are $\geq
20$ $h^{-1}$Mpc, $\geq 35$ $h^{-1}$Mpc and $\geq 65$ $h^{-1}$Mpc in the self-similar models with index
$n=0$, $-1$, $-1.5$.

\item An analytical description of the collapse failure rate of the embedded voids is obtained. The collapse
failure rate is defined as the change of probability of a void that will not collapse at a given redshift
interval or size interval. According to this, the collapse failure rate of embedded voids is given by the
negative derivative of the collapse probability distribution in terms of the collapse barrier/redshift
$\delta_{c}\approx z_{c}$.

Consequently, it is shown that $5$ $h^{-1}$Mpc size embedded voids around $z_{v}=1.4$ have more chance to collapse compared to large size $>5$ $h^{-1}$Mpc embedded voids. In the self-similar models, this tendency of collapsing of small size embedded voids reduces towards lower spectral indices. This may indicate that the
collapse failure, or risk of an embedded void not collapsing, increases with increasing size. Given the definition of growing voids which do not collapse, this is an expected result and
agrees with \cite{sw,eszrarussell,Ceccarelli}.

The failure rates of growing and embedded voids represent completely different behavior. In
Fig.~\ref{fig:failurecomparison}, the comparison of the two different failure rate behaviors is given.
According to this, the failure rate of
growing voids measures the change in the probability distribution of a void which starts its evolution at
merging barrier $\delta_{v_{1}}\approx z_{v_{1}}$ and later on incorporates its double
size at a merging barrier $\delta_{v_{2}}\approx z_{v_{2}}$. The failure rate of embedded voids
measures the change in the probability distribution of a void that starts merging at
$\tilde{\delta_{v}}\approx z_{v}$ which then incorporates its double size and collapses at a collapse
barrier $\delta_{c}\approx z_{c}$. Also, as
is seen in Fig.~\ref{fig:failurecomparison}, the shape of the collapse failure rate of the embedded voids is
related to the lognormal survival model, while only growing voids show the shape of another special failure
rate called the Weibull survival model \citep{Kleinbaum} as is shown by \cite{eszrarussell}. In addition,
the failure rates for embedded voids
increase up to a turnaround point at a certain radius at a given redshift. After these particular radii are
reached, the failure rate starts decreasing. These turnaround void sizes in different models indicate the
voids that have the most chance to survive without collapsing. Then the turn around points of the failure
rates give the size value of dominant embedded voids at a given redshift in the volume distribution.

\begin{figure}
\begin{tabular}{cc}
\includegraphics[width=0.45\textwidth]{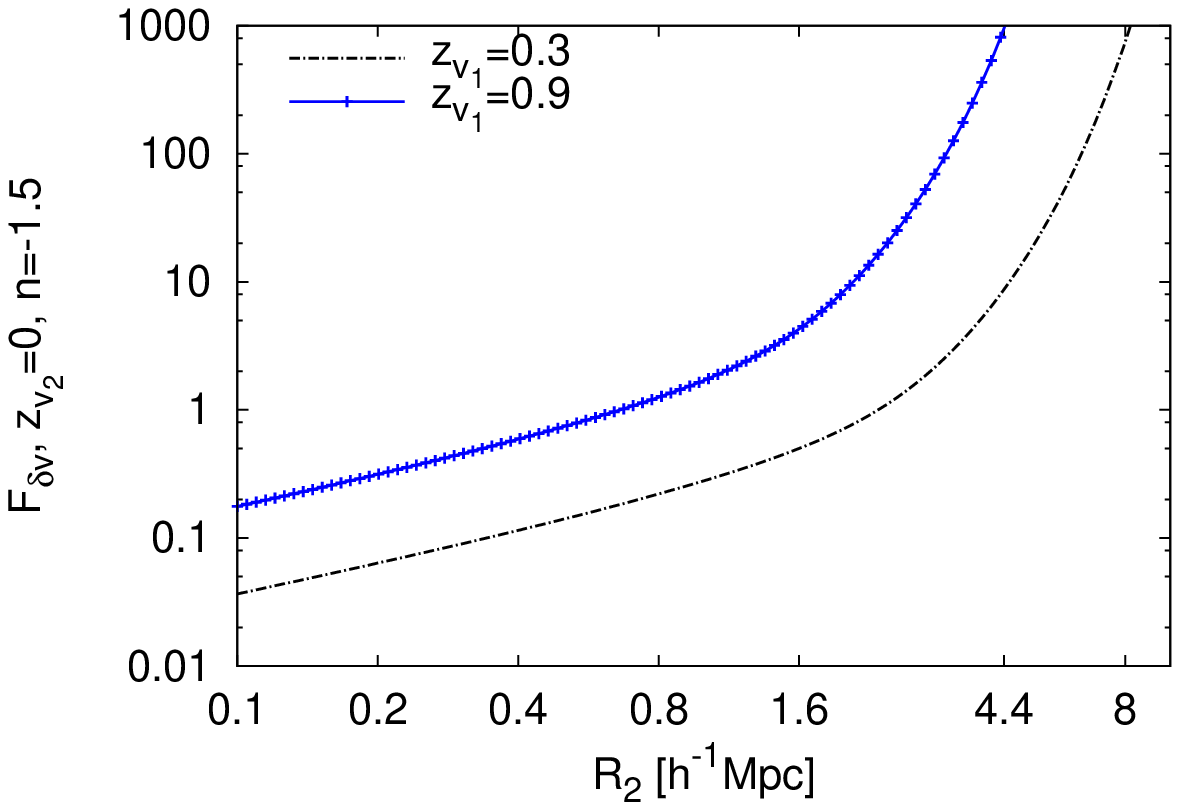}
\includegraphics[width=0.45\textwidth]{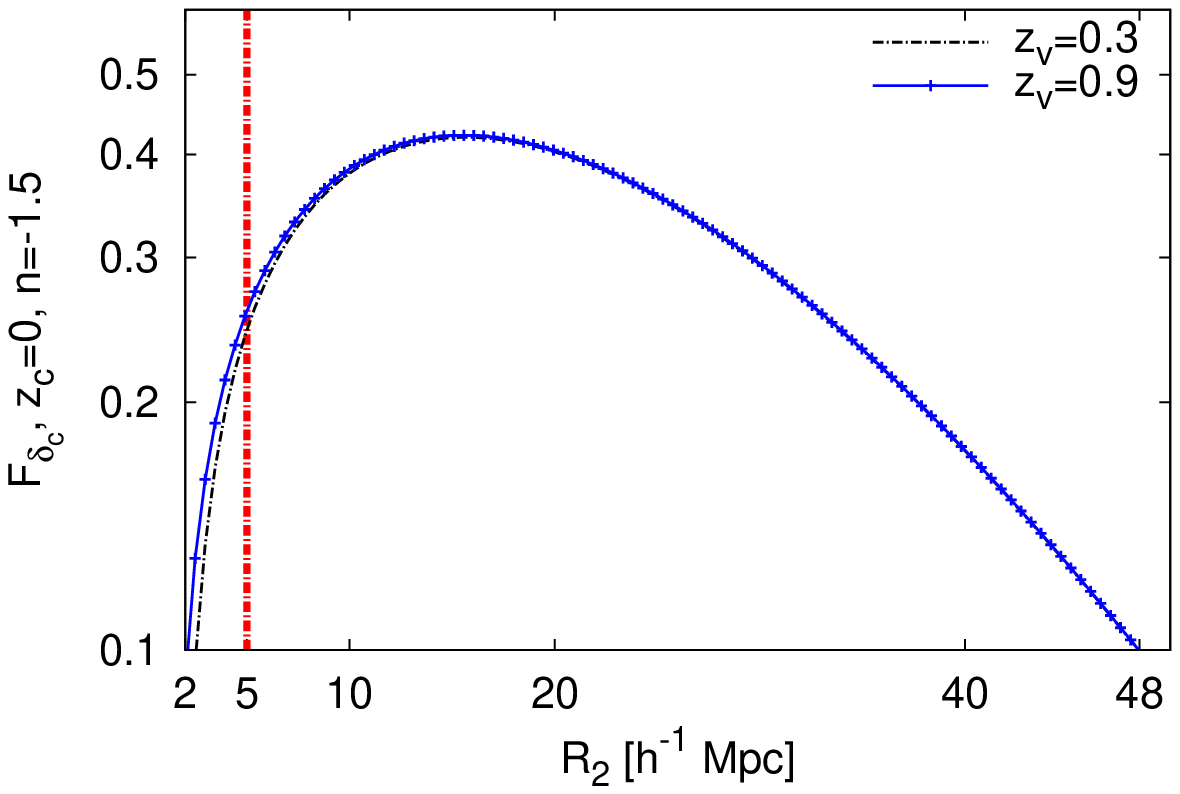}
\end{tabular}
\caption{Comparison of the failure rates of growing (left) and embedded (right) voids for the self-similar
model $n=$ $-1.5$. These panels represent the failure rates of voids in terms of the one-barrier (left) and
the
two-barrier (right) EPS formalisms.}
\label{fig:failurecomparison}
\end{figure}

\item It is shown that although there are analytical solutions for the probability
functions of void formation times in the one-barrier approach of \cite{eszrarussell}, there are no
analytical solutions for the expected void distribution in terms of the formation barrier based on the
two-barrier formalism. Fig. \ref{fig:probcomparison} shows the comparison of the two- and one-barrier EPS formalism of
void formation probability distributions for three different self-similar models. The two-barrier approach
provides a better solution of the hierarchical void evolution than the
one-barrier EPS formalism. The reason for this is the probability function of void formation times in the
one-barrier EPS formalism has negative values for small formation barrier $\delta_{f}$ or redshift $z_{f}$,
values which is not acceptable due to the definition of the probability function. However the two-barrier
approach does not have this problem. Nevertheless, the formation time distribution of the two-barrier
approach is lower than the one-barrier one. This indicates that there are more void structures in the
one-barrier EPS formalism than the two-barrier formalism.

\begin{figure}
\centering
\begin{tabular}{cc}
\includegraphics[width=0.45\textwidth]{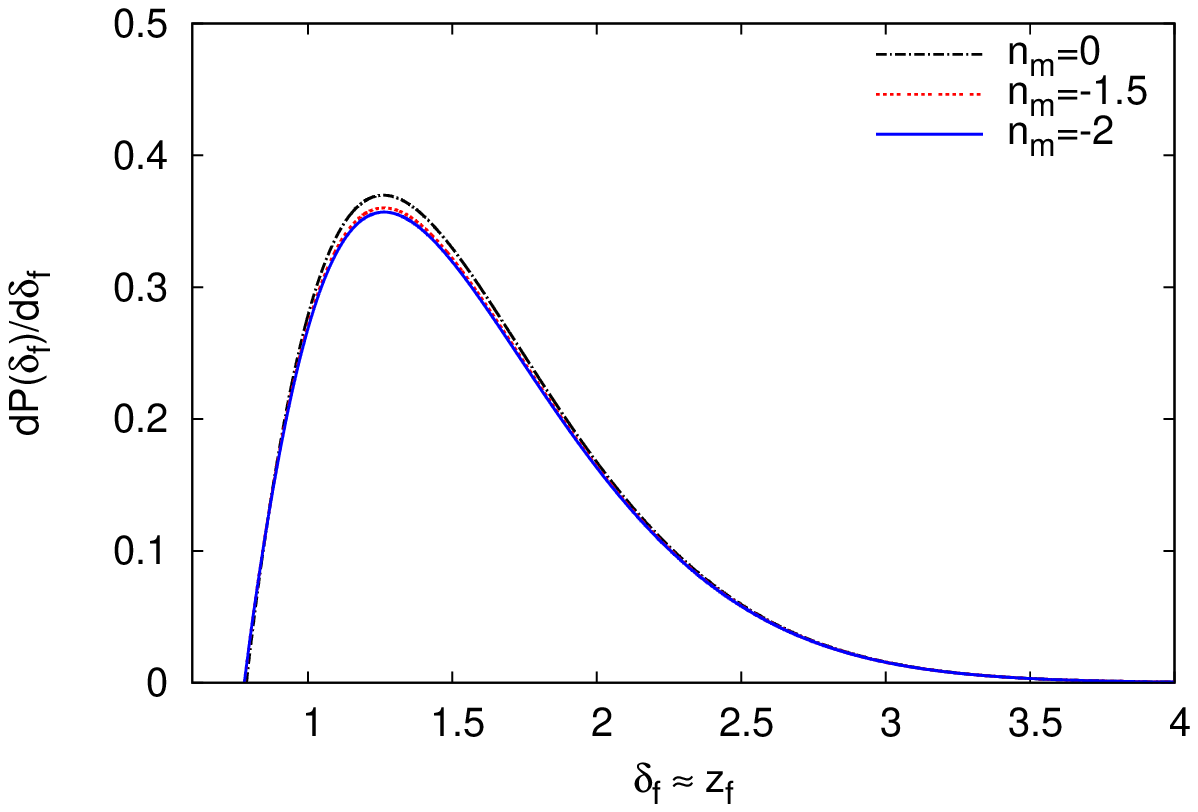}
\includegraphics[width=0.45\textwidth]{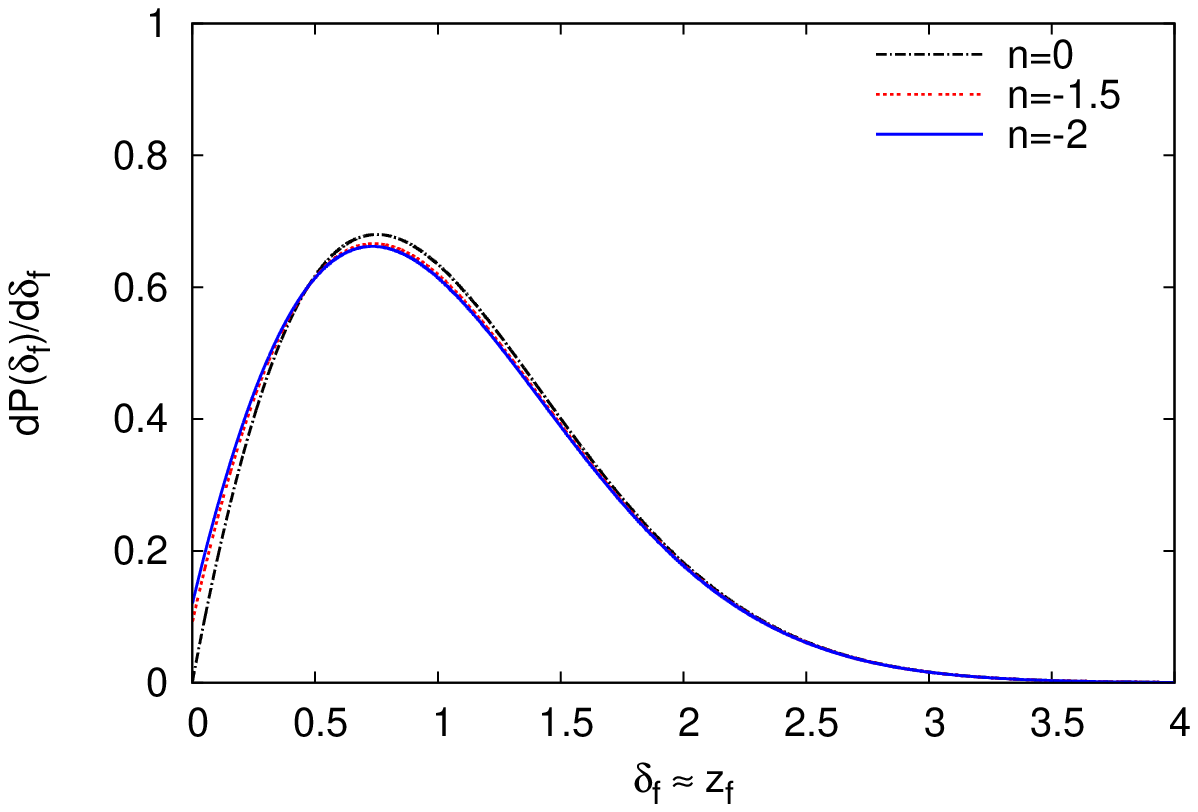}
\end{tabular}
\caption{Comparison of numerical solutions of the probability distribution of formation times of voids in
the two-barrier (left) and one-barrier (right) EPS formalisms in the self-similar models with spectral
index $n=0$, $-1.5$, $-2$. In each plot, when the index decreases the
probability distribution decreases as well. As can be seen, the difference between self-similar models in
each plot is small.}
\label{fig:probcomparison}
\end{figure}
\end{enumerate}
As an extension of the one-barrier void merging tree algorithm of \cite{eszrarussell}, here merging,
survival, failure and formation time distributions of voids, that are embedded in overdense regions, are
obtained. Also, this study only focuses on the self-similar models, which are approximated by a
$\Lambda$CDM model by following \cite{sw}. Unlike
the one-barrier void merging tree algorithm, the algorithm of the two-barrier
void merging tree has complex mathematical derivations. Although its mathematical derivation is complex, it can
be seen that its structure leads to the extension of the one-barrier void merging algorithm of
\cite{eszrarussell}.

The main goal of this paper is to provide a treatment of the void in cloud problem of the EPS formalism, and
constructing an embedded void merging algorithm based on this new method. That is why, this extended void
merging algorithm, focused on only the self-similar models in the case of spherical voids in the EdS
Universe. As a result of this, collapse and expansion of a void are described in terms of linear theory, and
the non-linear effects for smaller void sizes ($< 5$ $h^{-1}$Mpc) in the void merging algorithm is neglected.
However, it should be pointed out that when voids evolve in time, they tend to be
more spherical \citep{centme,fuji,be85} and this tendency is described by the Bubble Theorem of \cite{icke}.
Moreover, numerical studies show this tendency \citep{Tavasoli2013} even for small scale subvoids
\citep{2013MNRAS.434.1192R}. On the other hand, \cite{Jennings2013MNRAS} obtain the number density of
spherical voids using a linearly extrapolated density $\delta_{v}=-2.7$ in the dark matter distribution from N-body simulations in a
$\Lambda$CDM cosmology. They show that the abundance of voids at redshift $z = 0$ does not match the abundance
given by \cite{sw}. \cite{Jennings2013MNRAS} stress the fact that the EPS formalism of \cite{sw} does not
take into account merging of voids as they expand. Note that here we investigate only the self-similar spectra
to construct a void merging algorithm which is based on the LC93 dark matter merging algorithm. LC93 merging
tree algorithm of dark matter haloes is limited to the self-similar models and provides an approximated dark
matter solutions. That is why, investigating the extended void merging algorithm in different $\Lambda$CDM
models by using different $CDM$ power spectra and taking into account non-linear effects on small size voids
seem essential in order to achieve a proper full understanding of the formation and dynamics of the Cosmic Web
from a void based structure formation. As a result, the extended and growing void merging algorithms can be
tested and then may provide a realistic framework. Therefore, the $\Lambda$CDM models and non-linear effects on
small size voids in these void merging algorithms will be discussed in detail in a follow-up paper.

\section*{Acknowledgments}
Russell would like to sincerely thank the referee, Dr. Spyros Basilakos for his insightful comments and
suggestions which improved this study constructively.
\bibliographystyle{mn2e}
\bibliography{bib}

\begin{thebibliography}{}

\bibitem[\protect\citeauthoryear{{Aragon-Calvo} \& {Szalay}}{{Aragon-Calvo} \&
  {Szalay}}{2013}]{2013MNRAS.428.3409A}
{Aragon-Calvo} M.~A.,  {Szalay} A.~S.,  2013, \mnras, 428, 3409

\bibitem[\protect\citeauthoryear{{Benson}, {Hoyle}, {Torres} \&
  {Vogeley}}{{Benson} et~al.}{2003}]{benson}
{Benson} A.~J.,  {Hoyle} F.,  {Torres} F.,    {Vogeley} M.~S.,  2003, \mnras,
  340, 160

\bibitem[\protect\citeauthoryear{{Bertschinger}}{{Bertschinger}}{1985}]{be85}
{Bertschinger} E.,  1985, \apjs, 58, 1

\bibitem[\protect\citeauthoryear{{Beygu}, {Kreckel}, {van de Weygaert}, {van
  der Hulst} \& {van Gorkom}}{{Beygu} et~al.}{2013}]{Beygu2013}
{Beygu} B.,  {Kreckel} K.,  {van de Weygaert} R.,  {van der Hulst} J.~M.,
  {van Gorkom} J.~H.,  2013, \aj, 145, 120

\bibitem[\protect\citeauthoryear{{Blumenthal}, {da Costa}, {Goldwirth}, {Lecar}
  \& {Piran}}{{Blumenthal} et~al.}{1992}]{blumenthal}
{Blumenthal} G.~R.,  {da Costa} L.~N.,  {Goldwirth} D.~S.,  {Lecar} M.,
  {Piran} T.,  1992, \apj, 388, 234

\bibitem[\protect\citeauthoryear{{Bond}, {Cole}, {Efstathiou} \&
  {Kaiser}}{{Bond} et~al.}{1991}]{bo}
{Bond} J.~R.,  {Cole} S.,  {Efstathiou} G.,    {Kaiser} N.,  1991, \apj, 379,
  440

\bibitem[\protect\citeauthoryear{{Bond}, {Kofman} \& {Pogosyan}}{{Bond}
  et~al.}{1996}]{bond}
{Bond} J.~R.,  {Kofman} L.,    {Pogosyan} D.,  1996, \nat, 380, 603

\bibitem[\protect\citeauthoryear{{Bos}, {van de Weygaert}, {Dolag} \&
  {Pettorino}}{{Bos} et~al.}{2012}]{Bos2012}
{Bos} E.~G.~P.,  {van de Weygaert} R.,  {Dolag} K.,    {Pettorino} V.,  2012,
  \mnras, 426, 440

\bibitem[\protect\citeauthoryear{{Brookfield}, {van de Bruck} \&
  {Hall}}{{Brookfield} et~al.}{2008}]{Brookfield2008}
{Brookfield} A.~W.,  {van de Bruck} C.,    {Hall} L.~M.~H.,  2008, \prd, 77,
  043006

\bibitem[\protect\citeauthoryear{{Ceccarelli}, {Paz}, {Lares}, {Padilla} \&
  {Lambas}}{{Ceccarelli} et~al.}{2013}]{Ceccarelli}
{Ceccarelli} L.,  {Paz} D.,  {Lares} M.,  {Padilla} N.,    {Lambas} D.~G.,
  2013, \mnras, 434, 1435

\bibitem[\protect\citeauthoryear{{Centrella} \& {Melott}}{{Centrella} \&
  {Melott}}{1983}]{centme}
{Centrella} J.,  {Melott} A.~L.,  1983, \nat, 305, 196

\bibitem[\protect\citeauthoryear{{Clampitt}, {Cai} \& {Li}}{{Clampitt}
  et~al.}{2012}]{Clampitt2012}
{Clampitt} J.,  {Cai} Y.-C.,    {Li} B.,  2012, ArXiv e-prints

\bibitem[\protect\citeauthoryear{{Colberg} \& {et al.}}{{Colberg} \& {et
  al.}}{2008}]{2008MNRAS.387..933C}
{Colberg} J.~M.,  {et al.} 2008, \mnras, 387, 933

\bibitem[\protect\citeauthoryear{{Colberg}, {Sheth}, {Diaferio}, {Gao} \&
  {Yoshida}}{{Colberg} et~al.}{2005}]{colb}
{Colberg} J.~M.,  {Sheth} R.~K.,  {Diaferio} A.,  {Gao} L.,    {Yoshida} N.,
  2005, \mnras, 360, 216

\bibitem[\protect\citeauthoryear{{da Costa}, {Geller}, {Pellegrini}, {Latham},
  {Fairall}, {Marzke}, {Willmer}, {Huchra}, {Calderon}, {Ramella} \&
  {Kurtz}}{{da Costa} et~al.}{1994}]{dac}
{da Costa} L.~N.,  {Geller} M.~J.,  {Pellegrini} P.~S.,  {Latham} D.~W.,
  {Fairall} A.~P.,  {Marzke} R.~O.,  {Willmer} C.~N.~A.,  {Huchra} J.~P.,
  {Calderon} J.~H.,  {Ramella} M.,    {Kurtz} M.~J.,  1994, \apjl, 424, L1

\bibitem[\protect\citeauthoryear{{D'Aloisio} \& {Furlanetto}}{{D'Aloisio} \&
  {Furlanetto}}{2007}]{D'Aloisio2007}
{D'Aloisio} A.,  {Furlanetto} S.~R.,  2007, \mnras, 382, 860

\bibitem[\protect\citeauthoryear{{D'Amico}, {Musso}, {Nore{\~n}a} \&
  {Paranjape}}{{D'Amico} et~al.}{2011}]{Damico2011}
{D'Amico} G.,  {Musso} M.,  {Nore{\~n}a} J.,    {Paranjape} A.,  2011, \prd,
  83, 023521

\bibitem[\protect\citeauthoryear{{Dubinski}, {da Costa}, {Goldwirth}, {Lecar}
  \& {Piran}}{{Dubinski} et~al.}{1993}]{dubinski1993}
{Dubinski} J.,  {da Costa} L.~N.,  {Goldwirth} D.~S.,  {Lecar} M.,    {Piran}
  T.,  1993, in {Chincarini} G.~L.,  {Iovino} A.,  {Maccacaro} T.,   {Maccagni}
  D.,  eds, Observational Cosmology Vol.~51 of Astronomical Society of the
  Pacific Conference Series, {Void Evolution and the Large-Scale Structure}.
p.~188

\bibitem[\protect\citeauthoryear{{Einasto}, {Tago}, {Jaaniste}, {Einasto} \&
  {Andernach}}{{Einasto} et~al.}{1997}]{ei}
{Einasto} M.,  {Tago} E.,  {Jaaniste} J.,  {Einasto} J.,    {Andernach} H.,
  1997, \aaps, 123, 119

\bibitem[\protect\citeauthoryear{{El-Ad}, {Piran} \& {Dacosta}}{{El-Ad}
  et~al.}{1997}]{El-Ad1997}
{El-Ad} H.,  {Piran} T.,    {Dacosta} L.~N.,  1997, \mnras, 287, 790

\bibitem[\protect\citeauthoryear{{Farrar} \& {Peebles}}{{Farrar} \&
  {Peebles}}{2004}]{FarrarP2004}
{Farrar} G.~R.,  {Peebles} P.~J.~E.,  2004, \apj, 604, 1

\bibitem[\protect\citeauthoryear{{Farrar} \& {Rosen}}{{Farrar} \&
  {Rosen}}{2007}]{Farrar2007}
{Farrar} G.~R.,  {Rosen} R.~A.,  2007, Physical Review Letters, 98, 171302

\bibitem[\protect\citeauthoryear{{Fisher}, {Huchra}, {Strauss}, {Davis},
  {Yahil} \& {Schlegel}}{{Fisher} et~al.}{1995}]{fisher}
{Fisher} K.~B.,  {Huchra} J.~P.,  {Strauss} M.~A.,  {Davis} M.,  {Yahil} A.,
  {Schlegel} D.,  1995, \apjs, 100, 69

\bibitem[\protect\citeauthoryear{{Fitzsimmons}, {Pitman} \&
  {Yor}}{{Fitzsimmons} et~al.}{1993}]{Fpitman93}
{Fitzsimmons} P.,  {Pitman} J.,    {Yor} M.,  1993, in {Cinler} E.,  {Chung}
  K.~L.,   J. S.~M.,  eds, Seminar on Stochastic Processes {Markovian bridges:
  construction, Palm interpretation, and splicing)}.
pp 101--134

\bibitem[\protect\citeauthoryear{{Fujimoto}}{{Fujimoto}}{1983}]{fuji}
{Fujimoto} M.,  1983, \pasj, 35, 159

\bibitem[\protect\citeauthoryear{{Furlanetto} \& {Piran}}{{Furlanetto} \&
  {Piran}}{2006}]{piran2006}
{Furlanetto} S.~R.,  {Piran} T.,  2006, \mnras, 366, 467

\bibitem[\protect\citeauthoryear{{Geller} \& {Huchra}}{{Geller} \&
  {Huchra}}{1989}]{gehu}
{Geller} M.~J.,  {Huchra} J.~P.,  1989, Science, 246, 897

\bibitem[\protect\citeauthoryear{{Gottl{\"o}ber}, {{\L}okas}, {Klypin} \&
  {Hoffman}}{{Gottl{\"o}ber} et~al.}{2003}]{gott}
{Gottl{\"o}ber} S.,  {{\L}okas} E.~L.,  {Klypin} A.,    {Hoffman} Y.,  2003,
  \mnras, 344, 715

\bibitem[\protect\citeauthoryear{{Gubser} \& {Peebles}}{{Gubser} \&
  {Peebles}}{2004a}]{GubserP2004b}
{Gubser} S.~S.,  {Peebles} P.~J.~E.,  2004a, \prd, 70, 123511

\bibitem[\protect\citeauthoryear{{Gubser} \& {Peebles}}{{Gubser} \&
  {Peebles}}{2004b}]{GubserP2004}
{Gubser} S.~S.,  {Peebles} P.~J.~E.,  2004b, \prd, 70, 123510

\bibitem[\protect\citeauthoryear{{Hellwing} \& {Juszkiewicz}}{{Hellwing} \&
  {Juszkiewicz}}{2009}]{Hellwing2009}
{Hellwing} W.~A.,  {Juszkiewicz} R.,  2009, \prd, 80, 083522

\bibitem[\protect\citeauthoryear{{Hoyle} \& {Vogeley}}{{Hoyle} \&
  {Vogeley}}{2002}]{Hoyle02}
{Hoyle} F.,  {Vogeley} M.~S.,  2002, \apj, 566, 641

\bibitem[\protect\citeauthoryear{{Hoyle} \& {Vogeley}}{{Hoyle} \&
  {Vogeley}}{2004}]{hoylevogley2004}
{Hoyle} F.,  {Vogeley} M.~S.,  2004, \apj, 607, 751

\bibitem[\protect\citeauthoryear{{Icke}}{{Icke}}{1984}]{icke}
{Icke} V.,  1984, \mnras, 206, 1P

\bibitem[\protect\citeauthoryear{{J{\~o}eveer}, {Einasto} \&
  {Tago}}{{J{\~o}eveer} et~al.}{1978}]{joveer}
{J{\~o}eveer} M.,  {Einasto} J.,    {Tago} E.,  1978, \mnras, 185, 357

\bibitem[\protect\citeauthoryear{{Jennings}, {Li} \& {Hu}}{{Jennings}
  et~al.}{2013}]{Jennings2013MNRAS}
{Jennings} E.,  {Li} Y.,    {Hu} W.,  2013, \mnras, 434, 2167

\bibitem[\protect\citeauthoryear{{Jones} \& {et al.}}{{Jones} \& {et
  al.}}{2004}]{2004MNRAS.355..747J}
{Jones} D.~H.,  {et al.} 2004, \mnras, 355, 747

\bibitem[\protect\citeauthoryear{{Karachentsev}, {Karachentseva}, {Huchtmeier}
  \& {Makarov}}{{Karachentsev} et~al.}{2004}]{Karachentsev2004}
{Karachentsev} I.~D.,  {Karachentseva} V.~E.,  {Huchtmeier} W.~K.,    {Makarov}
  D.~I.,  2004, \aj, 127, 2031

\bibitem[\protect\citeauthoryear{{Kirshner}, {Oemler} Jr., {Schechter} \&
  {Shectman}}{{Kirshner} et~al.}{1981}]{kir}
{Kirshner} R.~P.,  {Oemler} Jr. A.,  {Schechter} P.~L.,    {Shectman} S.~A.,
  1981, \apjl, 248, L57

\bibitem[\protect\citeauthoryear{{Kleinbaum} \& {Klein}}{{Kleinbaum} \&
  {Klein}}{2011}]{Kleinbaum}
{Kleinbaum} D.~G.,  {Klein} M.,  2011, {Survival Analysis: A Self-Learning Text
  }

\bibitem[\protect\citeauthoryear{{Kraan-Korteweg}, {Shafi}, {Koribalski},
  {Staveley-Smith}, {Buckland}, {Henning} \& {Fairall}}{{Kraan-Korteweg}
  et~al.}{2008}]{kraan}
{Kraan-Korteweg} R.~C.,  {Shafi} N.,  {Koribalski} B.~S.,  {Staveley-Smith} L.,
   {Buckland} P.,  {Henning} P.~A.,    {Fairall} A.~P.,  2008, {Outlining the
  Local Void with the Parkes HI ZOA and Galactic Bulge Surveys}.
p.~13

\bibitem[\protect\citeauthoryear{{Kreckel}, {Joung} \& {Cen}}{{Kreckel}
  et~al.}{2011}]{kreckel2011}
{Kreckel} K.,  {Joung} M.~R.,    {Cen} R.,  2011, \apj, 735, 132

\bibitem[\protect\citeauthoryear{{Kreckel}, {Platen}, {Arag{\'o}n-Calvo}, {van
  Gorkom}, {van de Weygaert}, {van der Hulst} \& {Beygu}}{{Kreckel}
  et~al.}{2012}]{Kreckel2012AJ}
{Kreckel} K.,  {Platen} E.,  {Arag{\'o}n-Calvo} M.~A.,  {van Gorkom} J.~H.,
  {van de Weygaert} R.,  {van der Hulst} J.~M.,    {Beygu} B.,  2012, \aj, 144,
  16

\bibitem[\protect\citeauthoryear{{Lacey} \& {Cole}}{{Lacey} \&
  {Cole}}{1993}]{lace}
{Lacey} C.,  {Cole} S.,  1993, \mnras, 262, 627

\bibitem[\protect\citeauthoryear{{Lavaux} \& {Wandelt}}{{Lavaux} \&
  {Wandelt}}{2010}]{Lavaux2010}
{Lavaux} G.,  {Wandelt} B.~D.,  2010, \mnras, 403, 1392

\bibitem[\protect\citeauthoryear{{Lavaux} \& {Wandelt}}{{Lavaux} \&
  {Wandelt}}{2012}]{Lavaux2012}
{Lavaux} G.,  {Wandelt} B.~D.,  2012, \apj, 754, 109

\bibitem[\protect\citeauthoryear{Marchal}{Marchal}{2003}]{philippe}
Marchal P.,  2003, in Banderier C.,  Krattenthaler C.,  eds, Discrete Random
  Walks, DRW'03 Vol.~AC of DMTCS Proceedings, {Constructing a sequence of
  random walks strongly converging to Brownian motion}.
Discrete Mathematics and Theoretical Computer Science, pp 181--190

\bibitem[\protect\citeauthoryear{{Mathis} \& {White}}{{Mathis} \&
  {White}}{2002}]{mathis}
{Mathis} H.,  {White} S.~D.~M.,  2002, \mnras, 337, 1193

\bibitem[\protect\citeauthoryear{{Maurogordato}, {Schaeffer} \& {da
  Costa}}{{Maurogordato} et~al.}{1992}]{mauro}
{Maurogordato} S.,  {Schaeffer} R.,    {da Costa} L.~N.,  1992, \apj, 390, 17

\bibitem[\protect\citeauthoryear{{Pan}, {Vogeley}, {Hoyle}, {Choi} \&
  {Park}}{{Pan} et~al.}{2012}]{Panvogeley2012}
{Pan} D.~C.,  {Vogeley} M.~S.,  {Hoyle} F.,  {Choi} Y.-Y.,    {Park} C.,  2012,
  \mnras, 421, 926

\bibitem[\protect\citeauthoryear{{Paranjape}, {Lam} \& {Sheth}}{{Paranjape}
  et~al.}{2012}]{paranjape}
{Paranjape} A.,  {Lam} T.~Y.,    {Sheth} R.~K.,  2012, \mnras, 420, 1648

\bibitem[\protect\citeauthoryear{{Pitman}}{{Pitman}}{1999}]{pitman99}
{Pitman} J.,  1999, ELECTRON. J. PROBAB, 4, 1

\bibitem[\protect\citeauthoryear{{Plionis} \& {Basilakos}}{{Plionis} \&
  {Basilakos}}{2002}]{plib}
{Plionis} M.,  {Basilakos} S.,  2002, \mnras, 330, 399

\bibitem[\protect\citeauthoryear{{Press} \& {Schechter}}{{Press} \&
  {Schechter}}{1974}]{presc}
{Press} W.~H.,  {Schechter} P.,  1974, \apj, 187, 425

\bibitem[\protect\citeauthoryear{{Regos} \& {Geller}}{{Regos} \&
  {Geller}}{1991}]{regge}
{Regos} E.,  {Geller} M.~J.,  1991, \apj, 377, 14

\bibitem[\protect\citeauthoryear{{Ricciardelli}, {Quilis} \&
  {Planelles}}{{Ricciardelli} et~al.}{2013}]{2013MNRAS.434.1192R}
{Ricciardelli} E.,  {Quilis} V.,    {Planelles} S.,  2013, \mnras, 434, 1192

\bibitem[\protect\citeauthoryear{{Russell}}{{Russell}}{2013}]{eszrarussell}
{Russell} E.,  2013, \mnras

\bibitem[\protect\citeauthoryear{{Sahni}, {Sathyaprakah} \&
  {Shandarin}}{{Sahni} et~al.}{1994}]{sahni1994}
{Sahni} V.,  {Sathyaprakah} B.~S.,    {Shandarin} S.~F.,  1994, \apj, 431, 20

\bibitem[\protect\citeauthoryear{{Saunders} \& {et al.}}{{Saunders} \& {et
  al.}}{2000}]{saunders}
{Saunders} W.,  {et al.} 2000, \mnras, 317, 55

\bibitem[\protect\citeauthoryear{{Shang}, {Crotts} \& {Haiman}}{{Shang}
  et~al.}{2007}]{Shang2007}
{Shang} C.,  {Crotts} A.,    {Haiman} Z.,  2007, \apj, 671, 136

\bibitem[\protect\citeauthoryear{{Shectman}, {Landy}, {Oemler}, {Tucker},
  {Lin}, {Kirshner} \& {Schechter}}{{Shectman} et~al.}{1996}]{shect}
{Shectman} S.~A.,  {Landy} S.~D.,  {Oemler} A.,  {Tucker} D.~L.,  {Lin} H.,
  {Kirshner} R.~P.,    {Schechter} P.~L.,  1996, \apj, 470, 172

\bibitem[\protect\citeauthoryear{{Sheth} \& {van de Weygaert}}{{Sheth} \& {van
  de Weygaert}}{2004}]{sw}
{Sheth} R.~K.,  {van de Weygaert} R.,  2004, \mnras, 350, 517

\bibitem[\protect\citeauthoryear{{Strauss}, {Davis}, {Yahil} \&
  {Huchra}}{{Strauss} et~al.}{1992}]{strauss}
{Strauss} M.~A.,  {Davis} M.,  {Yahil} A.,    {Huchra} J.~P.,  1992, \apj, 385,
  421

\bibitem[\protect\citeauthoryear{{Sutter}, {Lavaux}, {Wandelt}, {Hamaus},
  {Weinberg} \& {Warren}}{{Sutter} et~al.}{2013}]{sutter2013}
{Sutter} P.~M.,  {Lavaux} G.,  {Wandelt} B.~D.,  {Hamaus} N.,  {Weinberg}
  D.~H.,    {Warren} M.~S.,  2013, arXiv e-prints

\bibitem[\protect\citeauthoryear{{Sutter}, {Lavaux}, {Wandelt} \&
  {Weinberg}}{{Sutter} et~al.}{2012a}]{2012ApJ...761..187S}
{Sutter} P.~M.,  {Lavaux} G.,  {Wandelt} B.~D.,    {Weinberg} D.~H.,  2012a,
  \apj, 761, 187

\bibitem[\protect\citeauthoryear{{Sutter}, {Lavaux}, {Wandelt} \&
  {Weinberg}}{{Sutter} et~al.}{2012b}]{sutter}
{Sutter} P.~M.,  {Lavaux} G.,  {Wandelt} B.~D.,    {Weinberg} D.~H.,  2012b,
  \apj, 761, 44

\bibitem[\protect\citeauthoryear{{Tavasoli}, {Vasei} \& {Mohayaee}}{{Tavasoli}
  et~al.}{2013}]{Tavasoli2013}
{Tavasoli} S.,  {Vasei} K.,    {Mohayaee} R.,  2013, \aap, 553, A15

\bibitem[\protect\citeauthoryear{{Tikhonov}, {Gottl{\"o}ber}, {Yepes} \&
  {Hoffman}}{{Tikhonov} et~al.}{2009}]{tikhonovstefan2009}
{Tikhonov} A.~V.,  {Gottl{\"o}ber} S.,  {Yepes} G.,    {Hoffman} Y.,  2009,
  \mnras, 399, 1611

\bibitem[\protect\citeauthoryear{{Tikhonov} \& {Karachentsev}}{{Tikhonov} \&
  {Karachentsev}}{2006}]{tk2008}
{Tikhonov} A.~V.,  {Karachentsev} I.~D.,  2006, \apj, 653, 969

\bibitem[\protect\citeauthoryear{{Tikhonov} \& {Klypin}}{{Tikhonov} \&
  {Klypin}}{2009}]{TikhonovKlypin2009}
{Tikhonov} A.~V.,  {Klypin} A.,  2009, \mnras, 395, 1915

\bibitem[\protect\citeauthoryear{{Tinker} \& {Conroy}}{{Tinker} \&
  {Conroy}}{2009}]{tinker2009}
{Tinker} J.~L.,  {Conroy} C.,  2009, \apj, 691, 633

\bibitem[\protect\citeauthoryear{{Tully}, {Shaya}, {Karachentsev}, {Courtois},
  {Kocevski}, {Rizzi} \& {Peel}}{{Tully} et~al.}{2008}]{tully2008}
{Tully} R.~B.,  {Shaya} E.~J.,  {Karachentsev} I.~D.,  {Courtois} H.~M.,
  {Kocevski} D.~D.,  {Rizzi} L.,    {Peel} A.,  2008, \apj, 676, 184

\bibitem[\protect\citeauthoryear{{van de Weygaert}}{{van de
  Weygaert}}{1991}]{PhDWeygaert}
{van de Weygaert} M.~A.~M.,  1991, PhD thesis, Ph.~D.~thesis, University of
  Leiden (1991)

\bibitem[\protect\citeauthoryear{{van de Weygaert}}{{van de
  Weygaert}}{2003}]{weygaert2003}
{van de Weygaert} R.,  2003, {The cosmic foam: stochastic geometry and spatial
  clustering across the universe}.
pp 175--196

\bibitem[\protect\citeauthoryear{{van de Weygaert} \& {Bond}}{{van de Weygaert}
  \& {Bond}}{2008}]{vdWbond}
{van de Weygaert} R.,  {Bond} J.~R.,  2008, in {Plionis} M.,  {L{\'o}pez-Cruz}
  O.,   {Hughes} D.,  eds, A Pan-Chromatic View of Clusters of Galaxies and the
  Large-Scale Structure Vol.~740 of Lecture Notes in Physics, Berlin Springer
  Verlag, {Observations and Morphology of the Cosmic Web}.
p.~409

\bibitem[\protect\citeauthoryear{{van de Weygaert} \& {Platen}}{{van de
  Weygaert} \& {Platen}}{2011}]{rienerwin}
{van de Weygaert} R.,  {Platen} E.,  2011, Int. J. Mod. Phys., 1, 41

\bibitem[\protect\citeauthoryear{{van de Weygaert} \& {van Kampen}}{{van de
  Weygaert} \& {van Kampen}}{1993}]{wk}
{van de Weygaert} R.,  {van Kampen} E.,  1993, \mnras, 263, 481

\bibitem[\protect\citeauthoryear{{Vervaat}}{{Vervaat}}{1979}]{wim}
{Vervaat} W.,  1979, Annals of Probabability, 1, 143

\bibitem[\protect\citeauthoryear{{Viel}, {Colberg} \& {Kim}}{{Viel}
  et~al.}{2008}]{Viel2008}
{Viel} M.,  {Colberg} J.~M.,    {Kim} T.-S.,  2008, \mnras, 386, 1285

\end{thebibliography}
\appendix
\section{Analytical Derivations In the Extended Void Merging Tree}\label{appendix:analyticsolutionsprob}
The probability of an embedded void forming at $\delta_{f}\approx z_{f}$, that later on doubles its volume and
collapses at the barrier $\delta_{c}\approx z_{c}$, is given by,
\begin{eqnarray}
P\left(\delta_{f} < {{\tilde{\delta}_{v}}}| V_{2},{{\delta_{c}}}\right)&=&P\left(V_{1}< V_{2}/2
{{\tilde{\delta}_{v}}}|V_{2},{{\delta_{c}}}\right)\nonumber\\
&=&\int^{S_{h}=S_{2}({V_{2}}/2)}_{S_{2}}\left(\frac{V_{2}}
{V_{1}}\right)f_{S_{1}}\left(S_{1},\tilde{\delta}_{v}|S_{2},\delta_{c}\right)d S_{1},
\nonumber
\end{eqnarray}
\noindent
In self-similar models with index  $n=1$, $0$, $-1.5$, $-2$, the exact solutions of the probability function are
obtained for the spherical model as,
\small
\begin{eqnarray}
P_{n=1}&=&\tilde{\Omega}_{1}\frac{5}{6}\left[\text{erf}\left(\sqrt{\frac{5}{6}}\frac{\tilde{k}}{\sqrt{S_h}}\right)-
\text{erf}\left(\sqrt{\frac{5}{6}}\frac{\tilde{k}}{\sqrt{S_2}}\right)\right]\\
P_{n=0}&=&\tilde{\Omega}_{0}(S_{2})\frac{\tilde{k}}{S_{2}}\left[\text{erf}\left(\frac{\tilde{k}}{\sqrt{2}
\sqrt{S_h-S_2}}\right)
\left(\tilde{k}-\frac{S_2}{\tilde{k}}\right)-\sqrt{\frac{2}{\pi }} \sqrt{S_{h}-S_2}
e^{\frac{-\tilde{k}^2}{\left(S_h-S_2\right)}}\right]\\
P_{n=-1.5}&=&\tilde{\Omega}_{-1.5}(S_{2})\frac{\tilde{k}}{S^{2}_{2}}\left(\text{erf}\left(\frac{\tilde{k}}{\sqrt{2}
\sqrt{S_{h}-S_2}}\right)
\left(2 \tilde{k} S_{2}-\frac{\tilde{k}^3}{3} -\frac{S^{2}_{2}}
{\tilde{k}}\right)+\frac{1}{3}\sqrt{\frac{2}{\pi}}\sqrt{S_{h}-S_{2}}\left(S_{h}+5
S_{2}-\tilde{k}^2\right)e^{-
\frac{\tilde{k}^2}{2\left(S_{h}-S_{2}\right)}}\right)\\
P_{n=-2}&=&\tilde{\Omega}_{-2}(S_{2})\frac{\tilde{k}}{S^{3}_{2}}\left(\text{erf}\left(\frac{\tilde{k}}{\sqrt{2}
\sqrt{S_{h}-S_2}}\right)
\left(\frac{\tilde{k}^5}{15}-{\tilde{k}^3} S_{2}+3 \tilde{k} S^{2}_{2}-\frac{S^{3}_{2}}
{\tilde{k}}\right)+\frac{1}{15}\sqrt{\frac{2}{\pi}}\sqrt{S_{h}-S_{2}}\right.\times\nonumber\\
&&\left.\left(\tilde{k}^4-\tilde{k}^2 S_{h}+3 S^{2}_{h}+
S_{2}(9S_{h}-14\tilde{k}^2)+
33 S^2_{2}\right)e^{- \frac{\tilde{k}^2}{2\left(S_{h}-S_{2}\right)}}\right),
\label{probanalyticsolsmall}
\end{eqnarray}
\normalsize
\noindent
where $e$ stands for the exponential function while $\tilde{k}$ and $\tilde{\Omega}(S_{2})$ are defined as,
\begin{eqnarray}
\tilde{k}&\equiv&\tilde{\delta}_{v}-\delta_{c},\nonumber\\
\tilde{\Omega}_{1}(S_{2})&\equiv&
e^{-\frac{4}{3}\frac{S_{2}}{\delta^{2}_{v}}-14.222\frac{S^2_{2}}{\delta^{4}_{v}}},\phantom{a}
\tilde{\Omega}_{0}(S_{2})\equiv
e^{-\frac{4}{3}\frac{S_{2}}{\delta^{2}_{v}}-11.061\frac{S^2_{2}}{\delta^{4}_{v}}},\nonumber\\
\tilde{\Omega}_{-1.5}(S_{2})&\equiv &
e^{-\frac{4}{3}\frac{S_{2}}{\delta^{2}_{v}}-6.614\frac{S^2_{2}}{\delta^{4}_{v}}},
\phantom{a}\tilde{\Omega}_{-2}(S_{2})\equiv
e^{-\frac{4}{3}\frac{S_{2}}{\delta^{2}_{v}}-6.121\frac{S^2_{2}}{\delta^{4}_{v}}}.
\label{deltaparameters}
\end{eqnarray}
\noindent
\bsp
\label{lastpage}
\end{document}